The Evolution of Low Mass Helium Stars towards Supernova

Type I Explosion

Thesis submitted for the degree of

"Doctor of Philosophy"

By

Roni Waldman

Submitted to the Senate of the Hebrew University

December 2005

This work was carried out under the supervision of

Prof. Zalman Barkat

# A B S T R A C T

In this work we set up to explore the hypothesis, that helium stars in a certain mass range can evolve to a carbon core explosion similar to what is widely accepted as an explanation for the supernova type I phenomenon. This should happen when their carbon – oxygen core grows thanks to the helium shell burning above the core. For this purpose we undertook a thorough examination of helium stars in the mass range from Chandrasekhar's (*1.4 $M_\odot$*) up to about *2.5 $M_\odot$* (above which the fate of the star is already different).

We found that in the mass range of about *1.7 – 2.2 $M_\odot$*, indeed this can happen. The exact mass range depends on several physical uncertainties, such as the nuclear reaction rate and the resulting carbon mass fraction in the core, the treatment of convection, the *Urca* process and so on.

The main new insight we believe we gained is the crucial importance of an "early" off-center ignition of carbon. The problem this solves is, that since the temperature in the vicinity of the outward advancing helium burning shell increases to the point where the freshly produced carbon ignites, an aggressive carbon burning shell is created, which moves inward, and most probably exhausts all the carbon. Thanks to the previous episode of off-center carbon burning, however, a carbon depleted region stops this new shell from advancing towards the center. After the helium burning shell reaches *Chandrasekhar's* mass, the now *super-Chandrasekhar* mass carbon – oxygen core contracts, and the residual degenerate carbon at the center is ignited, resulting in a runaway similar to the classical *SN I* scenario.

Since the structure of the carbon – oxygen core of the helium stars of our interest is very similar to that of a carbon – oxygen star of the same mass, and their behavior is very similar to that of a mass accreting carbon – oxygen star, we also thoroughly examined the behavior of carbon – oxygen stars. We discovered that the models which ignite carbon off-center (in the mass range of about $1.05 – 1.18\ M_{\odot}$, depending on the carbon mass fraction) present an interesting $SN\ I$ progenitor scenario of their own, since whereas in the standard scenario runaway always takes place at the same density of about $2\ X\ 10^9\ gr/cm^3$, in our case, due to the small amount of carbon ignited, we get a whole range of densities from $1\ X\ 10^9$ up to $6\ X\ 10^9\ gr/cm^3$.

In comparison to the standard $SN\ I$ scenario, our helium star scenario has the following distinctions:

1.  The accretion rate is not a free parameter.

2.  At the moment of explosion, if it will indeed occur, the star will possess an envelope of non-negligible mass, consisting of helium and carbon burning products, and possibly some oxygen burning products.

3.  Similar to the case of carbon – oxygen stars above, they feature runaway at a variety of densities.

We emphasize, that we regard this scenario as a basis for an astrophysical model. If our results are indeed sound, their main contribution could be to help resolve the emerging recognition that at least some diversity among $SNe\ I$ exists, since runaway at various central densities is expected to yield various outcomes in terms of the velocities and composition of the ejecta, which should be modeled and compared to observations.

Several issues which were beyond the scope of this work call for further investigation:

1. The question of mass loss from the envelope, which might decrease the mass of the star below *Chandrasekhar's* before helium burning ceases, and thus undermine our scenario, has been affirmatively answered using various available mass loss formulae. However, all the existing formulae have been observationally calibrated with stars of mass range and/or evolutionary stage different than those of our interest, and might not be suitable to our case. We also feel that the complicated issue of mass loss driven by pulsations should be more thoroughly investigated.

2. The hot bottom burning is a complex phenomenon, which has been extensively dealt with in the literature for hydrogen burning stars. The key question is if and how do the bottom of the convective envelope together with the burning shell move outward, and its answer lies in an reliable modeling of the convective mixing process, which is evidently a three dimensional phenomenon. A further complication involves the double (helium and carbon burning) shell, which follows the ignition of freshly produced carbon, and the question of its stability. As is well known, in the case of hot bottom burning with a double hydrogen and helium burning shell, thermal pulsing arises.

3. For both the helium and carbon – oxygen case, a more thorough treatment of the question whether thermal *Urca* can hinder the formation of a convective region when electron capture on $Mg^{24}$ sets in is needed.

4. Our work provides initial models, which should be used for detailed simulations of the explosive runaway. The real value of our results would be judged by fitting the results of the dynamical simulations to the observational data.

5. The mass distribution function of suitable candidates for our helium star scenario has yet to be established. As for our carbon – oxygen star scenario, their incidence among the white dwarf population should be in the order of magnitude of 1%.

# C O N T E N T S

# 1. Introduction

The purpose of this work was to check the hypothesis, that the evolution of helium stars in a certain mass range may lead them to a carbon core explosion similar to what is widely accepted as an explanation for the supernova type I (*SN I*) phenomenon.

In order to clarify the aforementioned statement, we will briefly review the classical evolutionary scenario of the *SN I* progenitor.

The fact, that the evolutionary path of a mass accreting white – dwarf like degenerate star or core, either through accretion from a binary companion or through an advancing burning shell, depends almost* only on the accretion rate and the neutrino loss rate, has been known for a long time (*Arnett 1969*, *Paczynski 1970*) and has been explained by *Barkat 1971*. This fact has been employed mainly† in the research of mass accreting carbon – oxygen stars as a basis for explaining the *SN I* phenomenon as an explosion following the ignition of carbon, which is highly degenerate due to the high density – $\rho \geq 2 \times 10^9 \ gr/cm^3$. Accretion is caused by proximity to a companion star in a binary system, and as such depends on a complex variety of parameters.

We are aware of the difficulties of this scenario, involving the retention of the common envelope and other questions (*Livio & Riess 2003*). In this work we didn't attempt to found one scenario or the other, but to learn the expected outcome of such mass accretion.

---

* Note however that the lower the temperature at the beginning of accretion the later is the point of convergence. Thus for low enough temperature convergence may not be reached before ignition.

† A similar technique is employed for the investigation of novae.

In this scenario we have a white dwarf below Chandrasekhar's mass ($M_{ch}$), which accretes mass at an "appropriate" rate ($\dot{M}$). The increase in mass raises the density and temperature at the center, so that at a certain stage their values cross the "carbon ignition line"* (hereafter *"IG"*), and carbon is ignited. The immediate outcome is an increase of the entropy at the center, leading to the growth of a convective region. If the size of the convective region reaches a certain value, further increase in entropy leads to expansion of the center (a decrease of the density at the center – hereafter *"DEC"*), in opposing to the accretion tending to lead to contraction. Thus, between *IG* and *DEC* the density at the center increases, and after *DEC* it decreases. The temperature nevertheless continues to increase due to the nuclear reactions, and with it the nuclear reaction rate and the convective flux are also increasing. When the reaction rate reaches a point where the convection can no longer compete with the entropy production rate, nuclear burning continues at almost constant density, and could reach a "dynamic" regime, where the nuclear time (defined as the time needed to exhaust all the fuel, including oxygen, at constant density) is shorter than the dynamic time (which can be defined for example as a pressure scale height divided by the speed of sound). We will refer to this situation as a "runaway" (*RA*), and it is clear that under these circumstances hydrostatic equilibrium can no longer be assumed. It is worth noting, that the runaway doesn't necessarily lead to explosion, since electron capture behind the explosion front could produce a rarefaction wave which might convert the explosion into a collapse. Figure 1-1 shows the evolution of the density and temperature at the center of a carbon – oxygen star model of mass $M = 1.18\ M_\odot$ and carbon mass fraction of $X_c = 0.05$, displaying also the relevant carbon ignition line and the points *IG*, *DEC* and *RA* mentioned above.

---

* The ignition line is defined as the locus in the $\rho_c$, $T_c$ plane, where the energy production rate from nuclear reactions equals the neutrino losses. We will refer to this point as IG.

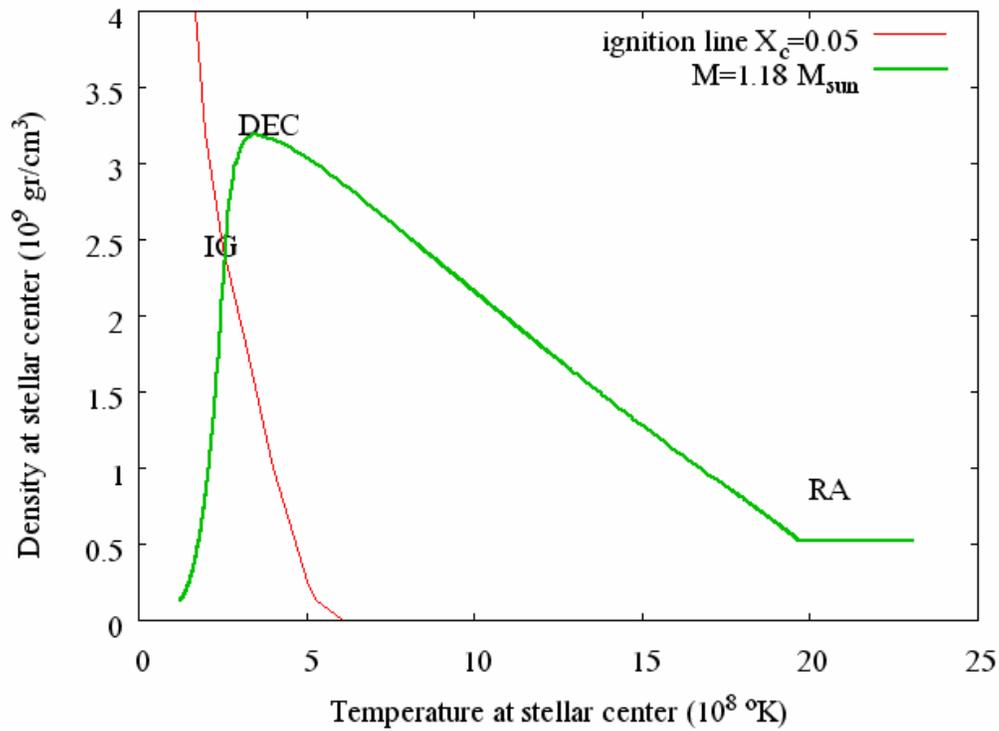

**Figure 1-1: The evolution of the temperature and density at the center of a typical carbon – oxygen star model, showing the various points along the path from ignition of carbon to runaway.**

Quantitatively speaking, it is clear that the *IG* point is dependent on the carbon mass fraction ($X_c$), and will be located at a higher density for a lower carbon mass fraction*.

---

* It also depends on the reaction rate, including the screening factor, and on the neutrino loss rate, but the sensitivity to these parameters is weaker.

The *DEC* point is also dependent on $X_c$. For a low enough $X_c$ the amount of carbon is insufficient to raise the temperature to the *RA* regime, and even not to the *DEC* point.

If we take into account the Q-value* of carbon burning ($\approx 4 \times 10^{17}$ *erg/gr*) and the specific heat in the region under discussion ( $(\partial T/\partial e)_\rho \approx 10^{-7}$ $^o K/erg$), we can estimate that the temperature will rise by about $4 \times 10^8$ $^o K$ for a mass fraction of one percent carbon ($X_c = 0.01$). Since for *RA* a temperature of about $1.4 \times 10^9$ $^o K$ is needed†, it is clear that if *IG* is reached at $T_c \approx 4 \times 10^8$ $^o K$, we need a carbon mass fraction of about *0.025* to reach *RA*. Clearly this is a rough estimate, since part of the energy produced by the burning of carbon is transported away, while on the other hand convection supplies fresh fuel. As we shall see later on, this estimate turns out to be pretty good.

To the abovementioned we should add the role which electron capture processes might play during evolution. A detailed discussion will be given in section 2.2.3; here we shall mention that the relevant processes are:

1. Electron capture on $Mg^{24}$, which is a product of carbon burning and is quite abundant, and $Na^{24}$, which cause a decrease of *Z/A*, and thus a decrease of the effective Chandrasekhar's mass, leading to an accelerated compression and local heating.

2. "*Urca*" processes of two kinds – thermal and convective, involving trace elements.

---

* The "Q-value" is defined as the amount of energy released by nuclear burning of a unit mass of matter through a specific nuclear reaction.

† This is the threshold temperature of oxygen ignition.

In this work we also checked the influence of these processes.

Regarding the astrophysical scenario, it is usual to deal with white dwarves that are the remnants of planetary nebulae (*PN*) formation. This is a situation when the growth of the carbon – oxygen core of the star causes a luminosity increase leading to envelope instability and ultimately to its ejection.

The typical mass of such white dwarves is about *0.6 $M_\odot$,* although more massive ones do exist. According to *Liebert et al. 2005*, some 6% of the white dwarves have masses above *1 $M_\odot$.* Regarding their composition, the carbon mass fraction generally lies within the range *$0.25 \leq X_c \leq 0.55$,* due to some uncertainties.

As mentioned before, the key point of the standard scenario is mass accretion from a binary companion. Clearly, the accretion rate depends on the structure and evolutionary history of the binary couple. Theoretical surveys have been made in the literature, where the accretion rate, as well as the initial mass of the accreting white dwarf and the composition of the accreted matter served as free parameters (e.g. *Nomoto & Sugimoto 1977*)

Before we set down to present and explain our hypothesis, we will stress an important detail. After the formation of a helium core of mass *$0.9 \leq M/M_\odot \leq 2.4$* (in stars of mass range *$3.0 \leq M/M_\odot \leq 10.0$*), and before the ignition of carbon, convection penetrates from the boundary of the star inward through the envelope and into the helium core (second dredge-up), thus mixing helium into the envelope and decreasing the mass of the helium core. The decrease of the helium core mass is very significant, as shown in Figure 1-2, taken from *Becker & Iben 1979* (page 842, figure 11).

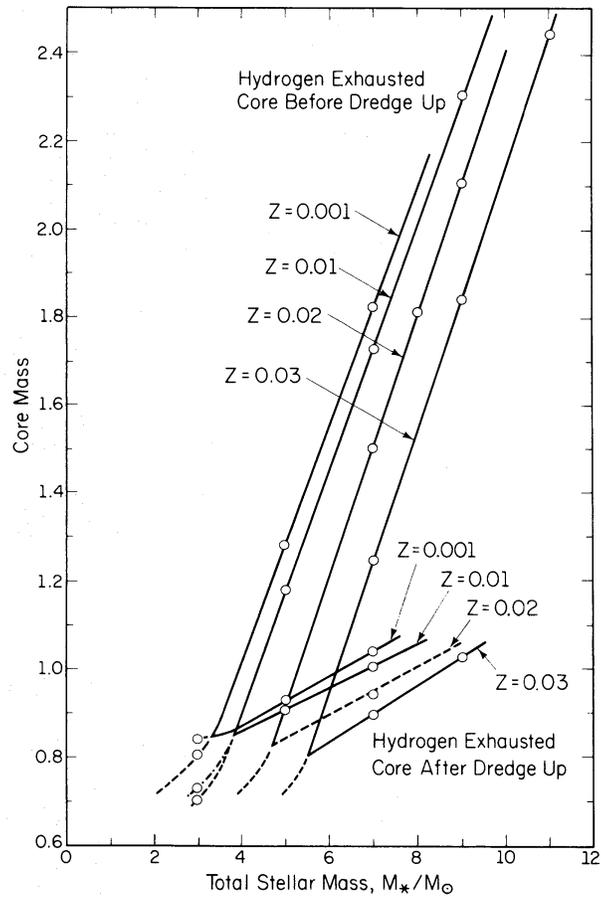

**Figure 1-2: The size of the hydrogen-exhausted core as a function of total stellar mass for both before and after the second dredge-up phase. In this figure, Y = 0.28 and Z is allowed to vary.**

The limits of the relevant mass range are not precisely defined, however a discontinuity appears to exist, above which there is no penetration at all prior to carbon ignition, and below it penetration is considerable. The works of (*Garcia-Berro & Iben 1994*, *Garcia-Berro & Iben 1999* and *Gil-Pons & Garcia-Berro 2001*) are dealing with this intermediate mass range lying around $8.0 \leq M_c/M_\odot \leq 10.0$, and as we shall see are relevant to our work.

Our hypothesis deals with helium stars. The evolution of helium stars was theoretically modeled since at least 1953 (*Crawford 1953*). The evolution in this case is relatively simple, and the possible outcomes have been covered a long time ago, so apparently it is difficult to arrive at new results. Many of those works have dealt with stars with mass above $2\,M_{\odot}$, and their final fate is what has been dubbed as "iron core collapse". They were expected to yield type II supernovae, however lately *Woosley et al. 1995* have suggested helium stars in the mass range $2.26 < M/M_{\odot} < 3.55$, which are formed as an end result of massive mass losing stars, may be the progenitors of type Ib supernovae. A comprehensive study, with extensive references, has been done by *Habets 1985*.

It is worth noting, that in studies dealing with massive helium stars it is assumed that such a star is the outcome of star that has lost its envelope **before** the onset of the penetration, so that the investigated masses are above the aforementioned penetration discontinuity.

In any helium star of large enough mass*, helium is ignited at the center, creates a convective core, and after helium is depleted leaves behind a carbon – oxygen mixture, its mass depending on the total mass of the star. Immediately following the cessation of helium burning at the center, a helium burning shell is ignited above the core. This shell gradually advances outward and increases the core. It can apparently be expected, that as the mass of the core will approach Chandrasekhar's mass, it will behave like a mass accreting carbon-oxygen star, and will reach the *RA* stage as mentioned before. There are at least two distinctions between this and the standard *SN I* scenario we presented before:

---

* This is certainly the case for helium stars with mass above Chandrasekhar's, in which we are interested.

1. The accretion rate is not a free parameter.

2. At the moment of explosion, if it will indeed occur, the star will possess an envelope of non-negligible mass, consisting of helium and carbon burning products, and possibly some oxygen burning products.

In this work we aimed to examine the feasibility of such a scenario (which we will call "*Super Chandrasekhar SN*"), by means of an extensive and thorough scan of the terms for its existence and of its details.

At this point we shall emphasize again, that we regard this scenario as a basis for an astrophysical model. In the discussion (chapter 4) we shall refer to the problems of this scenario as a model for *SN* progenitors.

As we shall show in section 3.2, we first scanned the relevant mass domain, and followed the outward moving helium shell. It is known, that the further the shell is from the center, the higher the reaction rate and the temperature in it are. It turns out, (as should have been expected), that well below $M_{ch}$, at about $m \approx 1.1 \, M_{\odot}$, the temperature is high enough to ignite the freshly produced carbon. It is easy to see, that in this case an inward advancing carbon burning shell is created. This result was pointed out by other authors (*Kawai et al. 1987*, *Timmes et al. 1994*), although for other circumstances.

The evolutionary picture in these cases exhibits the following sequence of stages:

1. A burning shell advances inward, while the convective region it creates above gradually increases.

2. Following the increase of the convective region, the inner parts expand and the temperature decreases, ultimately extinguishing the shell.

3. A subsequent contraction of the star ignites a burning shell below the point where the previous shell has been extinguished, and again an inward advancing carbon burning shell is created.

It is quite clear, that eventually the burning shell will reach the center of the star exhausting the carbon and producing an *O-Mg-Ne* core. However, thanks to neutrino losses, there is also a possibility for a minute fraction of carbon to be left over at the central area of the star (*Iben et al. 1996, Saio & Nomoto 1998*). With the increase in density, one expects that electron capture processes on *Mg* and *Ne* will induce a collapse to a neutron star or an explosion of some kind.

At this stage it occurred to us, that in case carbon burning has already taken place in a previous stage, so that below the current carbon ignition point there is a carbon depleted region, the shell can't advance to the center, and our scenario still has a chance. This is similar to the method of "fighting fire with fire".

In order to realize this idea, we referred to the mass range where the original carbon core is big enough to ignite carbon before it grows towards $M_{ch}$, but not too big so that ignition takes place off-center. In this case it is expected, that throughout an extensive region below the helium burning shell carbon will indeed be depleted.

Alas, investigating the required mass range is especially difficult, due to the need for extremely fine zoning in order to follow the burning shell and the convective region created above it. In section 3.1.2 we discuss this subject in detail.

The structure of the carbon – oxygen core of the helium stars of our interest is very similar to that of a carbon – oxygen star of the same mass, and the behavior of the core of our helium stars is very similar to that of a mass accreting carbon – oxygen star. Therefore, in order to lay the foundation for our understanding, we chose first to examine the behavior of carbon – oxygen stars. Finally we shall apply our findings for carbon – oxygen stars to investigate the relevant helium stars. We shall see that indeed there exists a mass range, where the suggested scenario is feasible.

In the following, we shall begin by first describing the computational methodology of our work, including the numerical algorithms and input physics in chapter 2.

Our results, both for carbon – oxygen and helium stars, are given in chapter 3, including a comparison with previously published works of other authors.

Finally, in chapter 4 we will discuss our results and present our conclusions, including observational predictions and suggestions for further research.

# 2. Methodology

## 2.1. *The evolutionary code*

The evolution was followed using a quasi-static *Lagrangian 1D* evolution code named *ASTRA*, which will briefly be described in the following. This code is an extensively improved version of the *ASTRA* evolution code first described by *Rakavy, Shaviv & Zinamon 1966*, and used with some modifications many times. The code being quasi-static means, that we deal with situations where the star is dynamically stable, and the time scales for the three main processes that govern the evolution of the star – namely the hydrodynamic motion, the convective mixing, and the exchange of energy due to thermonuclear reactions and radiative transport obey the relation: $t_{hydro} << t_{convec} << t_{thermo}$. This assumption is valid for most stages of stellar evolution, with the possible exception of very violent nuclear burning phases, where special care has to be taken. Since these stages are indeed in the scope of our interest, a special treatment of these stages was devised, and will be described in section 2.3. We will now describe the main features of the code; a detailed description will be given in Appendix A.

### 2.1.1. Structure of the code

The quasi-static assumption allows us to employ two major simplifications:

1. The two main equations of stellar evolution – the equation of energy conservation (represented by an equivalent equation for the rate of change of the entropy) and the equation of hydrostatic equilibrium, are solved separately, i.e. the change of entropy with time is solved for a given density profile, and then the equation of hydrostatic equilibrium is solved for a given entropy profile. For this purpose, the independent thermodynamic variables used are the entropy and density, rather than the density and temperature, as in most other evolution codes.

2.  Convective regions are treated as isentropic and fully mixed. This assumption greatly simplifies the program, as it eliminates the need to calculate the convective energy flux, and is valid (i.e. is in good agreement with the mixing length model of convection) wherever the pressure scale height is large compared to the size of the convective region of interest. In the stellar models of our interest, this kind of condition occurs throughout the stellar core. In the envelope this assumption is inaccurate, and thus we made an effort to validate our results using a different evolution code which comprises mixing length theory treatment (see further  section 3.5)

During each time step the code passes through the following stages:

1.  The boundaries of the convective regions are determined. For each convective region the composition is mixed and entropy is set to a common value, so that the total energy is conserved.

2.  Since the entropy profile has been changed, the density profile is changed to fulfill the equation of hydrostatic equilibrium.

3.  The size of the time step *dtime* is decided on, so that the thermodynamic parameters (density, temperature, entropy etc.) will not change by more than some preset fractions. This estimate is done based on the changes that had occurred during the previous step. Should the parameters change by more than some allowed fraction, the step is cancelled and repeated with reduced *dtime*.

4. For each zone the code calculates the energy production rate and the change of composition due to nuclear reactions and the energy loss rate due to neutrino production.

5. The entropy equation (i.e. energy conservation) is solved to get a new entropy profile, next the equation of state is used to calculate the pressure, temperature etc., from the density and entropy, and then the radiative energy flux is calculated. This is done iteratively until the obtained profile changes less than the required criterion. During this process the density is kept unchanged. Note that this means the reaction rates are calculated explicitly (i.e. set at the beginning of the time step), while the luminosity is calculated implicitly (i.e. solved for the values at the end of the time step).

6. After the entropy profile has been calculated, the density profile is changed to fulfill the equation of hydrostatic equilibrium.

### 2.1.2. The treatment of convection

As has been previously mentioned, we employ a somewhat simplified method for the treatment of convection in the code. This treatment is based on the aforementioned assumption, that the time scale of the convective mixing is much shorter than the time scale of the nuclear reactions and radiative transport that govern the entropy equation. Under the above assumption, we treat each convective region as a fully mixed isentropic zone, which is actually equivalent to using a very large mixing length coefficient in the widely popular mixing length theory treatment.

The actual algorithm consists of the following four stages:

1. Determine the boundaries of the convective regions, and mix the entropy and composition of each convective region accordingly.

2. Make a "*gedanken"* step. This is a regular time step going through the procedure of solving the entropy equation as described above, only that the entire star is assumed to be radiative. It is needed because otherwise a *Lagrangian* zone once declared convective could not turn to be radiative, since the criteria for convection are such, that a region of *Lagrangian* zones having the same entropy and composition would always be convective.

3. Determine new boundaries of the convective regions, and mix the entropy and composition of each convective region accordingly. This is done exactly as in stage 1, with the exception that a *Lagrangian* zone declared as radiative in stage 1 is not allowed to be convective, since the *gedanken* step is meant to allow convective regions to shrink and not to grow.

4. Restore the initial entropy, density and composition profiles (from prior to stage 1), in order to cancel the changes made by the *gedanken* step, and mix again the entropy and composition of each convective region according to the new convective region boundaries determined by the *gedanken* step.

The algorithm for the determination of the convective region boundaries works as follows:

1. Starting from the innermost *Lagrangian* zone upwards, we check each zone for stability against convection with its outer neighbor, until we find the first unstable pair. A new convective region is created, and the pair is marked as the inner and outer boundary of the convective region.

2. We mix the first zone with its outer neighbor. Mixing means that the composition is averaged, and a common value of the entropy is found, such that without changing the density of the individual zones, the total internal energy is conserved.

3. We go on checking the current outermost *Lagrangian* zone of the convective region for stability against convection with its outer neighboring *Lagrangian* zone. If they are unstable we add the outer neighbor to the convective region, and again mix all the *Lagrangian* zones included in the convective region. If they are stable we stop the process, and fix the outer boundary of the convective region.

4. Since during the process of finding the outer boundary of the convective region, the mixing process might have changed the composition and the entropy, it is possible that under the conditions the innermost *Lagrangian* zone we started from is now unstable against convection with its inner neighbor. Therefore we now go through a process similar to that in the previous stage, only going inwards from the inner boundary of the convective region, until we can fix the inner boundary.

5. After we fixed the inner and outer boundaries of the convective region, we repeat stages 1 through 4 starting from the *Lagrangian* zone immediately above the outer boundary of the previous convective region, until we have checked the entire star.

6. At last we again check for all the convective regions we found that the *Lagrangian* zones at their boundaries are stable against convection with their neighbors outside the convective region. This should be done due to the mixing, which can change the thermodynamic conditions and thus the stability against convection. If we find any instability, all the convective regions are canceled (but the changes done by the mixing are preserved), and the process is restarted from stage 1.

The checking of stability against convection follows the *Schwarzschild* criterion.

### 2.1.3.   The equations of hydrostatic equilibrium

Under our quasi-static assumption, we can assume that any change of the entropy profile will cause an immediate adjustment of the density profile, such that the velocity of each *Lagrangian* zone vanishes, and the equation of hydrostatic equilibrium is obeyed. Thus, in principle, we should solve the equation of hydrostatic equilibrium each time we alter the entropy profile – both during solution of the entropy equation and during the solution for the convective regions' boundaries. Since such a practice is both time-consuming and proves to have no effect on the results, we only solve the equation of hydrostatic equilibrium at the end of the time step (after solving the entropy equation), and after setting the boundaries of the convective regions (before rechecking the stability of the boundaries – step 6 in the description of the algorithm).

## *2.2.  Input physics*

### 2.2.1.  Nuclear reaction rates

Nuclear reactions were treated via two different sets of reaction rates, which were then compared to each other:

1.  Using an α-network of 13 elements from He$^4$ up to Ni$^{56}$, with reaction rates based on the "NON-SMOKER" tables by *Rauscher & Thielemann 2001*.

2.  Using the reaction rates according to *Caughlan & Fowler 1988* for multiple α-nuclei.

The screening factor was treated according to *Itoh et al. 1979* and *Itoh et al. 1980*, but a comparison was made with those of *Dewitt et al 1973*.

### 2.2.2.  Neutrino losses

Neutrino losses were taken according to *Itoh & Kohyama 1983*.

### 2.2.3.  Electron capture

Three different processes of electron capture are considered:

#### *2.2.3.1.  Thermal Urca (TU)*

Thermal *Urca* is a situation where at some *Lagrangian* mass point in the star ($m_{urc}$) the Fermi energy fulfills $E_f = E_{th}$, so that throughout a mass shell (henceforth "*Urca* shell" - *US*), lying in the range of $E_f = E_{th} \pm 1\ kT$, processes of emission and capture of electrons by some trace nuclei take place. Both processes are accompanied by production of neutrinos, which leave the star and create an effective local heat sink.

Since the star continues to shrink and $E_f$ increases, it is clear that $m_{urc}$ increases and the *US* moves outward. The magnitude of the process depends on the mass fraction of suitable nuclei (*Ergma & Paczynski 1974*). In our context, the dominant nuclei are $Na^{23}$, with a threshold of $E_{th} = 4.4\ MeV$ and consequently $\rho_{th} \approx 1.7\ X\ 10^9\ gr/cm^3$, and afterwards $Ne^{21}$, with a threshold of $E_{th} = 5.7\ MeV$, and thus $\rho_{th} \approx 3.5\ X\ 10^9\ gr/cm^3$.

We modeled the effect of the *TU* by means of the prescription by *Tsuruta & Cameron 1970*, while checking the sensitivity of the result to the reaction rate by varying the mass fraction of the relevant nuclei.

### 2.2.3.2. Convective Urca (CU)

In this case convection transfers nuclei through the *US*. These nuclei pass through a region where the difference $E_f - E_{th}$ is of the order of magnitude below *1 MeV*, and certainly $\Delta E >> kT$. Since in this case the average electron capture occurs below the Fermi level, a hole is created, which is filled by an electron from above the Fermi level, and the energy surplus is emitted as a $\gamma$ photon causing local heating.

The effect of *CU* on the relevant evolution has been discussed for a long time in the literature with contradictory conclusions (e.g. *Paczynski 1972*, *Bruenn 1973*, *Couch & Arnett 1975*, *Lazareff 1975*, *Regev & Shaviv 1975*, *Barkat & Wheeler 1990*, *Mochkovitch 1966*). Lately it turned out that the main effect of *CU* is to hinder the extension of the convective region above $m_{urc}$ (*Stein et al. 1999*, *Bisnovati-Kogan 2001*). Hence, in this work we tried to examine the practical significance of *CU*, by artificially forbidding convection above $m_{urc}$.

### *2.2.3.3. Electron capture on carbon burning products*

As *Miyaji et al. 1980* have already shown, processes of electron capture by certain burning product nuclei, when the Fermi level ($E_f$) exceeds a relevant threshold ($E_{th}$), can have a major importance during the stage preceding *RA*, when the density, and consequently the Fermi energy, are rising. In our context, the dominant nuclei from among the carbon burning products are* (*Miyaji et al. 1980* again) $Mg^{24}$, with a threshold of $E_{th} = 5.52\ MeV$ and consequently $\rho_{th} \approx 3.35\ X\ 10^9\ gr/cm^3$, and afterwards $Na^{24}$, which is the electron capture product of $Mg^{24}$, with a threshold of $E_{th} = 6.59\ MeV$, and thus $\rho_{th} \approx 5.25\ X\ 10^9\ gr/cm^3$.

Note that lately (*Gutierrez et al. 2005*) there have been claims that $Na^{23}$ is much more abundant than previously thought, and is more abundant than even $Mg^{24}$. Since the threshold for *EC* on $Na^{23}$ is lower than on $Mg^{24}$, it may have a considerable effect on the evolution. This probably means that the effects of *EC* should start to show up at a lower density, but we believe that other than affecting somewhat the density at *RA*, nothing qualitative should be expected. Anyhow, we considered $Na^{23}$ only in the context of thermal *Urca* (see section 2.2.3.1).

In order to calculate the effect of *EC*, it is necessary to know the mass fraction of the relevant nuclei, which is a result of carbon burning and the capture rate, as well as the neutrino loss rate and the resulting heating rate.

---

* Albeit the mass fraction of $Ne^{20}$ in the burning products of carbon is high, we didn't need to include *EC* on it, since the threshold in its case is high, and corresponds to $\rho > 6\ X\ 10^9\ gr/cm^3$, which is above the limit of our interest.

As will be evident from our results, this process has an important effect in our cases; therefore it was included as a standard in all our models, except when explicitly mentioned otherwise. Due to the existing uncertainties (cf. *Gutierrez et al. 2005*), in line with our basic approach, we checked the sensitivity of the results both to the mass fraction and to the capture rate by introducing "fudge factors".

We took the capture rates from *Miyaji et al. 1980*, but the range of variations we checked covers also the differences versus the rates given by *Oda et al. 1994*, which differ from the former by as much as an order of magnitude.

*Mochkovitch 1984* pointed out that in the presence of *EC*, due to its effect on the gradient of the electron mole number $Y_e$, the *Ledoux* criterion for convection has to be used, and hence the extension of the convective region is slower. In our case it turned out, that the effect of using the *Ledoux* criterion is quite small.

Another important caveat (*Stein 2005*): since it is clear that *EC* must start as *TU* (which is locally cooling), and can become exothermic only when occurring below the threshold by more than *kT*, the question whether a convective zone broad enough to ensure heating can be formed arises. In our case, where *EC* starts when significant carbon burning is already present, this problem may not be severe. However, more careful analysis is needed, to find out whether the local cooling might force the base of convection to move outward (similar to the hot bottom case in section 3.3). This is beyond the scope of this work.

### 2.2.4. Opacities

Radiative opacities were calculated according to the *OPAL* opacity tables (see *Iglesias & Rogers 1996*), using the tables and interpolation subroutine provided at the web site

*www-phys.llnl.gov/Research/OPAL.* The tables were extended to lower temperatures according to *Alexander & Ferguson 1994.*

Electron conduction was calculated according to *Iben 1975.*

### 2.2.5. Equation of state

The equation of state takes into account ionizations to all available levels of the different atom species in the composition. The distribution function of the various ionization levels is computed using a method similar to the one described in *Kovetz & Shaviv 1994.* The resulting electron density is then used together with the temperature in order to extract the pressure, energy, chemical potential and their derivatives respective to the electron density and temperature from a table computed in advance by solving the *Fermi-Dirac* integrals. The pressure, energy and entropy of the ions are then added as an ideal gas together with those of the radiation.

## 2.3. Treatment of deviations from the quasi-static assumption

As previously mentioned, our evolutionary code assumes, that the time scales for the three main processes that govern the evolution of the star – namely the hydrodynamic motion, the convective mixing, and the exchange of energy due to thermonuclear reactions and radiative transport, obey the relation: $t_{hydro} << t_{convec} << t_{thermo}$ .

However, as we shall see later, as the star approaches explosion, this relation between the timescales is no longer valid. As the reaction timescale $t_{thermo}$ becomes shorter, it first of all becomes comparable to the convective timescale $t_{convec}$, so in order to achieve a reasonable modeling, we have to appropriately restrict convection when this occurs. Hence, it is important to estimate these two timescales, and give an adequate treatment when the quasi-static assumption fails. In practice, we restrict the outer boundary of the convective zone (the inner boundary in our case is always at the center) to be not above the innermost *Lagrangian* zone, which fulfills $t_{convec}(r) > \alpha \, t_{thermo}$, where $r$ is the radius of the zone, and $\alpha$ is a "fudge factor" we use to check the sensitivity of the results. Note that $1/\alpha$ is in fact the number of convective turnover times during an interval of one thermonuclear timescale. The method for calculating the timescales $t_{thermo}$ and $t_{convec}$ is given below.

### 2.3.1. The convective timescale

The widely used mixing length theory gives an explicit relation between the convective luminosity and the convective velocity:

*(2-1)*
$$v_c = \left[ \frac{\lambda P \left( \dfrac{\partial \ln \rho}{\partial \ln T} \right)_P}{c_P T \rho^2} L_c \right]^{\frac{1}{3}}$$

Here $\lambda$ is the mixing length, $c_p = (\partial e/\partial T)_p$ is the specific heat at constant pressure, and $L_c$ is the convective flux through a unit of area.

We don't use the mixing length theory, but rather assume an isentropic convective region. Nevertheless, we can estimate the convective luminosity as follows:

Let $L_c$ be the convective luminosity and $L_r$ the radiative luminosity. At any point we have:

*(2-2)*
$$T \frac{\partial S}{\partial t} = q - \frac{\partial L_r}{\partial m} - \frac{\partial L_c}{\partial m}$$

However, we assume that $\partial s/\partial t$ is uniform throughout the convective region, thus:

(2-3) $$\int T \frac{\partial S}{\partial t} dm = \frac{\partial S}{\partial t} \int T dm = \int q \, dm - \Delta L_r - \Delta L_c$$

But $\Delta L_c = 0$ since the convective flux vanishes at the boundaries of the convective region, so we have:

(2-4) $$\frac{\partial S}{\partial t} = \frac{\int q \, dm - \Delta L_r}{\int T \, dm}$$

Together with (2-2) we have:

(2-5) $$\frac{\partial L_c}{\partial m} = q - \frac{\partial L_r}{\partial m} - T \frac{\int q \, dm - \Delta L_r}{\int T \, dm}$$

From the above equation we can derive the convective luminosity $L_c(m)$ for each point in the convective region, and then use (2-1) to get the convective velocity $v_c(m)$ at each point in the region.

The time needed for the convective flux to cover a distance $\Delta r$ in the vicinity of a *Lagrangian* point $m$ is of course:

(2-6) $$\Delta t = \frac{\Delta r(m)}{v_c(m)}$$

Consequently the time to reach a distance $r$ from the inner boundary of the convective region will be:

(2-7) $$\tau(r) = \int dt = \int \frac{dr}{v_c}$$

The "convective time scale" will be defined as the time $\tau(r)$ to the outer boundary of the convective region, i.e. the time needed to cross the entire length of the region.

It is clear that this is only a rough estimate; nevertheless a comparison with models of convective envelopes calculated for the same conditions with the mixing length theory using the method described by *Tuchman et al. 1978* gave an agreement better than a factor of 3.

### 2.3.2. The nuclear timescale

As the nuclear timescale we define the time needed to burn all the carbon and afterwards the oxygen at a given point (with initial temperature $T$, density $\rho$ and carbon mass fraction $X_c$), under the assumption of constant density. Clearly this is a lower limit to the time, since in reality there are also energy losses.

To get a quantitative estimate we used the analytical approximation of *Woosley & Weaver 1986* for the reaction rates of carbon and oxygen burning in the relevant range (i.e. $2 < T_9 < 6,\ \rho_9 < 4$):

*(2-8)*
$$\text{carbon}:\ q_c\left(\rho,T\right) = 8.25 \times 10^{15}\, X_c^2\, \rho_9^{2.79} T_9^{22}\ \text{(erg/sec)}$$
$$\text{oxygen}:\ q_o\left(\rho,T\right) = 8.25 \times 10^{15}\, X_o^2\, \rho_9^{2.79} T_9^{36}\ \text{(erg/sec)}$$

They also approximated the heat capacity using:

*(2-9)*
$$\frac{\partial T}{\partial e} = 1.57 \times 10^{16}\, \rho_9^{-0.26} T_9^{0.76}$$

Combining the last two equations we get (for carbon):

*(2-10)*
$$\frac{\partial T}{\partial t} = q\frac{\partial T}{\partial e} = 1.3 \times 10^{32}\, X_c^2\, \rho_9^{2.53} T_9^{22.76}$$

A similar equation can be obtained for the oxygen. Accordingly, the time for raising the temperature from $T_1$ to $T_2$ is given by:

*(2-11)*
$$t_{1 \to 2} = \frac{1}{1.3 \times 10^{32} X_c^2 \rho_9^{2.53}} \int_{T_1}^{T_2} \frac{dT}{T_9^{22.76}}$$

Given an initial temperature $T_1$, we can estimate the final temperature $T_2$ of carbon burnout, since:

*(2-12)*
$$\int_{T_1}^{T_2} \left( \frac{\partial e}{\partial T} \right)_\rho dT = X_c Q_v \qquad \textit{($Q_v$ is the Q-value)}$$

After we get the time $t_{1 \to 2}$ for carbon burnout and the final temperature $T_2$, we use $T_2$ as the initial temperature of oxygen burning, and through the same method we can get the time $t_{2 \to 3}$ of oxygen burnout.

Note that we do not take into account the fact, that $X_c$ is also a function of time, but it turns out to have a small effect. In fact, in our calculations we checked the validity of these estimates by artificially preventing the density from changing, and actually measuring these times. We found excellent agreement.

Estimating the timescale of oxygen burning is important, since in case the carbon mass fraction is very low, it might burn out without igniting the oxygen.

# 3. Results

## *3.1. Carbon – Oxygen stars*

### 3.1.1. Overview

The evolution of carbon – oxygen stars is determined by their mass ($M_c$) and composition. We assume the composition of the star is homogeneous – as a result of the helium burning producing it being convective. Therefore it is sufficient to specify the mass fractions of the relevant elements: $C^{12}$ and $O^{16}$, whereas the mass fractions of *Ne* and *Mg* are negligible. These mass fractions are determined at the end of helium burning, for which there is a well known uncertainty, due to uncertainties in the cross-section for the reaction $C(\alpha,\gamma)O$, which takes place during helium burning. It is common to assume the mass fraction of the carbon ranges between *0.25 $\leq X_c \leq$ 0.55*, and the mass fraction of the oxygen is its complementary to unity (*Umeda et al. 1998*).

The following figures show typical evolutions of carbon – oxygen stars of various masses on the $\rho_c$ (the density at the center), $T_c$ (the temperature at the center) plane. We can identify four fundamentally different ranges. Above Chandrasekhar's mass ($M_3 = M_{ch} \approx 1.4 M_\odot$) carbon is ignited at the center, and the stellar center evolves toward increasing temperatures and densities, igniting heavier fuels. Figure 3-1 shows a typical example of such a star, with a mass of *M = 2 $M_\odot$* and carbon mass fraction of *$X_c$ = 0.54*, together with an example of a star with *M = 1.4 $M_\odot$*, just below Chandrasekhar's mass.

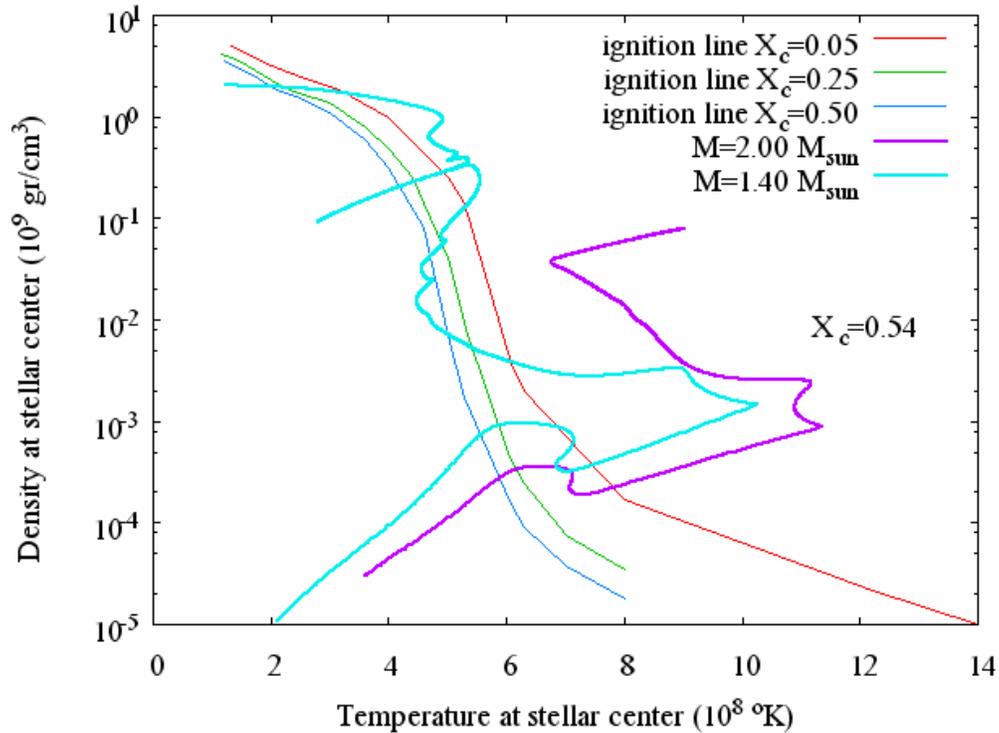

**Figure 3-1:** The evolution of the density vs. the temperature at the center of a 2 M$_\odot$ carbon star model above Chandrasekhar mass (M$_3$) and of a 1.4 M$_\odot$ carbon star model slightly below Chandrasekhar's mass. The carbon mass fraction is X$_c$ = 0.54 for both stars. The ignition lines are also plotted for various carbon mass fractions.

Below *M$_3$*, but above a certain limit *M$_2$*, carbon is ignited at the center, but finally evolution proceeds towards a white dwarf. Figure 3-2 shows a typical example of such a star, with a mass of *1.25 M$_\odot$*. The evolution of the same star with carbon burning turned off is shown for comparison, and we can see that both stars eventually evolve toward a white dwarf on the same path.

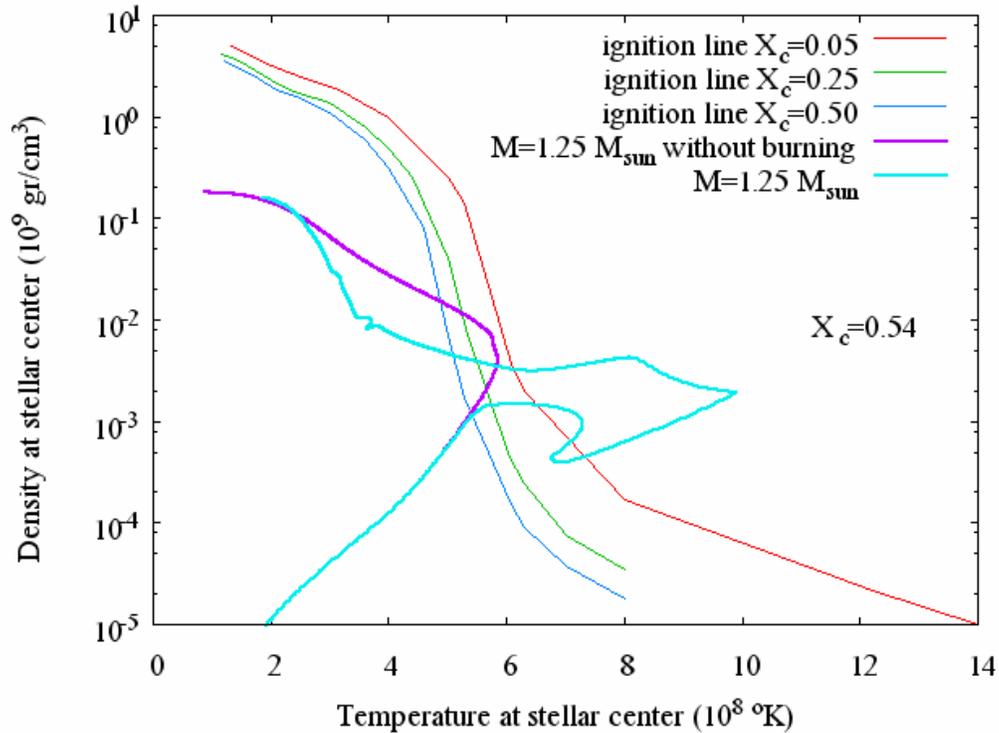

**Figure 3-2: The evolution of the density vs. the temperature at the center of a 1.25 M$_\odot$ carbon star model below Chandrasekhar mass (M$_3$) which ignites carbon at the center (M > M$_2$). The evolution of the same model with carbon burning turned off is shown for comparison. The carbon mass fraction is X$_c$ = 0.54. The ignition lines are also plotted for various carbon mass fractions.**

The value of $M_2$ depends on the carbon mass fraction; for our example of $X_c = 0.54$ it lies around $M_2 \approx 1.17 \, M_\odot$. Figure 3-3 shows the evolution of an $M = 1.17 \, M_\odot$ star, slightly below $M_2$. The star ignites carbon slightly above the center, but carbon burning immediately reaches the center, and the star continues its evolutionary path similar to a central-ignition star.

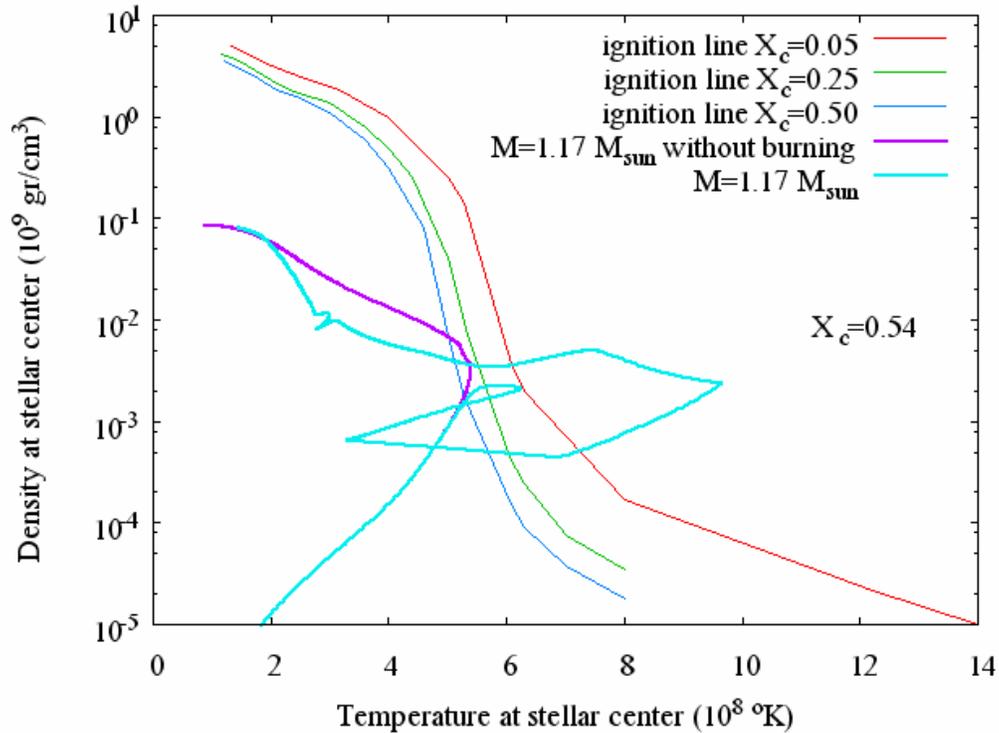

**Figure 3-3: The evolution of the density vs. the temperature at the center of a 1.17 M$_\odot$ carbon star model, which ignites carbon just above the center (M slightly smaller than M$_2$). The evolution of the same model with carbon burning turned off is shown for comparison. The carbon mass fraction is X$_c$ = 0.54. The ignition lines are also plotted for various carbon mass fractions.**

Below $M_2$, but above a lower limit $M_1$, carbon is ignited off-center, at a point depending both on the mass of the star and on the carbon mass fraction (see Figure 3-5 and Figure 3-6). Subsequently, carbon burning propagates both inwards toward the center and outwards, and after almost all the carbon in the core is exhausted, evolution returns to its original path towards a white dwarf.

As we already suggested in chapter 1, and will show in detail later, these stars are in the focus of our interest, since the off-center burning might leave behind enough carbon, which will "survive" the subsequent evolutionary phases, and finally ignite explosively after the mass reaches $M_{ch}$ through accretion. We will discuss the evolution of these stars in detail in section 3.1.2. Figure 3-4 shows a typical example of an off-center carbon igniting star with mass of $M = 1.10\ M_\odot$.

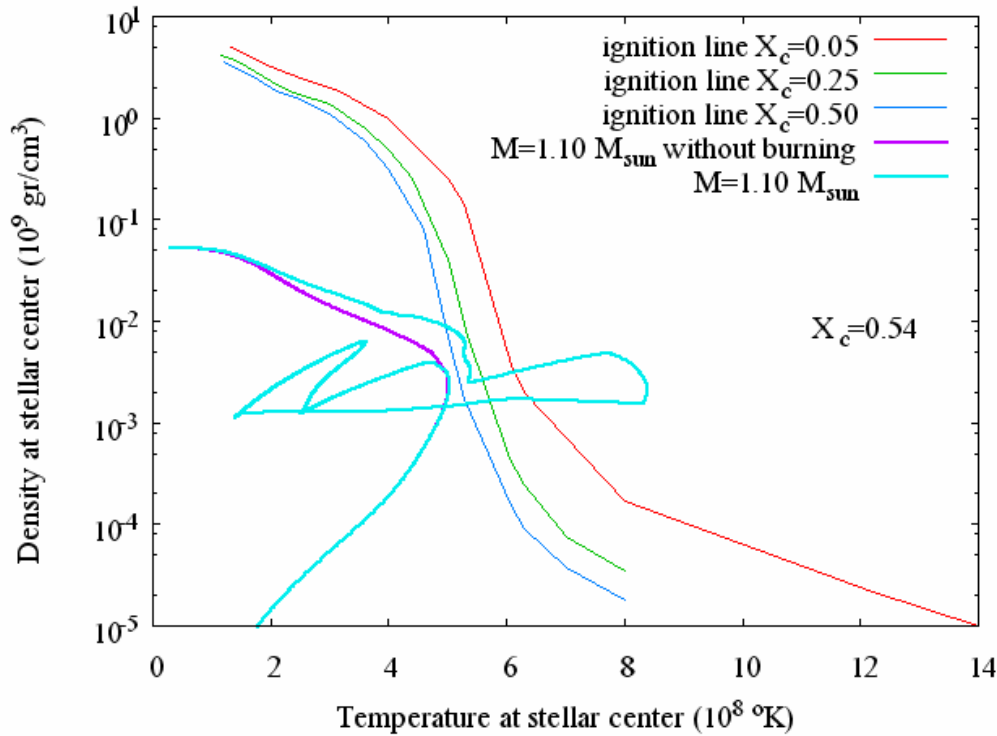

**Figure 3-4: The evolution of the density vs. the temperature at the center of a 1.10 M$_\odot$ carbon star model, which ignites carbon off center (M$_1$ < M < M$_2$). The evolution of the same model with carbon burning turned off is shown for comparison. The carbon mass fraction is X$_c$ = 0.54. The ignition lines are also plotted for various carbon mass fractions.**

We can see that at a the point of ignition the density and temperature at the center decrease, as a result of the expansion induced by the carbon burning shell above. As we will explain in detail in section 3.1.2, this shell extinguishes, subsequently causing the center to contract and heat again, but soon another off-center burning shell is ignited, causing the center to expand and cool a second time. Finally the burning reaches the center, causing it to rise to carbon burning temperature with almost no change in density. After the burning ceases, the center returns to its original path, as if no carbon burning had taken place.

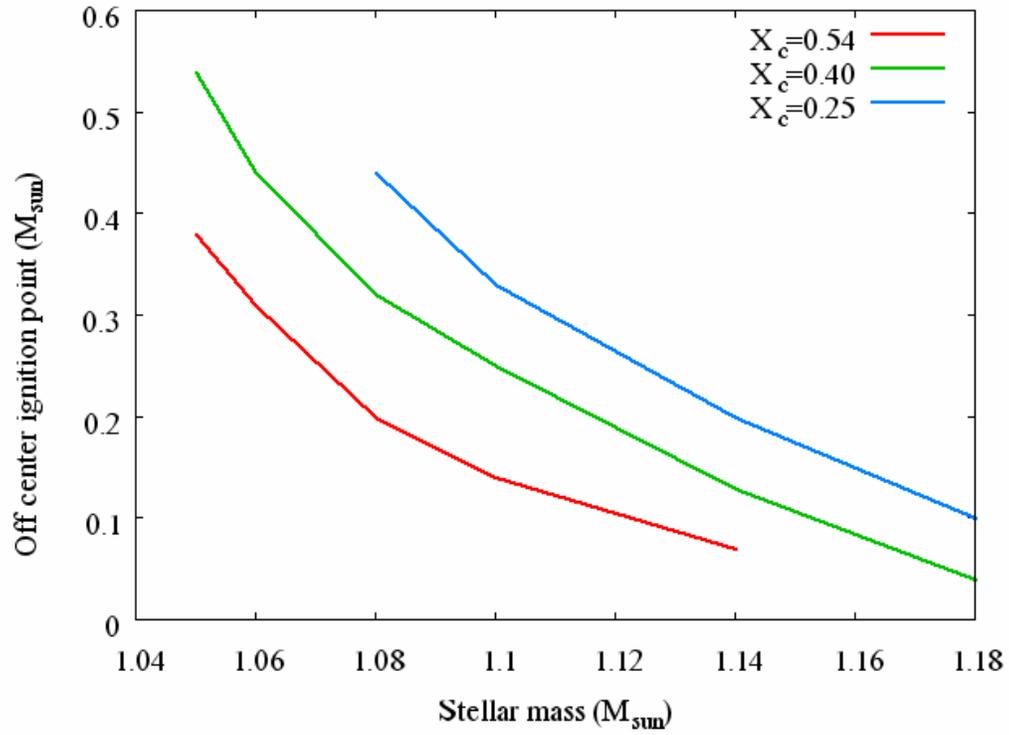

**Figure 3-5: The off-center ignition point vs. the stellar mass for various carbon mass fractions.**

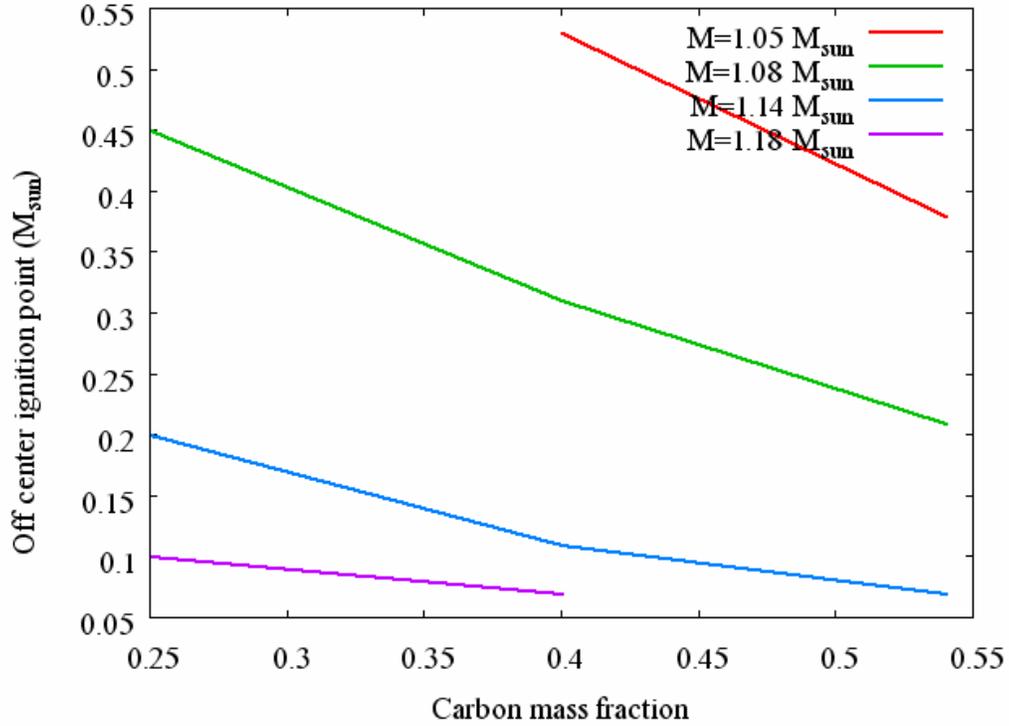

**Figure 3-6: The off-center ignition point vs. carbon mass fraction for various stellar masses.**

Below the limit $M_l$ the star evolves towards a white dwarf without the carbon being ignited. The value of $M_l$ depends of course on the carbon mass fraction. Figure 3-7 shows the evolution of the center of various carbon stars below and above the limit $M_l$, with a carbon mass fraction of $X_c = 0.54$, and we can see that the value of $M_l$ lies in the range $0.95\ M_\odot < M_l < 1.05\ M_\odot$. Figure 3-8 shows a similar plot for a lower carbon mass fraction of $X_c = 0.25$, and we can see that here the limit is slightly higher standing at $1.05\ M_\odot < M_l < 1.08\ M_\odot$.

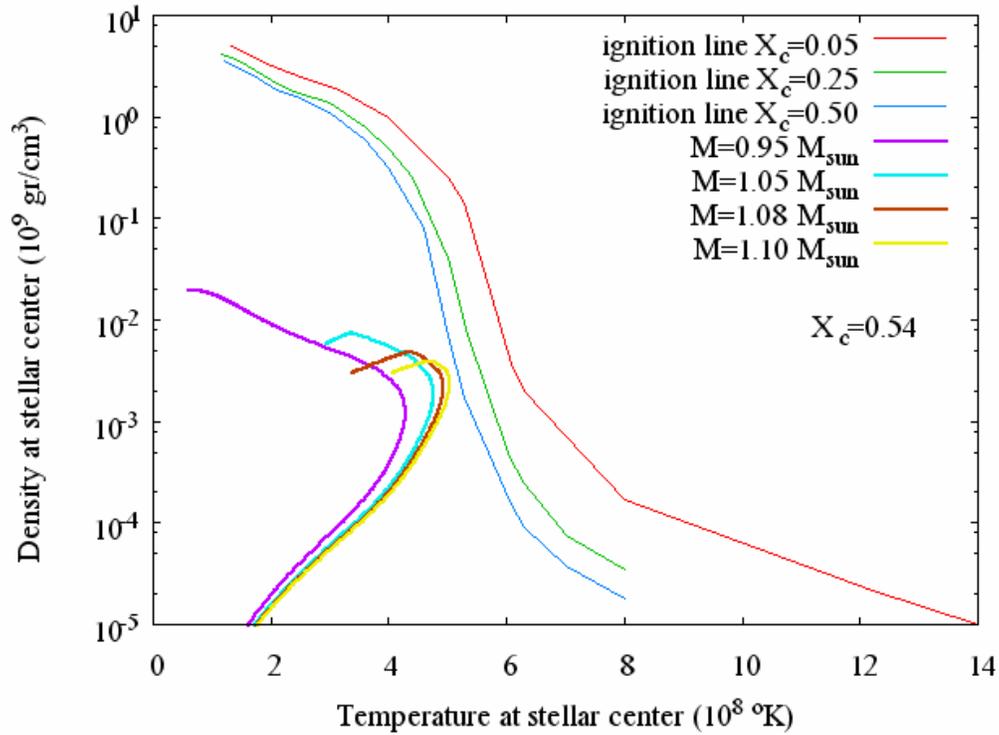

**Figure 3-7: The evolution of the density vs. the temperature at the center of carbon star models in the mass range between 0.95 and 1.10 M$_\odot$, below and above the carbon off-center ignition limit (M$_l$). The carbon mass fraction is X$_c$ = 0.54. The ignition lines are also plotted for various carbon mass fractions.**

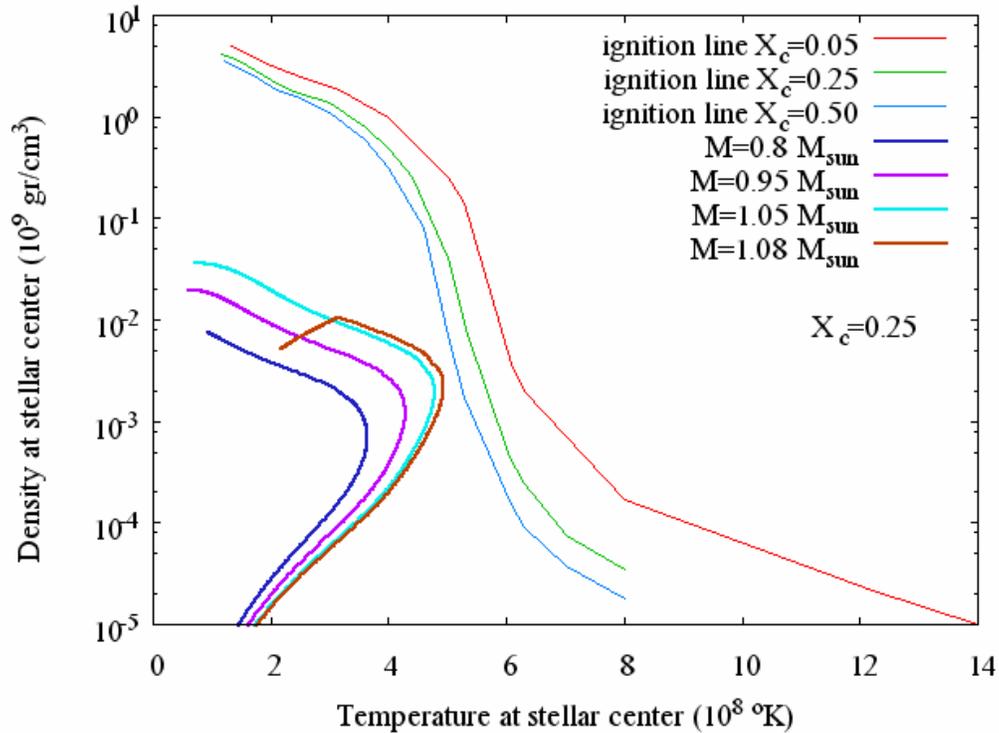

**Figure 3-8: The evolution of the density vs. the temperature at the center of carbon star models in the mass range between 0.95 and 1.10 M$_\odot$, below and above the carbon off center ignition limit (M$_1$). The carbon mass fraction is X$_c$ = 0.25. The ignition lines are also plotted for various carbon mass fractions.**

### 3.1.2. Off-center carbon burning

We will limit our interest to the range $M_1 \leq M_c \leq M_3$. In order to point out the major points of interest in this range, we will first describe a specific example.

We will take as an example a model with mass $M = 1.17\,M_\odot$, and a homogeneous mass fraction profile, i.e. $X_c(m) = X_c(0) = 0.26$. We will investigate the evolution by looking at various physical quantities. Figure 3-9 presents a Kippenhahn diagram, which shows the history of convective regions in the star. As can be seen, the star goes through various evolutionary stages, each one presenting a burning shell topped by a convective region.

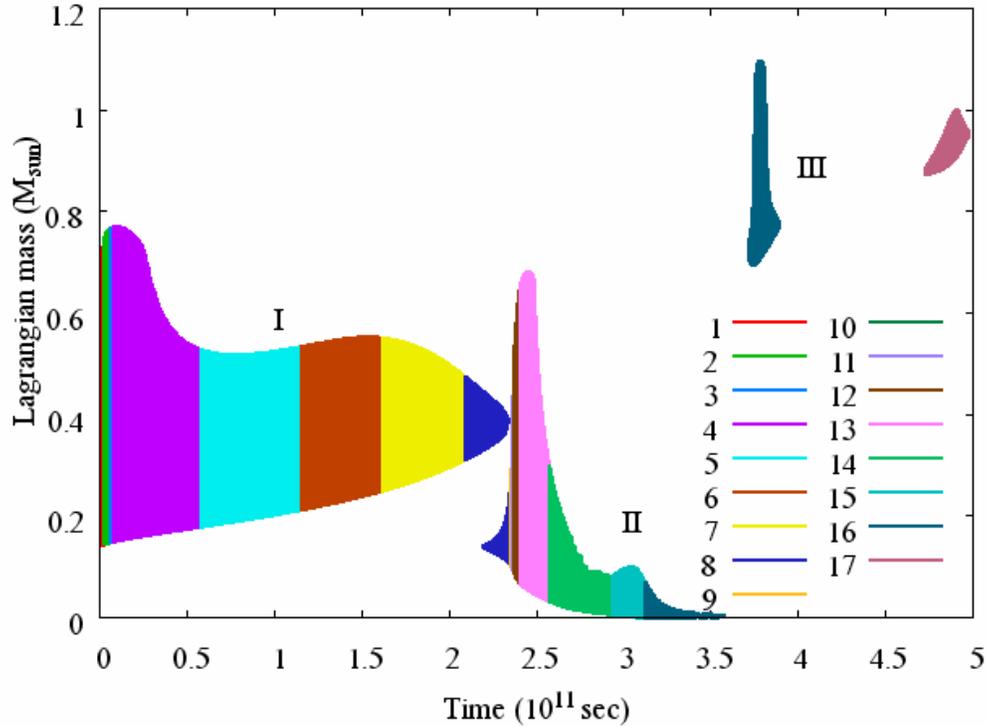

**Figure 3-9: The history of the convective regions during off-center carbon burning in a 1.17 M$_\odot$ carbon oxygen model with initial carbon mass fraction of X$_c$ = 0.26. The numbers I, II, III refer to the various stages of burning referred to in the text below. The colors correspond to the colors of the profiles in the following figures.**

We will describe the three burning stages shown above, by following the evolution of the carbon mass fraction profile.

### 3.1.2.1.    *Stage I*

The evolution of the carbon mass fraction is shown* in Figure 3-10, and we can see that in this case carbon ignites off-center, at $m \approx 0.145\ M_\odot$. The nuclear reaction rate ($q_n$) grows rapidly, and when it exceeds the energy loss rate by neutrino emission ($q_v$), a gradually growing convective burning zone develops, while the carbon mass fraction $X_c$ gradually decreases. At a certain stage (line 4 in Figure 3-10) the convective region begins to shrink, by retreating of both its inner and outer boundaries. This evidently leaves behind a gradient in the carbon mass fraction $X_c$. It is notable, that immediately below the inner boundary of the convective region lies a narrow radiative burning shell, which locally exhausts the carbon at a relatively higher rate. This shell slightly penetrates inward due to conduction. The decline of $X_c$, together with the expansion caused by the rise in entropy of the convective region, finally extinguishes $q_n$. It is important to realize, that due to $q_v$, the nuclear burning $q_n$ extinguishes, albeit $X_c$ has not completely vanished. This is a major point that will have important repercussions in what follows.

---

\* Note that *Gutierrez et al. 2005* has a very similar figure.

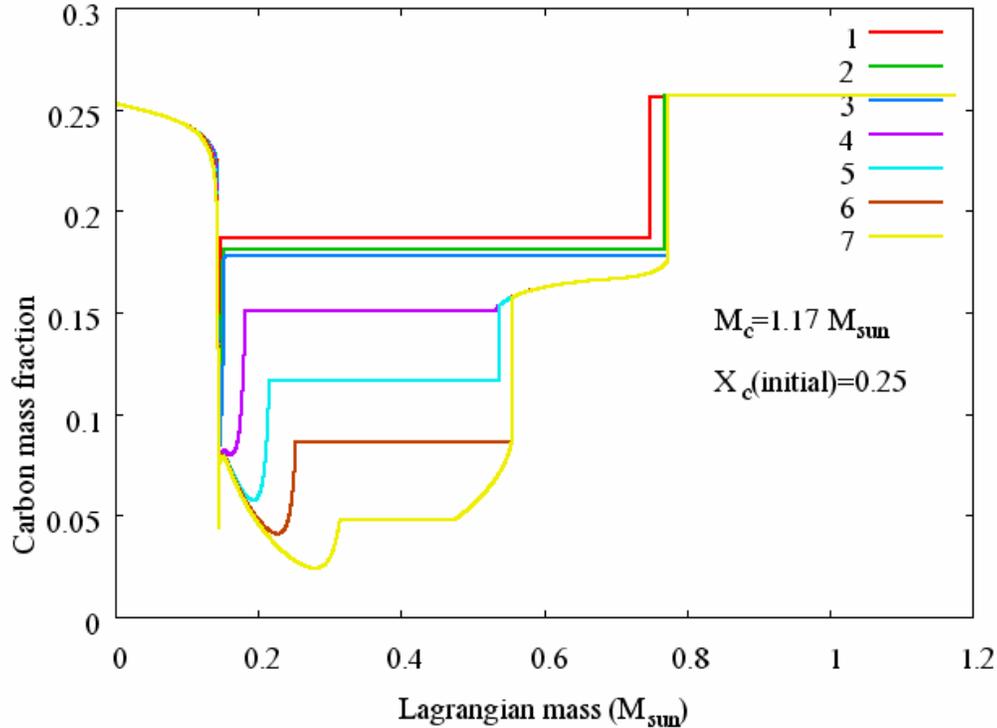

**Figure 3-10: The evolution of the carbon mass fraction profile in a 1.17 M$_\odot$ mass carbon star model during the stage I of off-center carbon burning. Each line represents the mass fraction profile at a different point in time, in the order of numbering in the legend, which corresponds to the numbering in Figure 3-9.**

### *3.1.2.2. Stage II*

After the nuclear burning extinguishes, the star continues to contract, leading momentarily to ignition of carbon in the region of the $X_c$ gradient remaining at the base of the former convective burning zone. Again $q_n$ rapidly rises above $q_v$, and a burning zone penetrates inward forming a convective region above it. The evolution of the carbon mass fraction during this stage is shown in Figure 3-12.

The behavior of $X_c$ in the convective region is complex. Although nuclear burning obviously decreases $X_c$, following the growth of the convective region, variations of $X_c$ due to incorporation of zones richer or poorer in carbon have to be taken into account. And indeed at certain stages $X_c$ in the convective region increases, while at other stages it decreases. This behavior varies from star to star, since it is dependent on the details of the preceding evolutionary stage. It repeats itself several times during subsequent evolutionary stages, and cannot be described in general terms, or as a phenomenon monotonically dependent on the stellar mass or the initial carbon mass fraction $X_c(t=0)$.

Several authors have described the case of an inward advancing carbon burning shell topped by a convective region. *Kawai et al. 1987* followed a carbon burning shell that has been ignited at *m = 1.07 $M_\odot$*, while mass was accreted on the white dwarf at a rate of *2.7 X 10$^{-6}$ $M_\odot$/yr*, while *Saio & Nomoto 1998* found a similar ignition at *m = 1.04 $M_\odot$* for an accretion rate of *1 X 10$^{-5}$ $M_\odot$/yr*. As already mentioned, the conclusion of these two papers is, that the shell penetrates inward only up to a certain point, where it diminishes as a result of the expansion induced by the rise of entropy in the convective region above the shell. After extinction of the shell, the star contracts, and a new shell is ignited at the point where the former one had extinguished, this possibly reoccurring several times. We didn't explicitly interest ourselves in shells ignited far from the center, since in our case the waning of the shell which ignites far from the center is a result of the lack of fuel depleted by previous burning episodes. However, also in the case described in section 3.1.2.1 as "stage I" the expansion contributes to extinguishing the shell and especially its radiative leading part. We will mention that in "stage II" the expansion is minimal since the extent of the convective region is small.

*Timmes et al. 1994* took the effort to estimate, based on reasonable assumptions, the velocity of the shell as a function of the density, temperature and carbon mass fraction, giving a tabulation of their results. A comparison with our results reveals a good agreement. Figure 3-11 shows the inward motion of the shell close to the center of the star. In this case $\rho = 6.4 \times 10^5 \ gr/cm^3$, $T = 7.3 \times 10^8 \ {}^oK$, $X_c = 0.32$. The figure shows that the velocity of the shell is about $1.3 \times 10^{-3} \ cm/sec$. A look at the relevant table in *Timmes et al. 1994* shows a reasonable agreement, since for $X_c = 0.30$ they have a velocity between $8.47 \times 10^{-4} \ cm/sec$ (for $T = 7 \times 10^8 \ {}^oK$) and $6.10 \times 10^{-3} \ cm/sec$ (for $T = 8 \times 10^8 \ {}^oK$). In the work of *Gil-Pons & Garcia-Berro 2001*, dealing with the formation of oxygen – neon dwarves during mass exchange in binaries, at a certain stage carbon is ignited off-center in a manner very much similar to that expected in our case (see figure 9 there), and even the resulting carbon profile is very much like ours (see figure 12 there).

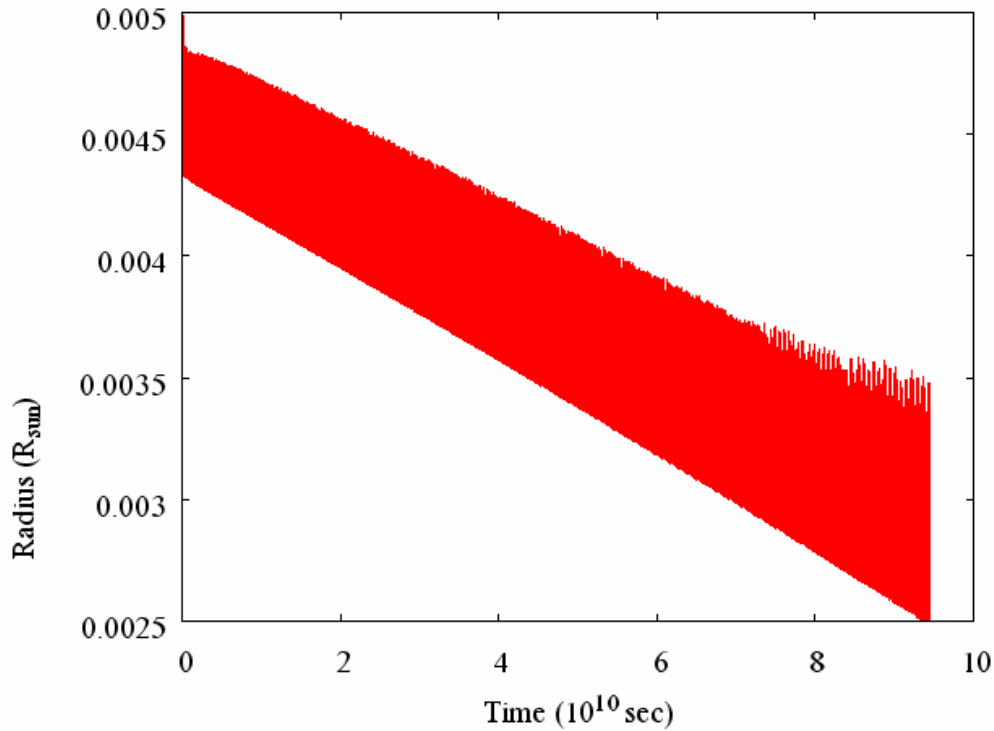

**Figure 3-11: The advancing off-center carbon burning shell – showing its velocity.**

After the nuclear burning flame reaches the center of the star (line 15 in Figure 3-12), the reaction rate starts to weaken with the decrease of $X_c$ and the reduction of the convective region.

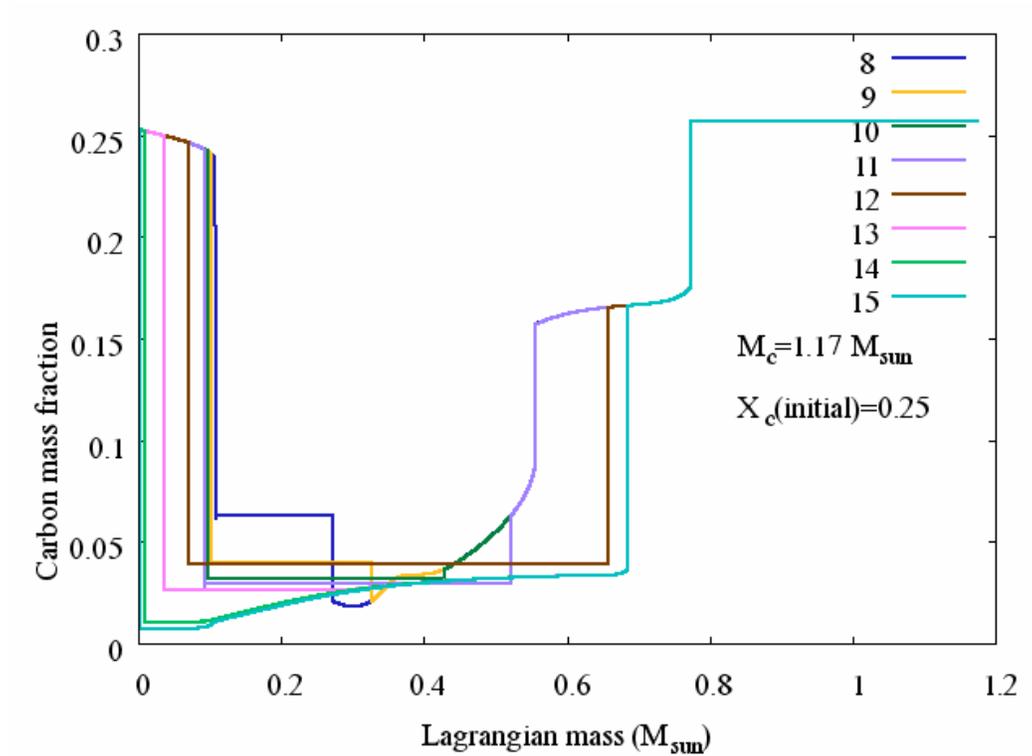

**Figure 3-12: The evolution of the carbon mass fraction profile in a 1.17 M$_{\odot}$ mass carbon star model during stage II of off-center carbon burning. Each line represents the mass fraction profile at a different point in time, in the order of numbering in the legend, which corresponds to the numbering in Figure 3-9.**

Figure 3-13 displays the advance of the flame inwards, and it is seen that it behaves like a regular almost self similar front. Note however, that close to the center the flame width narrows, and thus special care must be taken, otherwise the flame might incorrectly die out leaving an inner zone of unburned carbon.

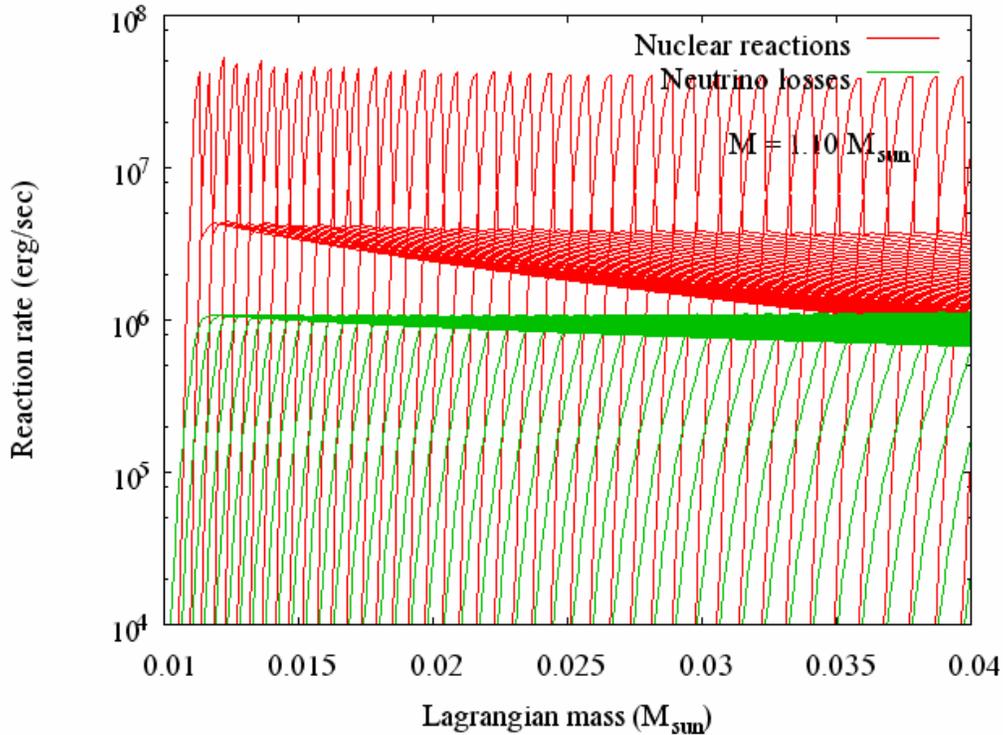

**Figure 3-13: The inward advancing carbon burning front during "stage II" in a 1.10 M$_\odot$ mass carbon – oxygen model. The nuclear reaction rate (red) and neutrino loss rate (green) are shown at various times.**

### 3.1.2.3.    *Stage III*

In the following stage, as a result of the contraction of the star, the carbon reignites, usually in a relatively carbon-rich zone, above the extent of the convective regions of the previous stages.

Also in this case a convective region develops, which grows up to a certain extent, and then the reactions extinguish again. At the same time a radiative burning front develops, which advances inward to the carbon-poor region along the composition gradient which has remained there due to the previous burning. We find (see Figure 3-14) that this flame gradually decays, and is only able to penetrate slightly.

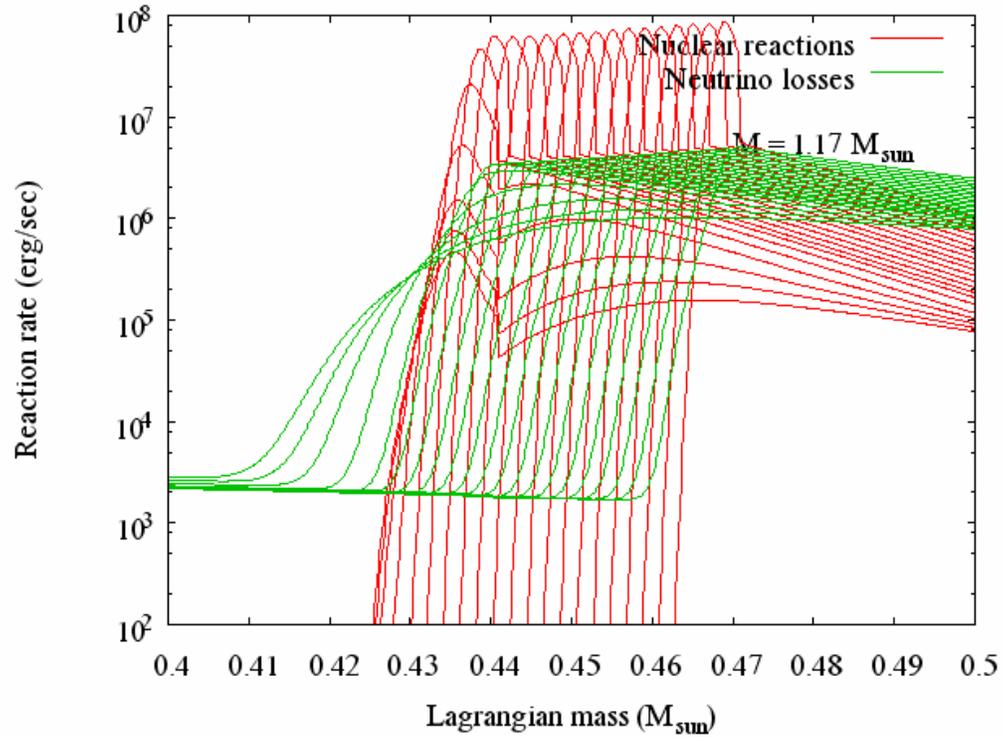

**Figure 3-14: The decay of the inward advancing carbon burning front during "stage III" in a 1.17 M$_\odot$ mass carbon – oxygen model. The nuclear reaction rate (red) and neutrino loss rate (green) are shown at various times.**

The evolution of the carbon mass fraction during this stage is shown in Figure 3-15.

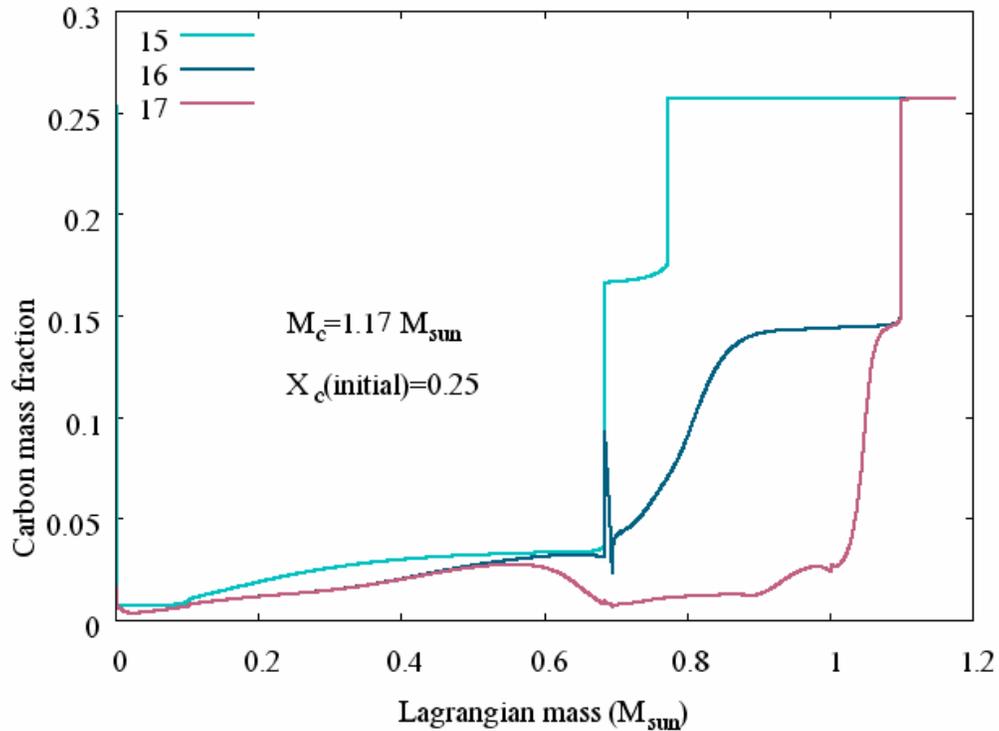

**Figure 3-15: The evolution of the carbon mass fraction profile in a 1.17 M$_\odot$ mass carbon star model during stage III of off-center carbon burning. Each line represents the mass fraction profile at a different point in time, in the order of numbering in the legend, which corresponds to the numbering in Figure 3-9.**

In some cases there are more stages where carbon is ignited in carbon-rich outer areas (Figure 3-9). Eventually, further contraction doesn't result in carbon ignition, and the star evolves toward a white dwarf, with a "frozen" carbon profile. Figure 3-16 displays this final profile for the main elements present in the star – carbon, oxygen, neon and magnesium.

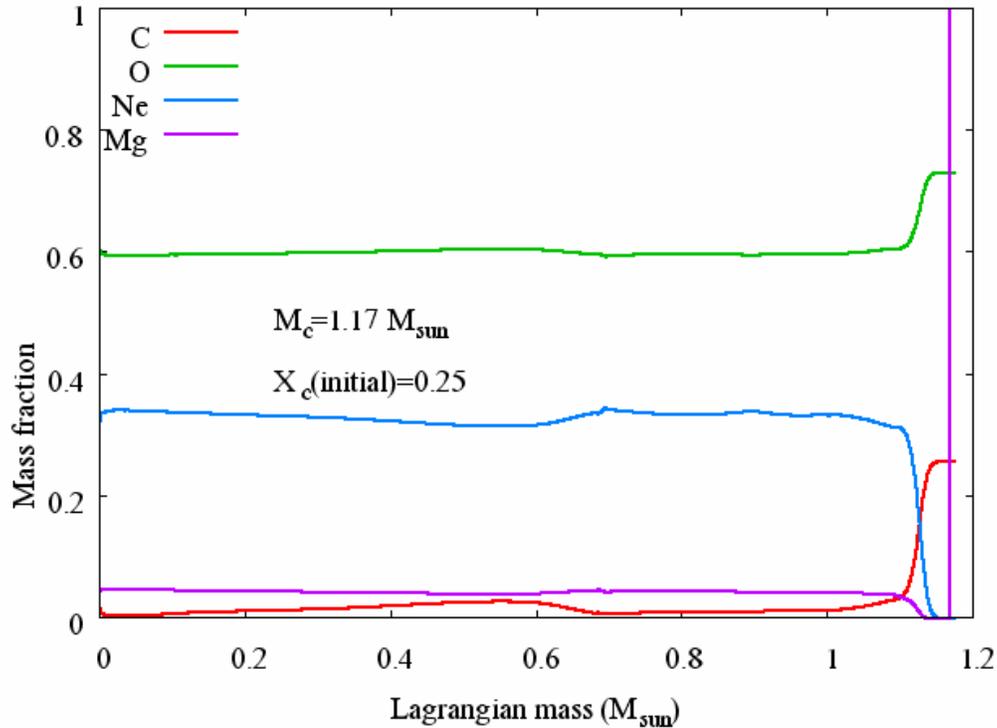

**Figure 3-16: The final element mass fraction profile in a 1.17 M$_\odot$ mass carbon star model after off-center carbon burning.**

Figure 3-17 displays* the carbon mass fraction profile at the end of carbon burning for various masses and initial carbon mass fraction of $X_c(t=0) = 0.54$. Figure 3-18 displays the carbon mass fraction profile at the end of carbon burning for various initial carbon mass fractions and total mass of $M_c = 1.22\ M_\odot$. We can see that the typical profile has a "bump" around the center, above it a region almost devoid of carbon, and in most cases a small carbon-rich zone near the outer boundary.

---

* Note the change of scale.

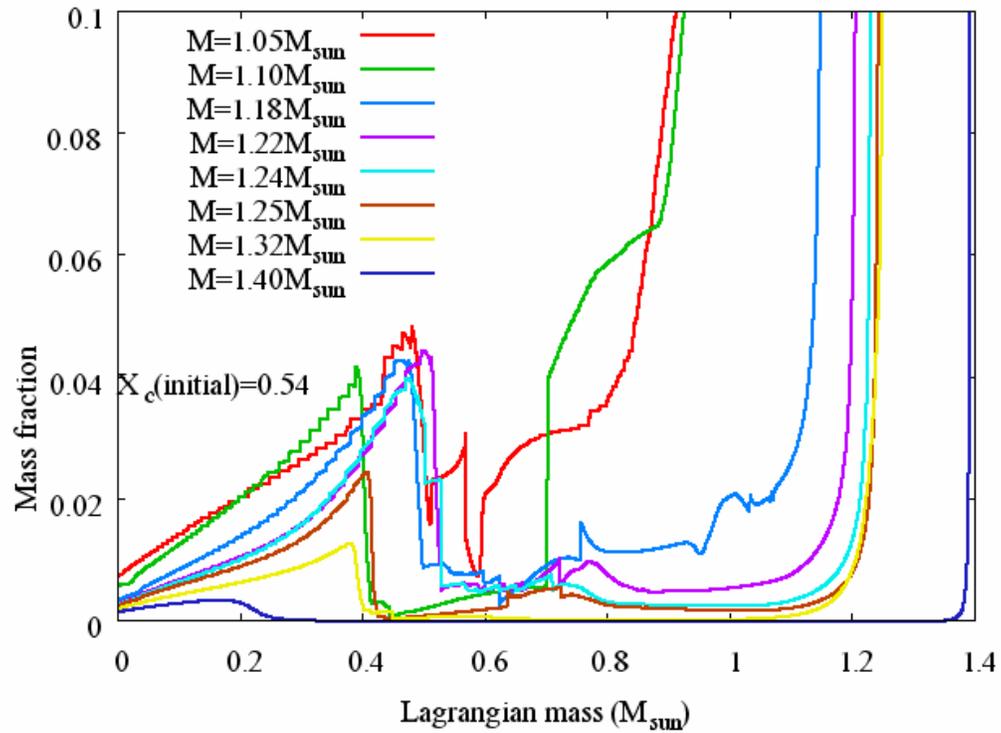

**Figure 3-17: The final carbon mass fraction profile in carbon star models of various masses and initial carbon mass fraction of $X_c$=0.54 after off-center carbon burning.**

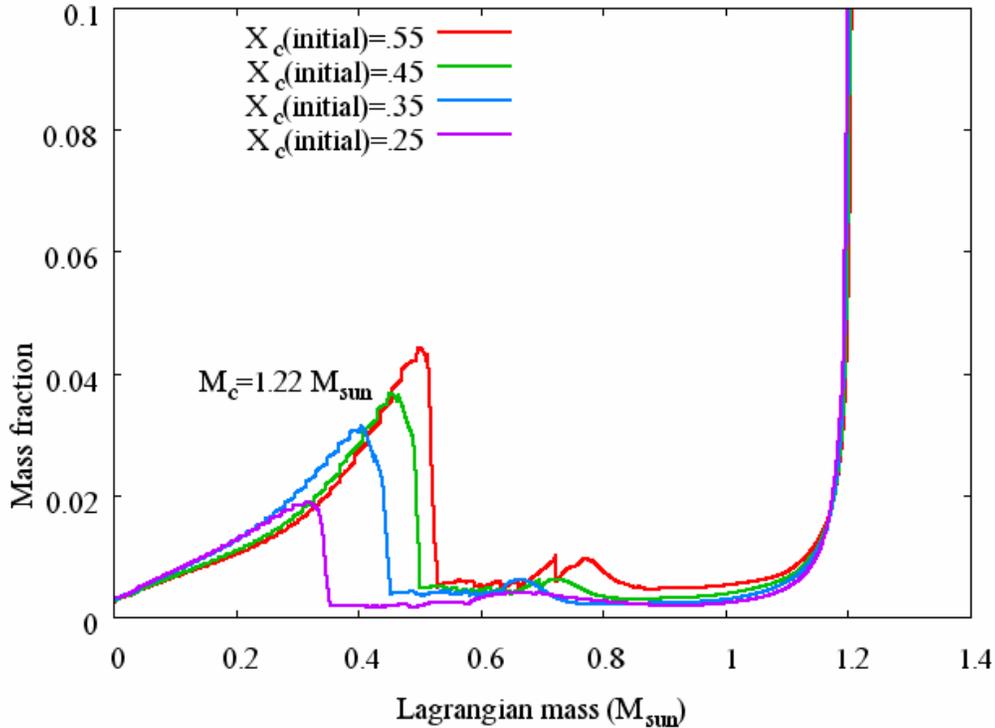

**Figure 3-18: The final carbon mass fraction profile in carbon star models of various initial carbon mass fractions with total mass of 1.22 M$_\odot$ after off-center carbon burning.**

The widely accepted scenario for the onset of a type I supernova explosion is accretion of matter by a white dwarf, therefore in the following we will discuss the various possibilities of accretion onto the white dwarves resulting from the evolutionary scenarios described above.

### 3.1.3. Accretion

For the purpose of understanding the various possibilities, we used the standard technique to investigate the outcome of accretion as a function of the relevant parameters. In a realistic case the accreted matter could be hydrogen, helium, carbon or a mixture of these. In the case of hydrogen or helium accretion, these also ignite, and form a burning shell advancing outward, leaving behind a newly-made layer of carbon – oxygen. Thus the composition of the accreted matter influences the effective rate of carbon – oxygen accretion. Variations in the entropy of the accreted matter can also be represented by dictating a suitable effective accretion rate. As we will show, the carbon mass fraction in the accreted matter can be of importance, but even then the main parameter is $\dot{M}$ .

It is known that (*Nomoto & Sugimoto 1977*, *Nomoto et al. 1984*) the relevant $\dot{M}$ for a supernova has to be:

*(3-1)* $$\dot{M} \geq 7.5 \times 10^{-7} \, M_{\Theta} \, / \, yr$$

This rate results from the mass-luminosity relation given originally by *Paczynski 1970* for a double shell of hydrogen and helium burning:

*(3-2)* $$L/L_{\Theta} = 59250 \left( M/M_{\Theta} - 0.52 \right)$$

Translating this relation to $\dot{M}$ gives:

*(3-3)* $$L = q = \dot{M} X Q_{\nu}$$

Here $X$ is the hydrogen mass fraction, and $Q_{\nu}$ is the *Q-value* of hydrogen burning.

It is worth to note, that in the case of helium burning, $\dot{M}$ is expected to be, and indeed is, higher by at least* a factor of about 10, since $Q_{\nu}$ is lower.

We decided to map the results for a wide range of $\dot{M}$ :

*(3-4)* $$7.5 \times 10^{-7} \, M_{\Theta} \, / \, yr \leq \dot{M} \leq 3 \times 10^{-4} \, M_{\Theta} \, / \, yr$$

---

* The luminosity itself is higher.

It is important to note, that for rates in this range, the initial mass of the white dwarf is insignificant, due to the "convergence feature" of the evolutionary paths. This argument is not to be confused with the fact, that when dealing with much smaller accretion rates, the initial mass is of much more importance.

In order to compare our results to the literature, we chose from among the many publications on this subject the extensive survey by *Nomoto & Sugimoto 1977*, followed by a study of the dependence on the accretion rate by *Sugimoto & Nomoto 1980*. It was important for us to verify that the evolutionary paths we get are similar to the literature, and thanks to the uniqueness of this path (for given accretion and neutrino loss rates) as mentioned above, this is indeed the case. It is obvious, that the onset of carbon burning depends also on the mass fraction, and we are especially interested in the cases where this mass fraction is particularly low.

### *3.1.3.1.    Homogeneous models*

In order to lay the foundations for understanding the outcome of accretion on the various models we got, we first studied the outcome of accretion on models having a homogeneous composition. We took as an example a model with mass of *1.18 $M_\odot$*, which lays in the range of interest regarding off-center burning, as explained in section 3.1.2.

Standard models in the literature mostly assume a carbon – oxygen mixture in the range *$0.25 \leq X_c \leq 0.55$*, as the product of helium burning. These values are appropriate for the stellar mass range where carbon is not ignited on the way towards the white dwarf state, i.e. for masses smaller than about *8 $M_\odot$*. As mentioned in section 3.1.2.2, the composition profile could extensively vary in the off-center ignition range, therefore we investigated the results in the range of *0.001* to *0.025* accordingly.

The accretion rate varied in the range mentioned in equation (3-4). For reasons which will be clarified in the following, the accreted matter didn't contain carbon at all.

Figure 3-19 shows the evolution up to explosion for several compositions with accretion rate of $\dot{M} = 7.5 \times 10^{-7} M_{\Theta} / yr$.

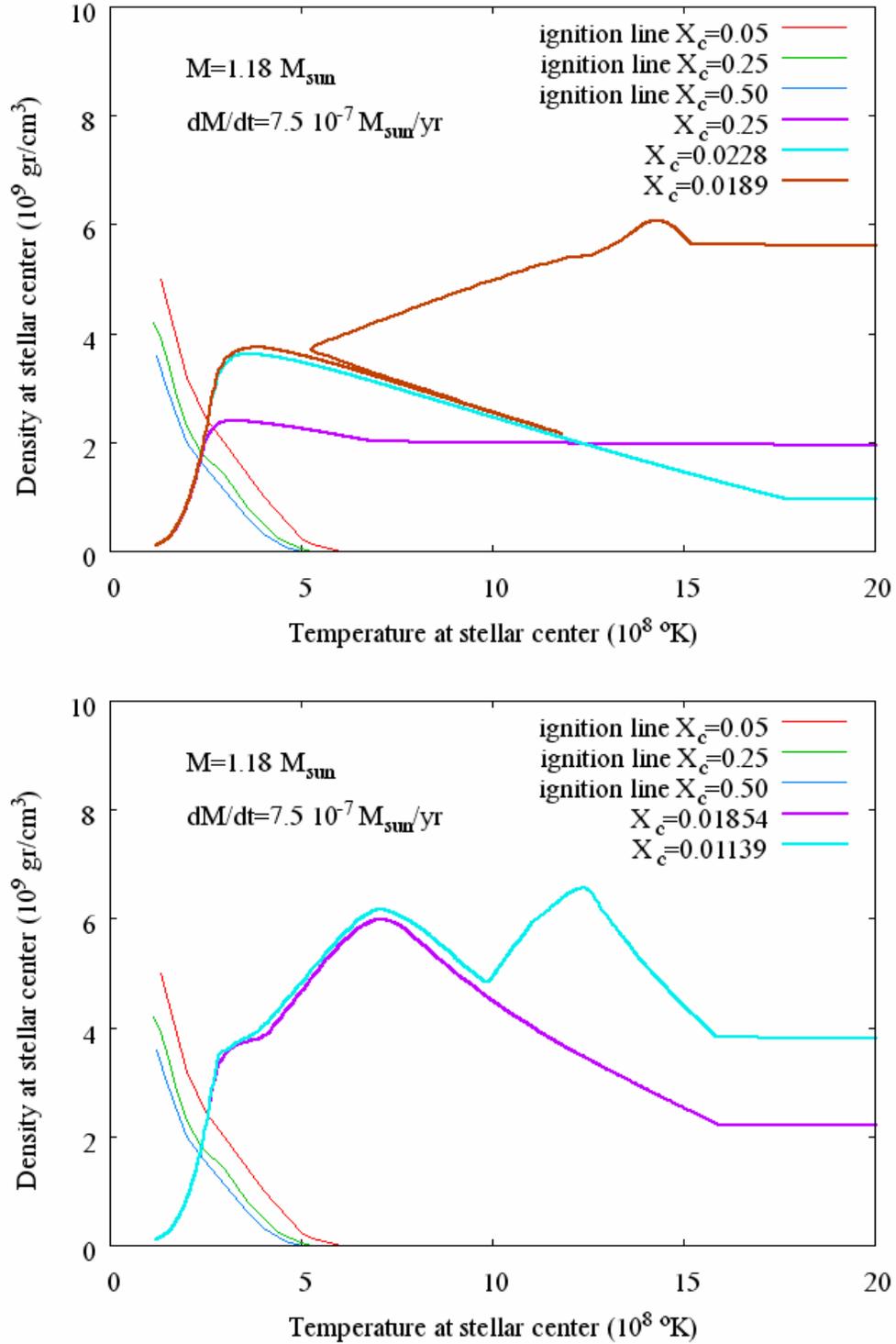

**Figure 3-19:** The effect of the initial composition on the evolution of accreting <span style="color:red">homogeneous composition</span> carbon star models towards explosive ignition of carbon. All models have an initial mass of 1.18 $M_\odot$, and accrete mass at a rate of 7.5 X $10^{-7}$ $M_\odot$/yr.

A more detailed discussion of what happens beyond ignition is given in 3.1.4. For clarity we describe here the results in general terms.

As expected, the case where $X_c = 0.25$ (hereafter "standard model") reconstructs the results known from the literature, i.e. $RA$ at about $\rho_c \approx 2 \, X \, 10^9 \, gr/cm^3$. For the small $X_c$ we are interested in, ignition occurs at higher densities (but at nearly the same temperature) as the "standard model", however runaway occurs at a higher temperature and at various densities.

For example for $X_c = 0.0228$ ignition indeed occurs at higher density, but runaway occurs at a density lower than the "standard model". A thorough observation shows the reason is, that while in the "standard model" runaway occurs as there is still carbon present, and therefore the runaway temperature is around $10^9 \, {}^oK$, in this case all the carbon has been exhausted, but has succeeded to raise the temperature to a level where oxygen burning succeeds to keep on raising the temperature up to the runaway temperature for oxygen – $1.8 \, X \, 10^9 \, {}^oK$, and therefore the relaxation continues longer.

Reducing the carbon mass fraction to e.g. $X_c = 0.01891$, we cross a threshold below which $DEC$ (which occurs below the density threshold for electron capture) fails to produce runaway before carbon is exhausted, and the center begins to cool and contract back. However, soon the threshold of electron capture is crossed, causing heating and accelerating contraction, finally ending in the ignition of oxygen and runaway.

For an even lower carbon mass fraction of e.g. $X_c = 0.01854$, we cross another threshold, below which the density threshold for electron capture is crossed before

*DEC*, so that it occurs at a considerably higher density. With the aid of *EC*, even though the carbon mass fraction was low, it did succeed in inducing runaway.

Finally, for an even lower carbon mass fraction of e.g. $X_c = 0.01139$ carbon is again exhausted before causing runaway, but subsequent electron capture rises the temperature and density enough to ignite oxygen, and thus leading to runaway.

Figure 3-20 demonstrates the effect of varying the accretion rate for the homogeneous model of *1.18 $M_\odot$* with a carbon mass fraction of $X_c = 0.0228$. We can see, that all models reach runaway at almost the same density and temperature. Figure 3-21 shows the same for a carbon mass fraction of $X_c = 0.01891$. Here we can see that for high enough accretion rate runaway occurs again at almost the same density and temperature as for the $X_c = 0.0228$ case, but for lower accretion rate runaway fails to arise before carbon is exhausted, and comes about only after contraction and heating due to electron capture.

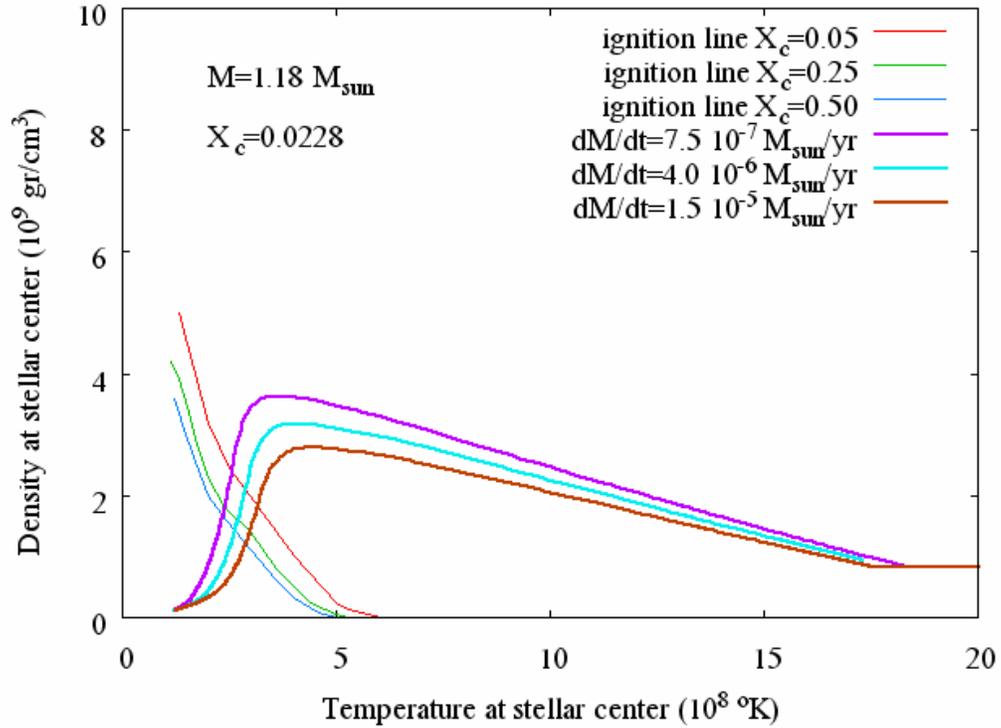

**Figure 3-20: The effect of accretion rate on the evolution of accreting <span style="color:red">homogeneous composition</span> carbon star models towards explosive ignition of carbon. All models have an initial mass 1.18 M$_\odot$ and a homogeneous carbon mass fraction of 0.0228.**

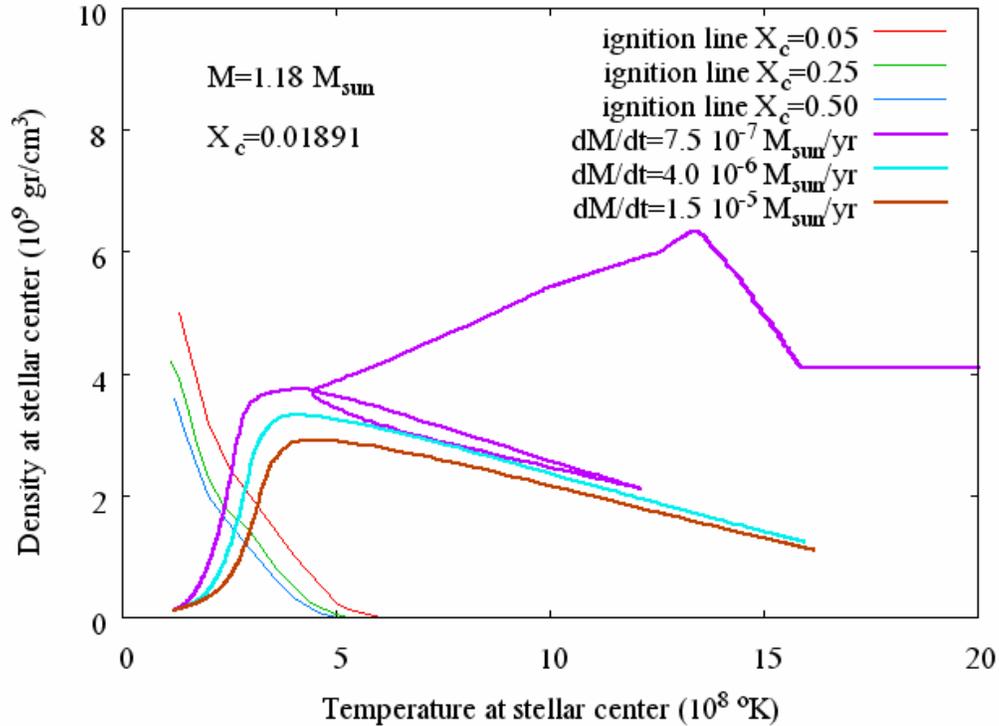

**Figure 3-21: The effect of accretion rate on the evolution of accreting <span style="color:red">homogeneous composition</span> carbon star models towards explosive ignition of carbon. All models have an initial mass 1.18 M$_\odot$ and a homogeneous carbon mass fraction of 0.01891.**

### 3.1.3.2.    Realistic models

We repeated the study of accretion on realistic models, i.e. the end products of evolution as described earlier. Figure 3-22 shows these results for a model with mass of *1.18 M$_\odot$*, for various accretion rates in the same range as above. As expected, the general behavior is similar to the homogeneous case, while quantitative differences exist due to non-homogeneous composition profiles.

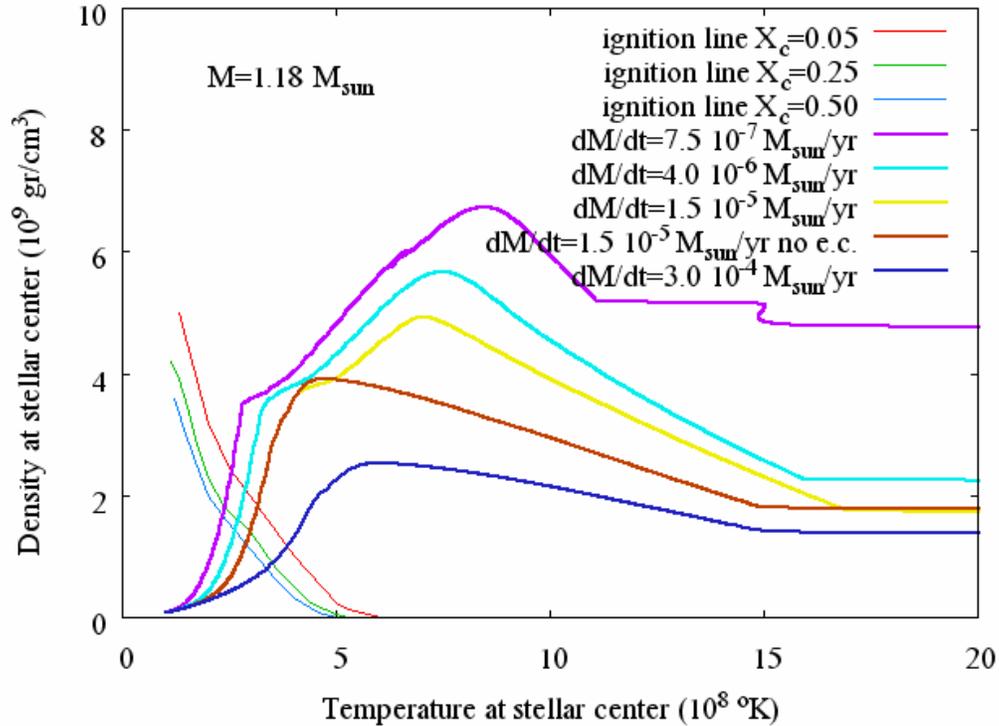

**Figure 3-22: The effect of accretion rate on the evolution of accreting realistic (inhomogeneous composition) carbon star models towards explosive ignition of carbon. All stars have an initial mass 1.18 M$_\odot$. All models have electron capture onto Mg$^{24}$ taken into account, except one case labeled "no e.c."**

As mentioned, all the above results are for accretion of matter devoid of carbon. This choice has two reasons:

1. We shall later show, that in the case of helium stars, the growth of the nucleus is indeed achieved by addition of matter almost devoid of carbon.

2. It turns out that, as previously mentioned, in the case of carbon accretion at high $\dot{M}$ above a certain threshold, which depends on the mass fraction of carbon in the accreted matter, carbon ignition occurs in the accreted layer, and a burning front develops, advances inward, and as already mentioned (section 3.1.2.2), might reach the center before explosion occurs. We didn't follow the front in this case, but we mapped the combinations of accretion rates and carbon mass fractions which lead to such a case. Figure 3-23 demonstrates this for a carbon – oxygen model of mass *1.12 M⊙*. We can see, that for an accretion rate of *7.5 X 10⁻⁷ M⊙/yr*, carbon does not ignite in the accreted layer, even if the carbon mass fraction in the accreted matter is as high as *0.5*. For a higher accretion rate of *7.5 X 10⁻⁶ M⊙/yr*, such ignition does not occur for a carbon mass fraction of *0.01*, but does occur for *0.05* and above.

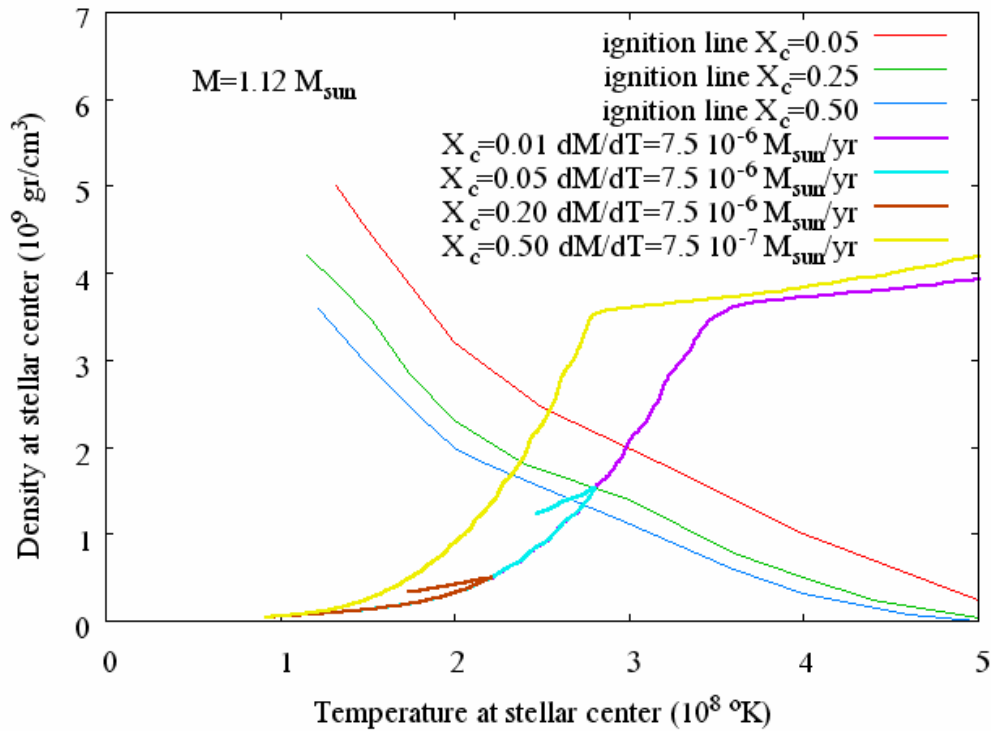

**Figure 3-23: The effect of the carbon mass fraction in the accreted matter, for various accretion rates, on a carbon – oxygen model of mass 1.12 M⊙.**

### 3.1.4. The influence of various parameters beyond ignition

Beyond ignition a convective region is formed, which goes on growing while supplying fuel to the nuclear flame, but at the same time also inducing expansion.

Clearly, convection will continue as long as the convective turn-over time-scale $t_{convec}$ is shorter than the fuel exhaustion time-scale $t_{thermo}$. Similarly, it is clear that the transition to a dynamic regime, i.e. to a situation where assuming hydrostatic equilibrium is no longer valid, will occur when the dynamic time-scale $t_{hydro}$ (which is of order of magnitude of a characteristic length divided by the speed of sound) is no longer short compared to $t_{thermo}$.

In section 2.3 above we discuss the subject of timescales in detail, including the definition of these quantities and treatment of the relevant scenarios. As we already mentioned there, our treatment implies comparing $t_{convec}$ and $t_{thermo}$, and turning convection off where $t_{convec} > \alpha\, t_{thermo}$, where $\alpha$ is a "fudge factor". Figure 3-24 shows the sensitivity to this factor $\alpha$, and we can see that the smaller it is (i.e. convection is turned off earlier) runaway occurs earlier, i.e. at higher $\rho_c$, and lower $T_c$, since at this stage convection causes expansion and thus hinders runaway.

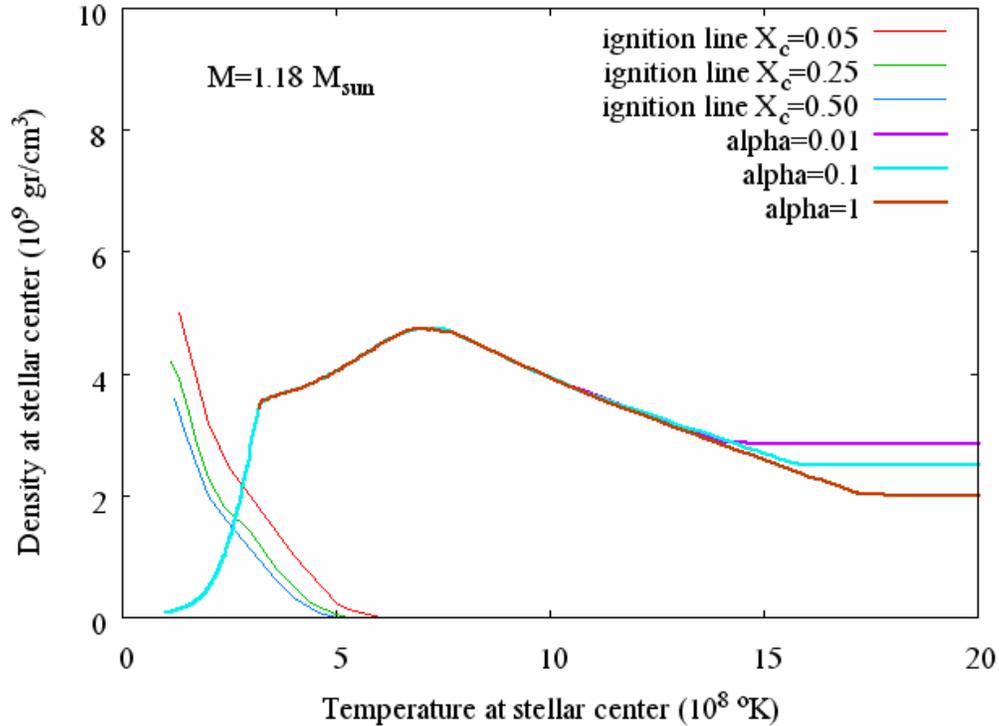

**Figure 3-24: The influence of the criterion for stopping the convection during runaway (the parameter α as defined in section 2.3) on accreting realistic (inhomogeneous composition) carbon star models. All models have an initial mass 1.18 M$_\odot$ and accrete matter at a rate of 4 X 10$^{-6}$ M$_\odot$/yr.**

### 3.1.5. The effect of electron capture processes

As we have seen, electron capture on the products of carbon burning has a major effect on the outcome of carbon ignition. The thermal *Urca* (*TU*) has a minor effect, and convective *Urca* (*CU*) has a reasonably small effect as we will now demonstrate. The method by which we modeled these processes is described in section 2.2.3.

#### 3.1.5.1. Electron capture on carbon burning products

Keeping in mind the existing uncertainties for this process (see section 2.2.3.3), we checked the sensitivity of the results by modifying the mass fraction of *Mg$^{24}$*, which in our standard models came out to be about *0.19*. Figure 3-25 shows the results, and we can see that lowering the mass fraction of *Mg²⁴* has a small effect causing earlier ignition, and thus runaway at lower $\rho_c$.

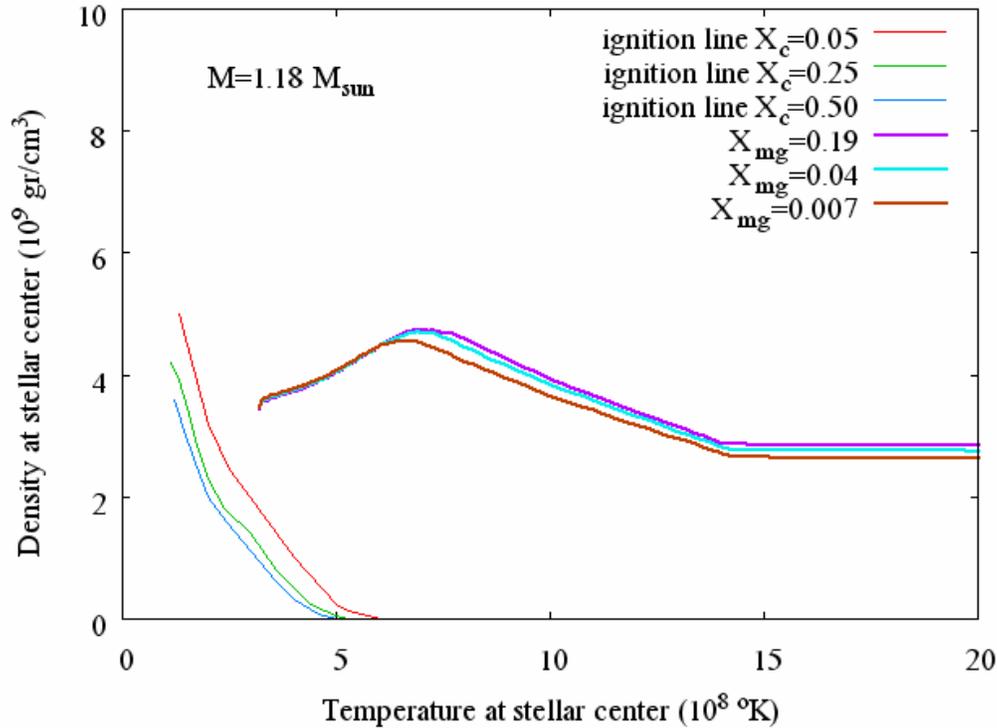

**Figure 3-25: The influence of the mass fraction of Mg²⁴ on the ignition and runaway of accreting realistic (inhomogeneous composition) carbon star models. All models have an initial mass 1.18 M⊙ and accrete matter at a rate of 4 X 10⁻⁶ M⊙/yr.**

### 3.1.5.2. Thermal Urca (TU)

Figure 3-26 shows the influence of the *TU* process for our "realistic" model of *1.18 M⊙*, for various mass fraction of the *Urca* nucleus *Na²³*. We get a minor perturbation only, which is manifested as a deviation in the $\rho_c$, $T_c$ path to lower temperatures due to the local cooling. As contraction continues, the *US* moves outward together with the local heat sink. When the *US* is far enough from the center, the density and temperature at the center start rising again, and shortly the path

coincides with that of the *TU*-less case. Comparison with results from the literature (*Gutierrez et al. 2005*) shows a satisfying agreement.

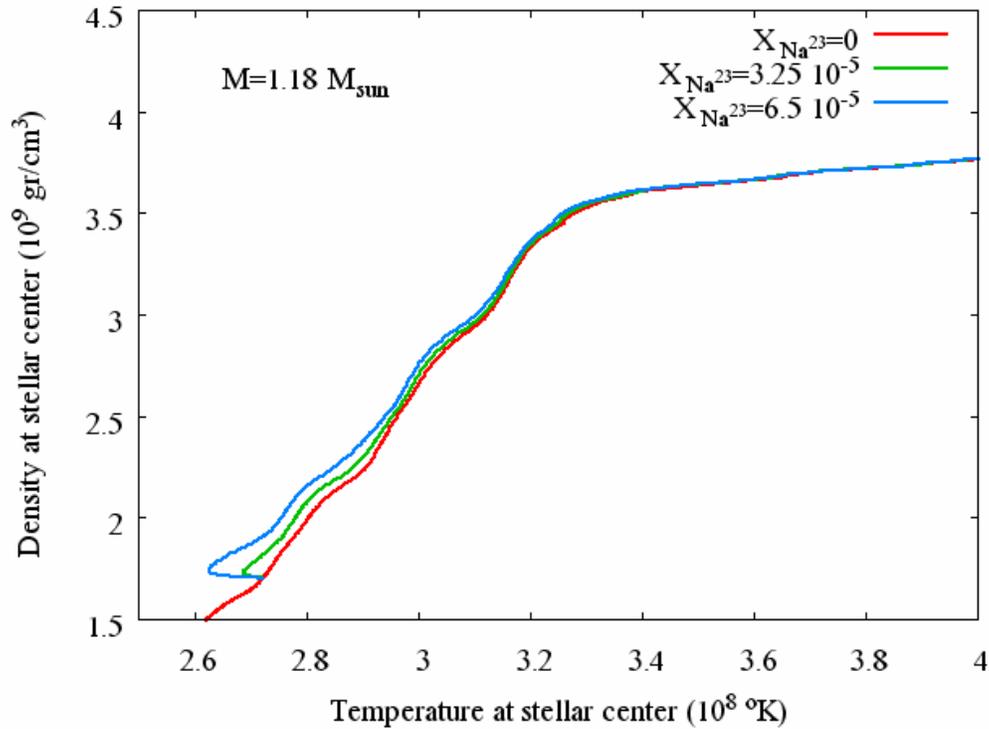

**Figure 3-26: The influence of the mass fraction of Na$^{23}$ for the thermal Urca process on the ignition and runaway of accreting realistic (inhomogeneous composition) carbon star models. All models have an initial mass 1.18 M$_{\odot}$ and accrete matter at a rate of 4 X 10$^{-6}$ M$_{\odot}$/yr.**

### 3.1.5.3.    *Convective Urca (CU)*

As we already mentioned, we modeled *CU* by artificially forbidding convection above the *Urca* shell, and examined its effect by varying the Fermi energy threshold at which the *Urca* shell is located.

This works in two opposite directions. On one hand the entropy and with it the temperature increase faster (which promotes approaching *RA*), but on the other hand there is less supply of fresh fuel, which can suppress burning, and might prevent the *RA*. Clearly, lack of fresh fuel supply might have cardinal importance in a case where the mass fraction of fuel is a-priori low.

We find that a feedback mechanism exists. When the star contracts, the *US* has to move farther away from the center (due to increase in $E_f$), thus the outer boundary of the convection moves outward as well, so that the effectiveness of limiting the extent of the convective region on diminishing the fuel supply is small in all the cases we checked. It is interesting to note, that when the star goes through an expansion phase, the situation is the opposite – the *US* approaches the center, and may limit the convective region. If this limiting is significant enough while the temperature is already high enough, it might induce an earlier runaway, which will thus occur at a higher density.

Figure 3-27 presents a comparison between the undisturbed case, where $E_f$ peaked at about *6.26 MeV*\*, with cases where convection was limited at $E_f = 4.4\ MeV$ (the threshold of $Na^{23}$), $E_f = 5.7\ MeV$ (the threshold of $Ne^{21}$), and $E_f = 5.2\ MeV$ (an intermediate value*)*.

We can see that the higher the threshold the earlier *RA* occurs, i.e. at a higher density and lower temperature. It is clear that this tendency is limited, since should we raise the threshold to the vicinity of *6.26 MeV*, the Fermi energy throughout our model will be below the threshold, and we will effectively be back to the case with no *CU* at all.

---

\* At the same time the US of $Ne^{21}$ was located at 0.09 $M_\odot$, and that of $Na^{23}$ at 0.45 $M_\odot$.

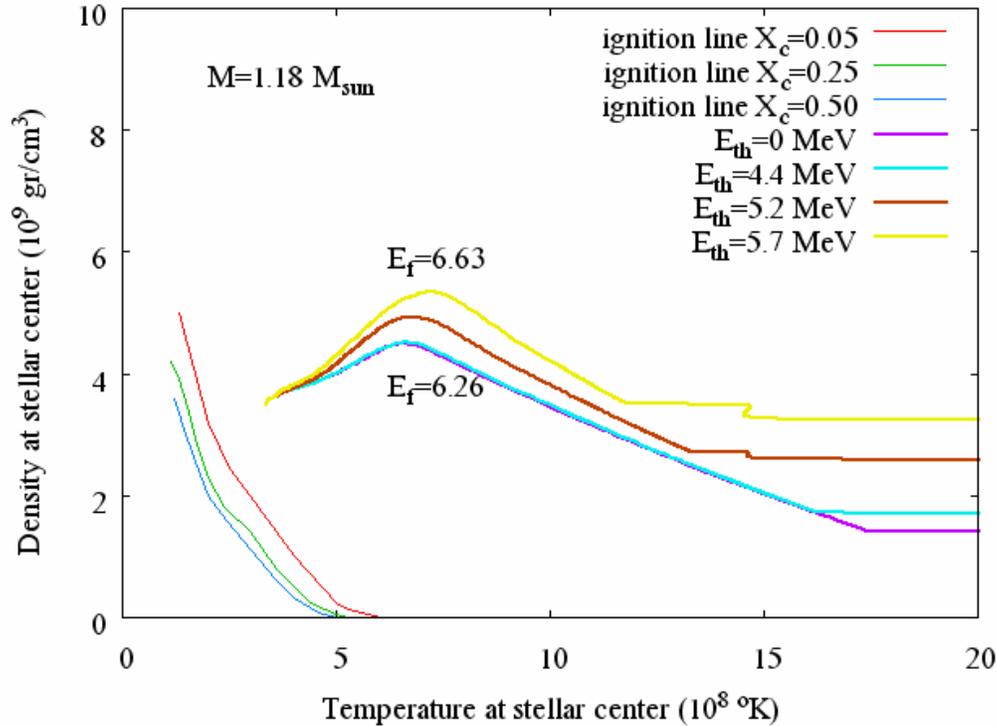

**Figure 3-27: The influence of the Fermi energy of the convective Urca shell on the ignition and runaway of accreting realistic (inhomogeneous composition) carbon star models. All models have an initial mass 1.18 M$_\odot$ and accrete matter at a rate of 4 X 10$^{-6}$ M$_\odot$/yr.**

Obviously the above conclusions are only a rough estimate, and should be in fact examined by means of multidimensional simulations, which are still difficult to undertake. However, we have reason to believe that the *CU* has no significant influence on the conclusions of this work.

### 3.1.6. Summary

In addition to the accepted scenario, where carbon – oxygen stars reach the white dwarf phase without carbon ignition, and then accrete matter until explosion, there exists a range where due to early ignition, white dwarves with a much smaller mass fraction of carbon are created. We mapped the behavior of these stars up to the explosion, and the main conclusion is, that a wide variety of models can be expected, where explosion can take place at various densities, mostly, but not exclusively, higher than usual, and with a much smaller amount of carbon. These two facts can have importance during explosion, both regarding the explosion mechanism and the nucleosynthesis.

## 3.2. Helium stars

### 3.2.1. Core formation

Immediately following helium ignition at the center, a helium star develops a convective core, which gradually grows until equilibrium is reached between the energy released by nuclear burning and energy loss by radiation transport. The mass of the core is a function of the stellar mass $M_{he}$, but also of the method by which the boundary of the convective region is determined. A variety of methods is available in the literature, which take into account in some way phenomena like semi-convection and overshooting. Since it is obvious that the evolution of the star following the extinction of nuclear burning is in fact dependent only on the mass of the core, we chose to allow for a range of values (by means of varying the numerical setup, e.g. the numerical zoning, the time step, etc.) for each case.

The composition of the helium burning residue is as well known dependent on the cross section $C(\alpha,\gamma)O$, for which an uncertainty range still exists. We studied a selection of cases, for which we varied the cross section by changing the value of the relevant parameter in our nuclear reaction rate library in the range of uncertainty, resulting in the residual carbon mass fraction varying in the range *0.2 – 0.6* accordingly. These values are known as typical for helium stars in the mass range of our interest, and are insensitive to stellar mass. It is worthwhile mentioning, that for much larger helium stars, carbon mass fraction will be significantly lower.

At the conclusion of helium burning in the core, convection retreats and burning extinguishes. Immediately following, a helium burning shell is ignited in the region above the core, where helium of course is still present. This shell extends at first over a wide region, gradually narrows, and advances outward radiatively, thus increasing the core. Regarding the residual composition, the picture is somewhat different than in the core burning stage, featuring a somewhat smaller mass fraction of carbon.

To compare our results with the literature we looked at the mass of the carbon – oxygen core as a function of the helium star mass. The thorough and detailed comparison carried out by *Habets 1985* shows a considerable scatter, e.g. for a star of *2 $M_\oplus$* we can see (table 4 in the above) that the mass of the carbon – oxygen core lies between *1.10* and *1.35 $M_\oplus$*. A meticulous look reveals, that it is difficult to point out the source of the discrepancies for each case, but beyond using different input physics (chiefly nuclear reaction rates and neutrino loss rates), it is clear that the algorithms used to determine the boundaries of the convective regions have a major consequence. Here we can list mainly the *Ledoux* vs. *Schwarzschild* criteria, and the various treatments of semi-convection and overshoot.

Similarly, the type of carbon ignition – i.e. center or off-center varies among the different authors, here the carbon mass fraction playing the main role.

We chose, as aforementioned, a simple as possible treatment, where we forced the carbon – oxygen core to achieve varying values in the abovementioned range, and checked the sensitivity of our results to variations in the diverse physical parameters.

We usually applied the *Schwarzschild* criterion, but in several cases we checked the difference with the *Ledoux* criterion. We are convinced that the overall picture we achieved is valid, although it can be "shifted" by using different methods or input physics.

For reasons which will be clarified later, we will discuss the behavior of the helium burning shell in two stages.

### 3.2.2.    The radiative helium burning shell

It turns out, that the luminosity of the shell, which of course determines the rate of growth of the core, rises gradually as the shell advances outward. As the shell moves outward the structure of the core below it gradually approaches that of a typical dwarf (see Figure 3-29). It is interesting that when we compare the behavior of the luminosity in various models it turns out (Figure 3-28), that while at first the luminosity depends on both the location of the shell and the mass of the star, we find that once the location of the shell is on top of the dwarf-like structure a convergence occurs, to a value depending only on the location of the shell, and not on the mass of the star.

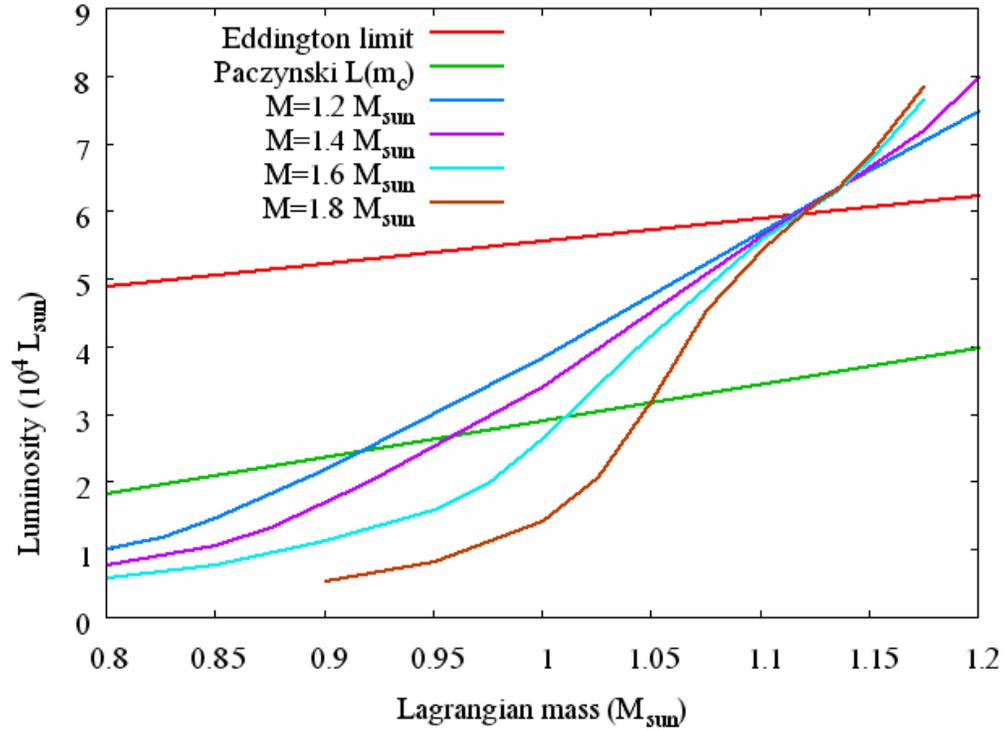

**Figure 3-28: The luminosity vs. carbon core mass in several helium star models. Also shown is the Eddington limit (red) and Paczynski's luminosity – core mass relation (green).**

This phenomenon is well known in hydrogen stars as the luminosity – core mass relation (*Paczynski 1970*), and occurs when the burning shell* is narrow, and located above a quasi white dwarf core in a region where a steep density gradient is present.

---

* The innermost of the two shells.

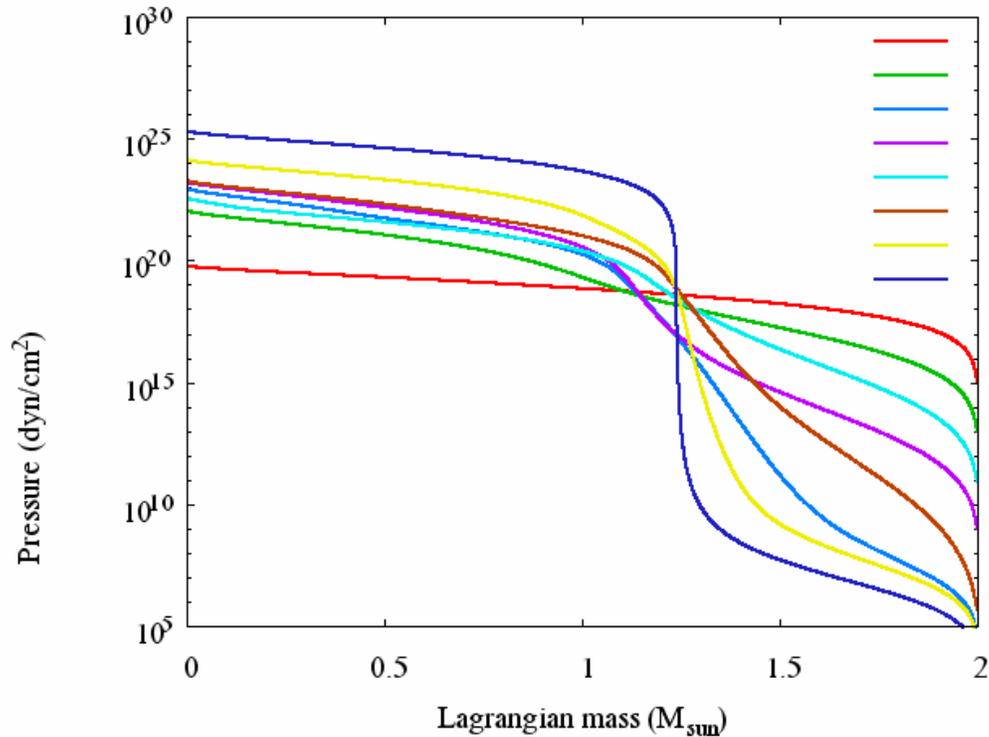

**Figure 3-29: The pressure profile in a helium star model of mass 2 M⊙ at different times. The time advances from top to bottom through the colors in the legend.**

Note that our case is simpler, since we deal with a single helium burning shell, while in the hydrogen star case a double shell exists, alternately burning hydrogen and helium. It is also clear that since the *Q-values* of hydrogen and of helium are very different (about a factor of 10); the relation between the luminosity and the core mass in our case is much different from the *Paczynski* relation (shown also in Figure 3-28).

In the vicinity of the convergence, the luminosity of the models also exceeds the *Eddington* luminosity, which will have a major importance as we shall see in section 3.2.3.

As we have seen in section 3.1.2, we can expect cases where carbon will not ignite, cases where it will ignite off-center, and cases where it will ignite at the center. It is clear that the mass and composition of the core will determine the occurrence of carbon ignition.

Figure 3-30 displays in the ($\rho_c$, $T_c$) plane the evolutionary path of a *1.8 M$_\odot$* helium star model, with an initial carbon core of *0.66 M$_\odot$*. The figure also displays the ignition lines for carbon, and the mass of the core along the path. This star does not ignite carbon until a later phase, when it ignites below the helium burning shell (see section 3.2.3).

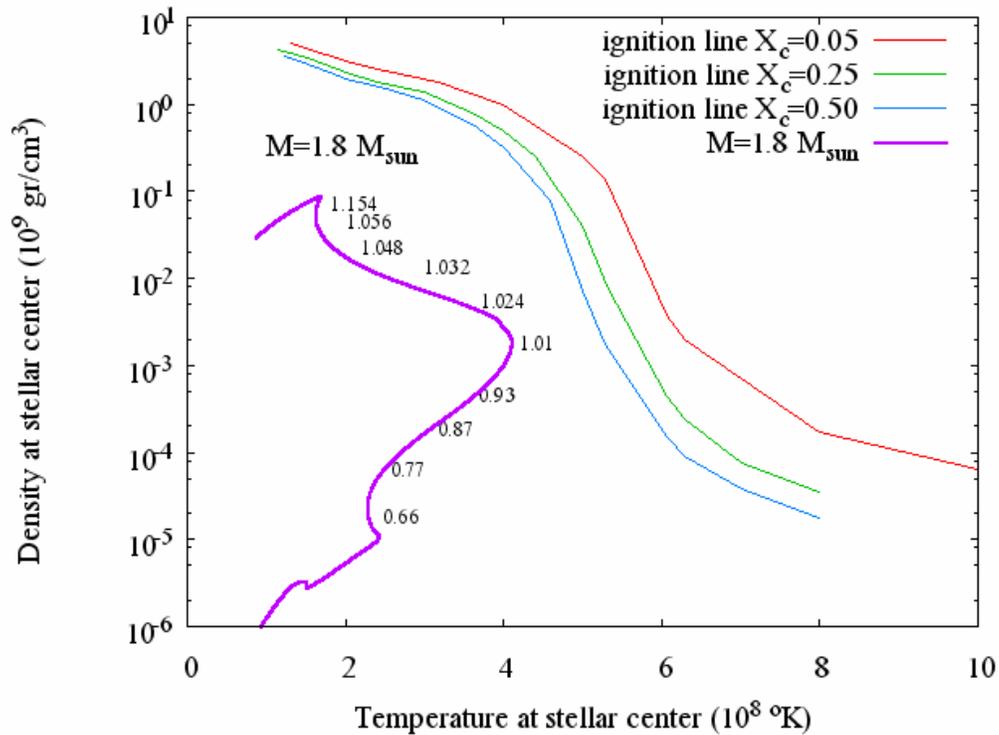

**Figure 3-30: The evolution of the density vs. temperature at the center of a helium star model of mass 1.8 M$_\oplus$. The ignition lines for carbon at various mass fractions are also shown. The mass of the core is demarcated along the path.**

Figure 3-31 shows the evolutionary path for a *2.0 M$_\odot$* helium star model, with an initial carbon core of *0.77 M$_\odot$*. In this case we can see that as the mass of the core

reaches *1.1 M⊙*, carbon ignites off-center, and due to the subsequent rise in entropy, the center expands. In this case ignition takes place at *m = 0.42 M⊙*.

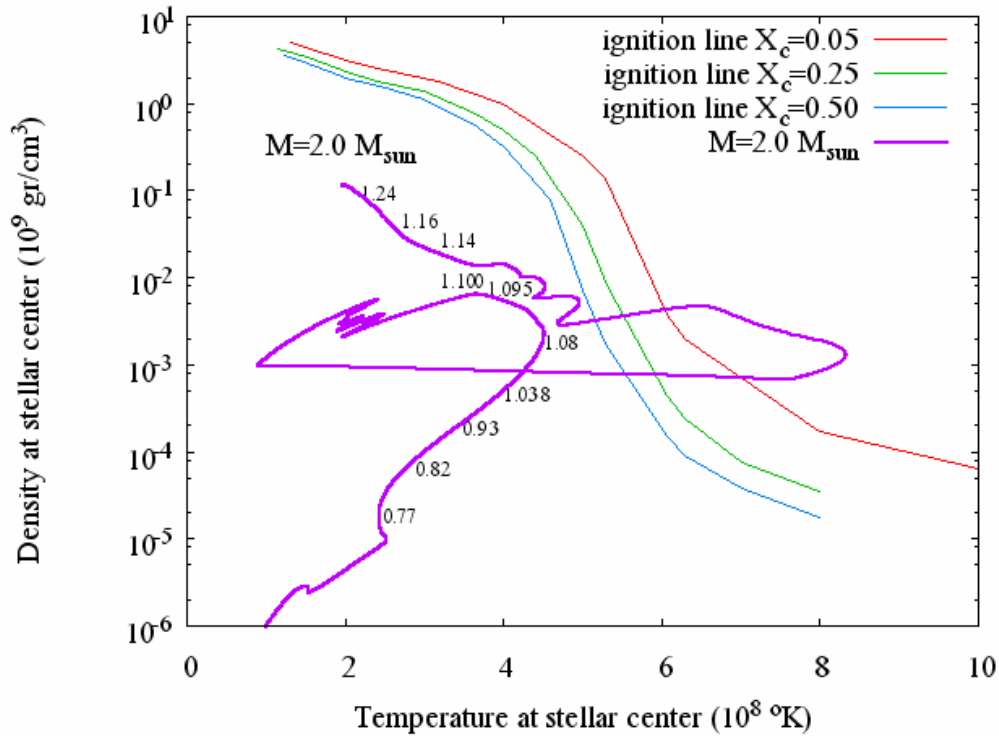

**Figure 3-31: The evolution of the density vs. temperature at the center of a helium star model of mass 2.0 M⊙. The ignition lines for carbon at various mass fractions are also shown. The mass of the core is demarcated along the path.**

It is important to note, that due to the neutrino losses which quickly increase at these stages, the rate of evolution of the core speeds up accordingly, and becomes large compared to the rate of its growth due to the burning shell. Thus, the stellar core behaves like a carbon – oxygen star with equal mass, and it is clear that its evolution will go through phases of creation and vanishing of convective regions, ignition and extinction of burning shells, consequently resulting in a complex composition profile similar to the carbon – oxygen star case, which has already been discussed earlier. An example of the composition profile in the core after the off-center carbon burning has ceased can be seen in Figure 3-32*.

As the temperature eventually drops (on the way to forming a white dwarf), neutrino losses diminish, and the core subsequently "recalls" that it has a burning shell above it. In fact we now have a core, which gradually accretes the residues of helium burning. In contrary to the accretion scenarios we investigated for carbon – oxygen stars, here we have no freedom in choosing the accretion rate or the composition, since these are determined by the behavior of the burning shell.

### 3.2.3.   The hot bottom helium burning shell

As mentioned before, the luminosity of the radiatively burning shell increases as it advances. As we have seen in Figure 3-28, the luminosity approaches and eventually even exceeds the *Eddington* luminosity. As has been already pointed out (cf. *Joss et al. 1973*), this situation requires the onset of a convective envelope, which indeed happens. The convection spreads from the outer stellar boundary inwards, and at a certain stage reaches the vicinity of the burning shell. From this moment the evolutionary picture dramatically changes. In this situation, which has been dubbed "hot bottom burning", there is a constant supply of fuel to the burning shell, and it turns out that two cases are possible. The key question† is whether or not the burning shell will continue advancing outward, or will rather stay fixed until the entire fuel supply will be exhausted. A full discussion is given in section 3.3.

---

* This figure actually represents a later stage of evolution, however the composition in the core (up to about m≈1.2 M$_\odot$) doesn't change after off-center burning.

† We will not take interest in the various compositions that are expected to be observed.

In any case, we have chosen to investigate the continuation of evolution via two parallel channels. In the first one we followed the advancing convective burning shell ("moving case"), and in the other one we forced it to remain static. The static case is initially simpler – the burning shell stays put until the fuel is almost exhausted. At some intermediate stage, as the fuel concentration has dropped, the luminosity begins to drop together with the nuclear burning rate. Since up to this point the core does not grow, the temperature and density hardly change. From now on the core quickly contracts, and we find that it grows at a rate of $\dot{m} \approx 1.5 \ X \ 10^{-5} \ M_\odot/yr$.

The temperature is now rising, and soon the leftover carbon below the former shell will ignite, resulting in a burning front moving inward. However, like in section 3.1.2.3, it dies out when a carbon poor zone is encountered.

The moving case requires special care, and will be discussed a bit later. Clearly, as the burning shell travels outward, the core grows, and we are in fact back to the case of a carbon – oxygen core with depleted carbon, which grows as a result of accretion toward Chandrasekhar's mass, so the manner in which it will explode (or not) depends on the conditions we are familiar with from section 3.1.

Note, that in both the static and the moving case, the growth rate of the core is essentially the same – $\dot{m} \approx 2 \ X \ 10^{-5} \ M_\odot/yr$.

As we have already mentioned, if runaway will occur, the helium star case has one fundamental difference compared to the carbon – oxygen star case – above the exploding core lays an envelope which might contain some tenths of solar mass of helium burning products. A typical composition of the star is shown in Figure 3-32.

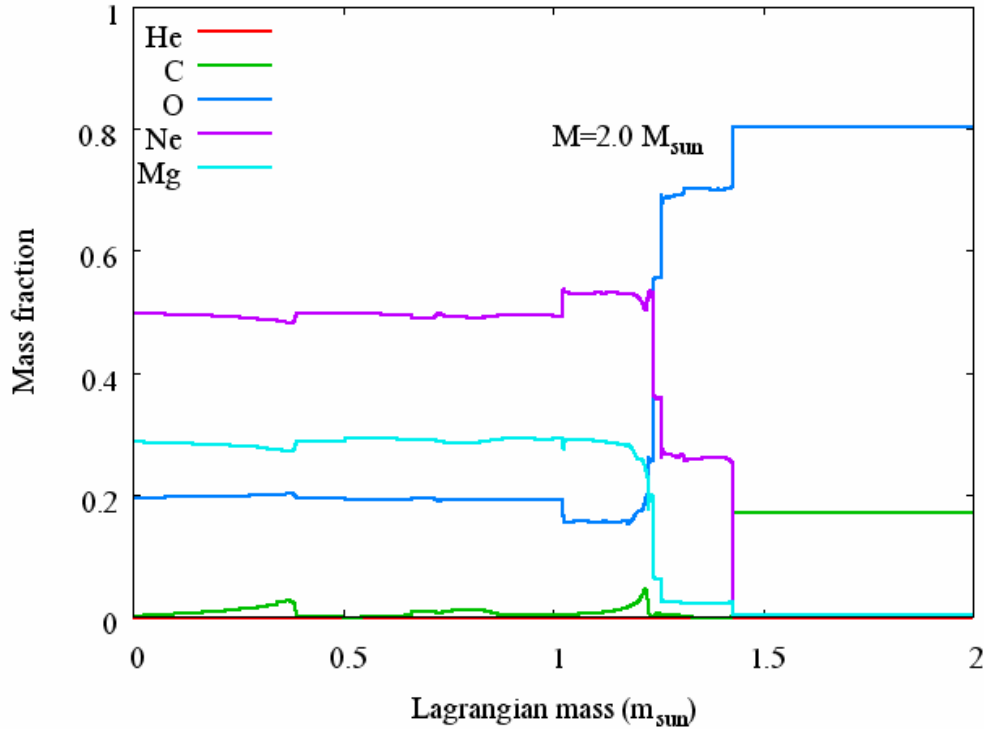

**Figure 3-32: The element mass fraction profile in a 2.0 M$_\odot$ mass helium star model after hot bottom helium burning.**

Like in the carbon – oxygen star case, here also electron capture (mainly on $Mg^{24}$) can be of importance, if evolution carries the center above a density of *3.7 X 10$^9$ gr/cm$^3$*.

Like in the carbon – oxygen star case, here also we can anticipate a large variety of results, depending on the details of the evolution history, chiefly through the amount of carbon remaining in the star, especially at its central region.

## 3.3. *More on the advancement of burning shells*

As we already mentioned, one of the main problems affecting the outcome of the simulated evolution is a reliable modeling of the advancement of the nuclear burning shells. In order to validate our results, we invested a great effort to check the sensitivity of our results, mainly to the numerical resolution, and to other factors as well.

The difficulty arises mainly in those cases, where the burning is at the bottom of a convective region, so that there is an interconnection between it and the size of the convective region.

In the previous chapters we encountered three situations where such kind of burning shell occurs. The first is the carbon burning shell which advances toward the center (see "stage II" in section 3.1.2), the second is the carbon burning shell which ignites on top of the partially depleted carbon zone and moves both inward and outward (see "stage III" in section 3.1.2), and the third is the helium burning shell at the base of the convective envelope advancing outward (hot bottom burning – *"HB"*), where at a later stage an inward advancing carbon burning shell is also ignited below the helium burning shell (see section 3.2.3). In each case it is especially important to find out whether the burning might die out. There are three factors which could in principle contribute to this matter – the expansion caused by the raise in the entropy in the convective region, the gradient in the fuel concentration, and the heat conduction.

The expansion is a global phenomenon, and its magnitude depends on the position and extent of the convective region, as well as on the rate of entropy increase through it – i.e. on the rate of nuclear burning. The heat conduction and the fuel concentration gradient are local factors, and an adequate numerical resolution is required to represent them correctly.

The behavior of the first two kinds of burning shells was already discussed in section 3.1.2. The hot bottom burning is a complex phenomenon, which has been extensively dealt with in the literature for hydrogen burning stars (e.g. *Lattanzio 1992*, *Boothroyd et al. 1993*, *Boothroyd et al. 1995*, *D'Antona & Mazzittelli 1996*, *Lattanzio et al. 1997*, *Driebe et al. 1998*, *Blöcker et al. 1999*). In the following we will discuss in detail the hot bottom helium burning shell.

As we remember, the onset of the *HB* stage is caused by the encounter between the inner boundary of the convective envelope advancing inward and the helium burning shell advancing outward.

Seemingly, since the rate of entropy production is at its maximum at the base of the burning shell, there is no reason for the base to be "released" from the convective region. However, the local entropy production rate is of course:

*(3-5)*
$$\dot{s} = \frac{\partial S}{\partial t} = q - \frac{\partial L}{\partial m}$$

Since the standard conditions for the $L(m_c)$ relation have been violated due to the presence of convection, we should not expect it to be obeyed. Indeed, because of the entropy production in the convective region together with the supply of fresh fuel, the temperature at the base of the region rises abruptly, and so does the luminosity. In fact the luminosity at the innermost zone of the convective region grows faster than at the zone below it, because the conductivity is not efficient enough to overcome this trend. Thus $\partial L/\partial m$ becomes positive, and at a certain stage the negative contribution of the term $-\partial L/\partial m$ becomes large enough to lower $\dot{s}$ at the innermost zone relative to the zones above it*. When this should happen, the zone in question will release itself from the convective region, so that the base of the convection will move outward. Usually at the moment of release there will still be nuclear burning going on in the released zone (since the fuel has not yet been exhausted), however once it becomes radiative it will diminish. We find that a stationary situation develops, featuring an outward moving burning front at the base of the convective region. Figure 3-33 shows the convective envelope and the helium burning shell on a Kippenhahn diagram for a *2.0 $M_\odot$* model.

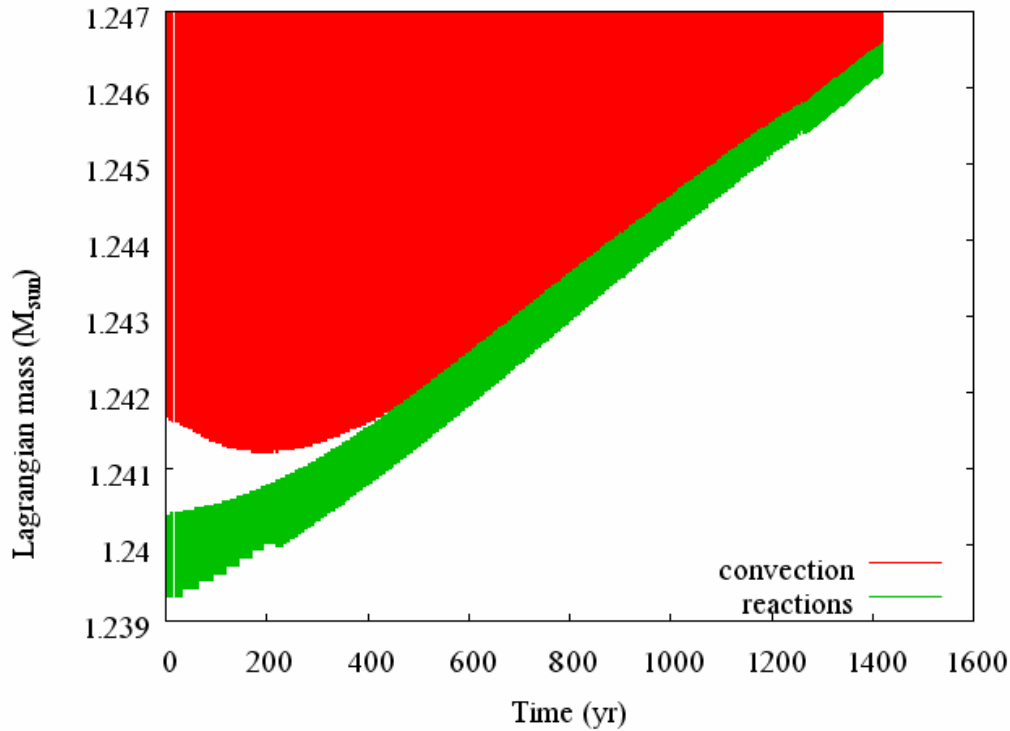

**Figure 3-33: The history of the convective envelope and the helium burning shell in a 2.0 M$_\odot$ helium star model at the beginning of the hot bottom burning stage.**

---

* We recall that in a convective region the luminosity monotonically decreases outward.

From this discussion it is clear, that correct numerical representation of the temperature and luminosity profiles in the relevant region is of crucial importance. This requirement considerably complicated the treatment and called for substantial effort. Indeed, we found that crude zoning might lead to a static situation. From a physical point of view the situation is even more severe. An accurate representation of convection, especially in a one-dimensional model, is non-existent. If and when conditions for "undershoot" are created, so that fuel will be supplied at a certain rate to the radiative burning zone, one might not rule out that the stationary situation might turn into a static one.

The outward advancement of the *HB* leaves behind carbon. After enough carbon is accumulated, it is ignited, and subsequently a double shell (*DSh*) structure is created (Figure 3-34).

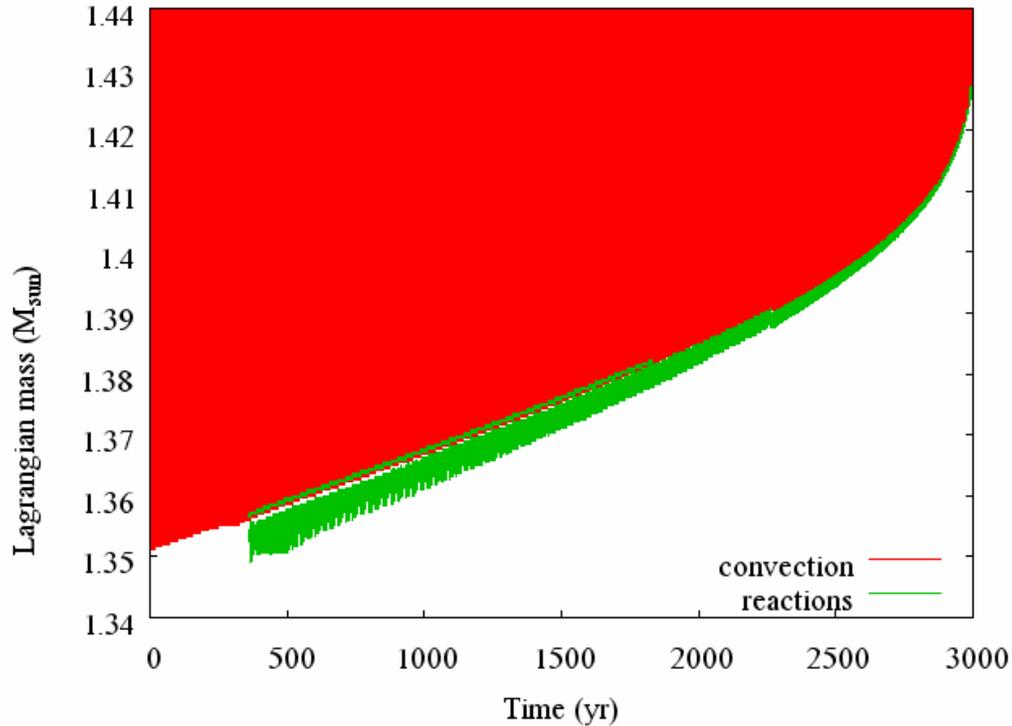

**Figure 3-34: The double shell - a carbon burning shell below the helium burning shell in a 2.0 M$_\odot$ helium star model during the hot bottom burning stage.**

An adequate numerical modeling shows that the *DSh* moves outward in a self-similar and stationary manner. During the outward motion of the *DSh* the concentration of helium is diminished, and even before the shells reach $M_{ch}$ it is exhausted. Subsequently, the *DSh* ceases and one shell burning carbon is left. For the case presented in Figure 3-34, the *DSh* ceases at about *m ≈ 1.38 M$_\odot$*. Figure 3-35 follows the same event by showing the reaction rate profile at different times. We can see that first the reaction rate in both shells rises, then as the helium is depleted, it drops and then vanishes in the helium burning shell, while continuing to rise in the carbon burning shell. The rise of the reaction rate in the burning shells is due to the rise of the burning temperature, which can be seen in Figure 3-36.

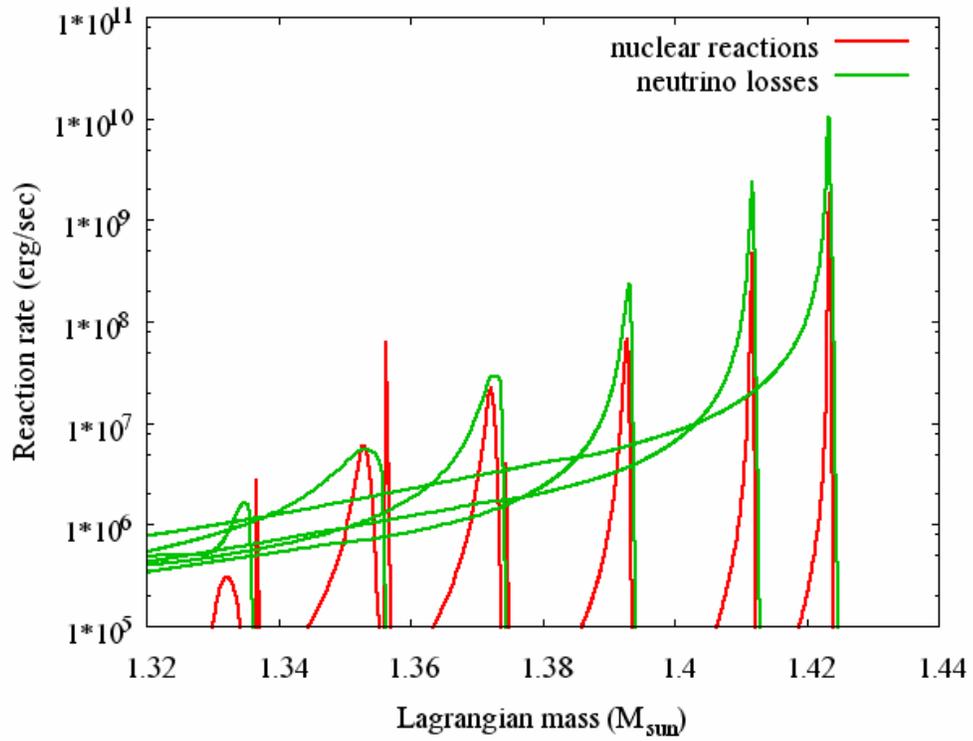

**Figure 3-35: The reaction rate profile in the double shell in a 2.0 M$_\odot$ helium star model during the hot bottom burning stage.**

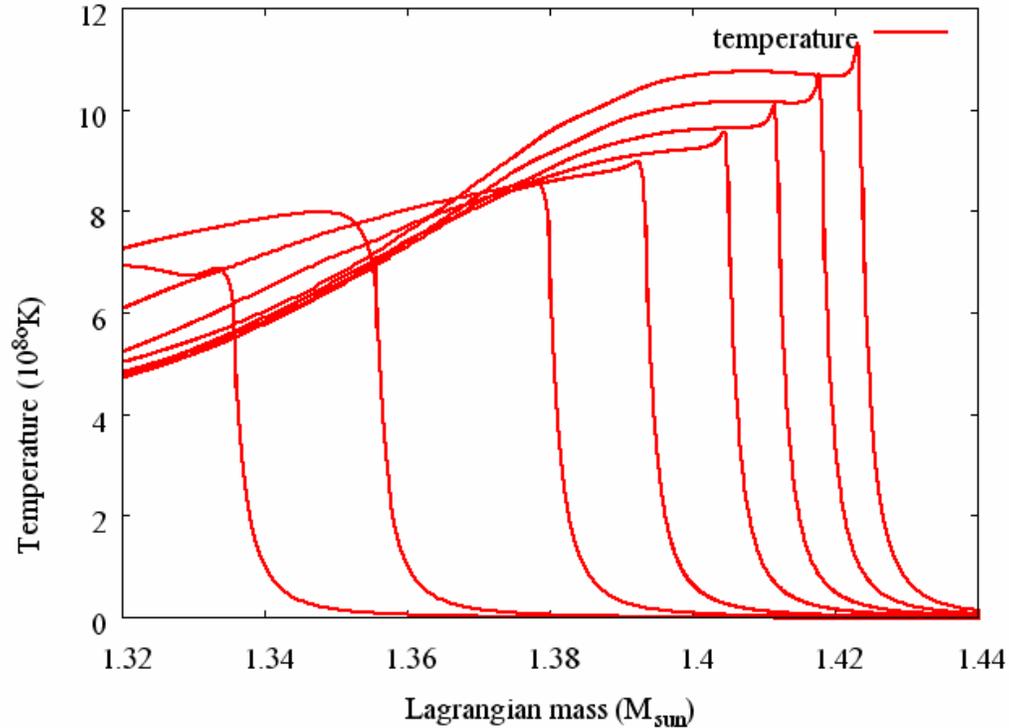

**Figure 3-36: The temperature profile in the double shell in a 2.0 M$_\odot$ helium star model during the hot bottom burning stage.**

Looking at the growth rate of the core (Figure 3-34), it turns out that during the *DSh* phase it is about *2 X 10$^{-5}$ M$_\odot$/yr*, after the *DSh* ceases it rapidly increases up to about *2 X 10$^{-4}$ M$_\odot$/yr* in the vicinity of *M$_{ch}$*. It is clear, that whether during these phases the star loses mass, it is important to compare the mass loss rate to the core growth rate, since the star will finally undergo central carbon ignition, only if at the relevant stage its mass is still above *M$_{ch}$*.

It is important to note, that carbon burning takes place in a situation, where throughout the carbon burning shell the nuclear energy production rate $q_n$ is below the neutrino losses $q_\nu$, whereas in the helium shell, while helium burning is strong, the situation is the opposite (Figure 3-37).

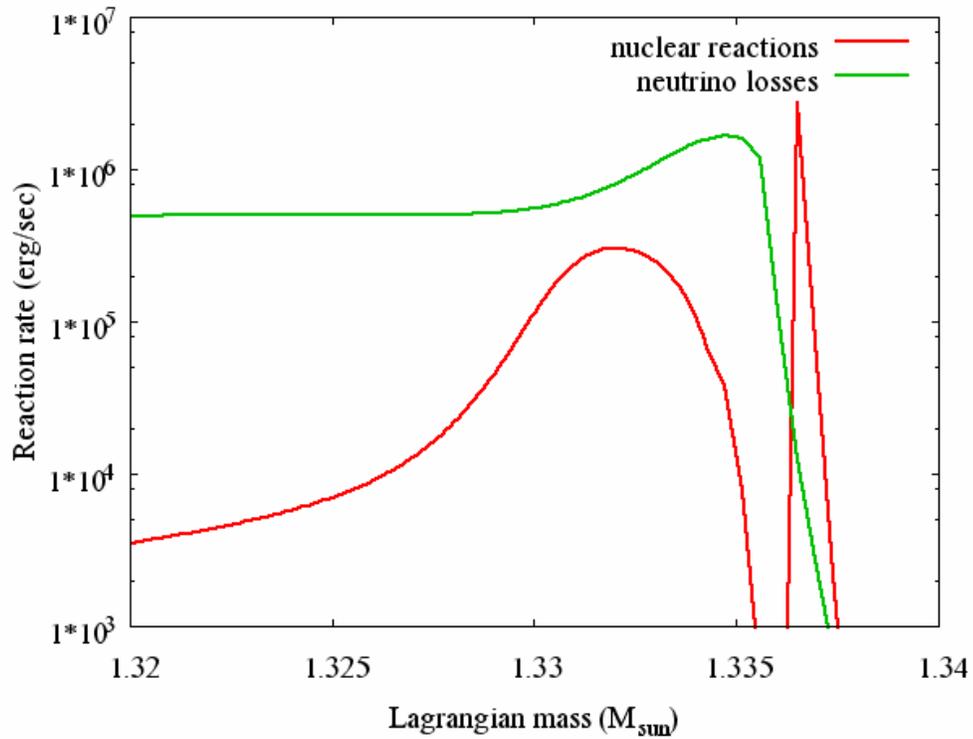

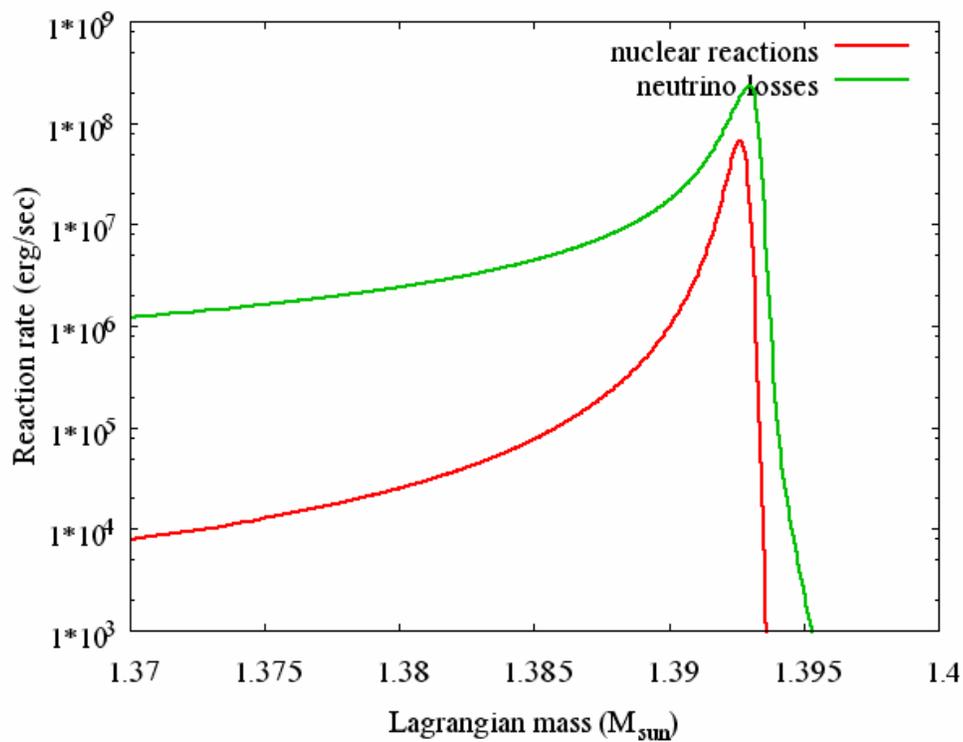

**Figure 3-37: The nuclear reaction and neutrino loss rate in the double shell, and after a single carbon burning shell is left, in a 2.0 M$_\odot$ helium star model during the hot bottom burning stage.**

As shown in Figure 3-36, towards $M_{ch}$ the carbon burning temperature reaches about *1.1 X 10$^9$ $^oK$*. Would the temperature rise a little higher (by about 10%), oxygen will be ignited. Since the concentration of oxygen is higher by about a factor of 4 than that of carbon, it is possible that the burning will boost, so that we will have $q_n > q_v$ in the shell. Should this happen, an inward advancing oxygen burning shell could be formed.

In our calculations we haven't encountered such a case, but we assess that similar to the case of carbon ignition at the edge of the core, also in this case should the oxygen ignite it will extinguish well before reaching the center due to expansion, and then during the contraction phase a competition between ignition of carbon at the center and re-ignition of the oxygen will take place. If in such a case explosion will take place, we will expect to find at the outer layers of the core not only carbon burning products, but also some oxygen burning products. Clearly, at this stage this is merely a speculation.

### 3.4. Mass loss

As previously mentioned, one of the delicate points of the scenario we are dealing with is the question of mass loss from the helium envelope, in the period during which the luminosity rises to several hundred thousand solar, and the radius also reaches hundreds of solar.

The mass loss from helium stars of mass in the range of our interest (*2-3 $M_\odot$*) has not been specifically treated in the literature. In order to address the issue of mass loss, the closest objects we can refer to are either *AGB* stars which have masses and radii comparable to our models, but over an order of magnitude lower luminosities and different envelope composition, or *Wolf-Rayet* stars, which are helium stars with luminosities similar to ours, but are more massive (above $8\,M_\odot$), and have radii smaller by more than one order of magnitude.

*AGB* mass loss is a complex process, driven by both shocks from pulsations and radiation pressure on dust (and molecules) formed in the cool extended atmosphere. Dynamical models of the extended envelopes do not yet convincingly predict mass loss rates. Instead, various observationally fitted formulae are used, giving the mass loss rate as a function of parameters such as the luminosity, mass, radius and pulsation period (*Zijlstra 2006*).

In this light, we chose to examine the issue of mass loss in two ways:

### 3.4.1. Limiting the stellar radius

This is also relevant of course to the binary case where the companion defines the *Roche lobe*.

The algorithm we used was to delete any *Lagrangian* zone that exceeds a predetermined radius $R_{loss}$.

Figure 3-38 shows the characteristic time dependence of the luminosity and radius in a typical case of a model with mass of $2\,M_\odot$*. In this case the luminosity rises from $10^5\,L_\odot$ to $7\times10^5\,L_\odot$ during a time interval of about $2\times10^4\,yr$. During the same time period the radius climbs from about $200\,R_\odot$ to around $450\,R_\odot$.

---

* For this purpose we chose the static case, since the moving case is too computer time consuming for this.

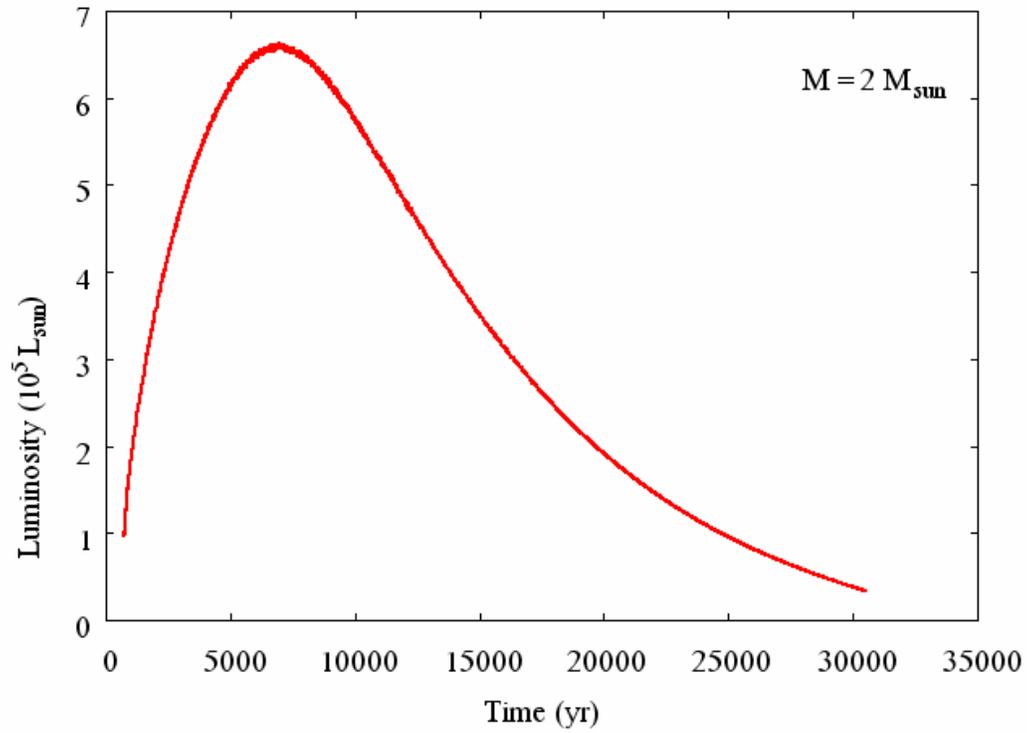

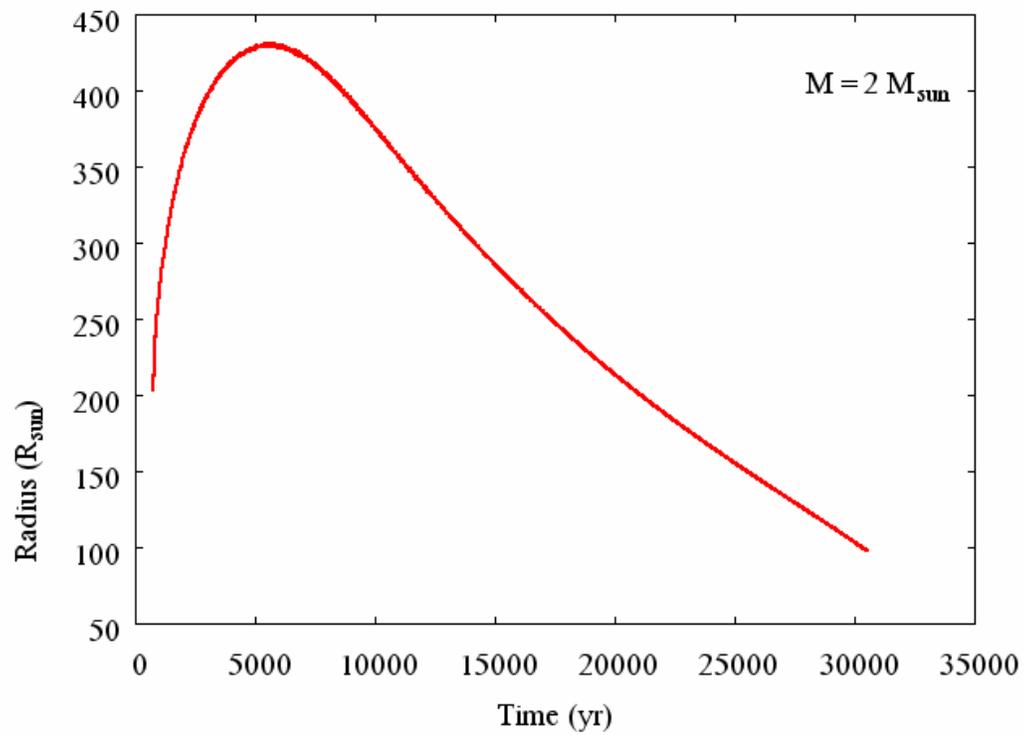

**Figure 3-38: The time dependence of the luminosity and radius of a 2 $M_\odot$ helium star model during the hot bottom helium burning phase.**

Figure 3-39 shows the time dependence of the luminosity and radius for radius limits of *320 $R_\odot$* and *220 $R_\odot$* respectively. In both cases the model indeed loses a considerable amount of mass, until the luminosity drops and mass loss ceases. In the first case the mass drops to *1.63 $M_\odot$*, in the second case to *1.46 $M_\odot$*. The effective mass loss rate in both cases is around *$10^{-4}$ $M_\odot$/yr*. In both cases the model eventually reaches runaway, and the difference between the two is only in the mass of the envelope.

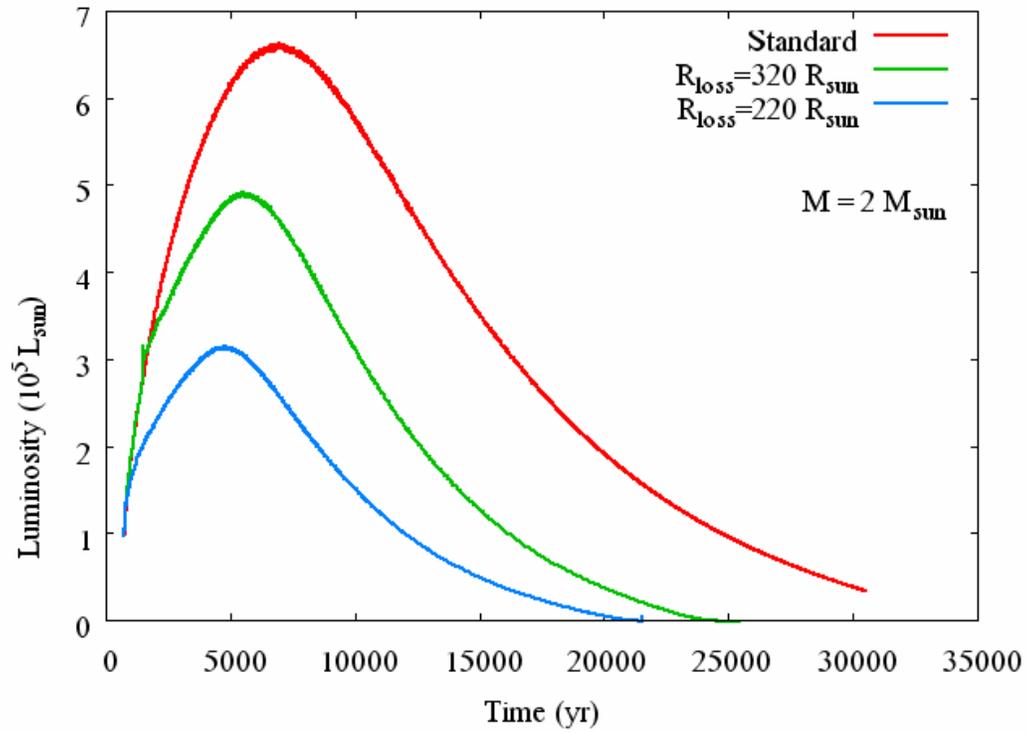

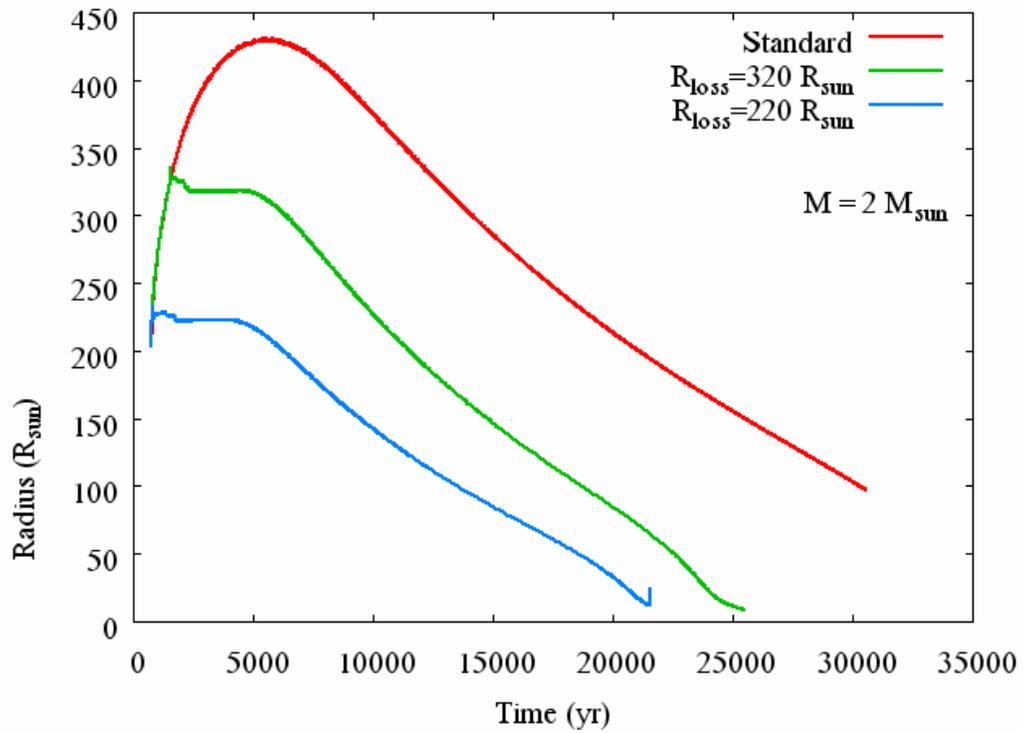

**Figure 3-39: The time dependence of the luminosity and radius of a 2 $M_\odot$ helium star model during the hot bottom helium burning phase, with radius limiting.**

### 3.4.2. Mass loss formulae

We used the most popular relation of *Reimers 1975* for red giant stars:

$$(3\text{-}6) \qquad \dot{M}(M_\Theta yr^{-1}) = -4\times10^{-13}\,\alpha\,\frac{{L}/{L_\Theta}\;{R}/{R_\Theta}}{{M}/{M_\Theta}}$$

Where α is an uncertainty factor in the range: *0.3≤α≤3*. We examined the sensitivity of the results to the factor α in this range.

Figure 3-40 compares the time dependence of the luminosity and radius for α = 1 to the standard case (without mass loss). We can see that due to the mass loss the decline of the luminosity and the radius is faster, and the maxima are lower, both of which alleviate the possibility of reducing the mass below *Chandrasekhar's*.

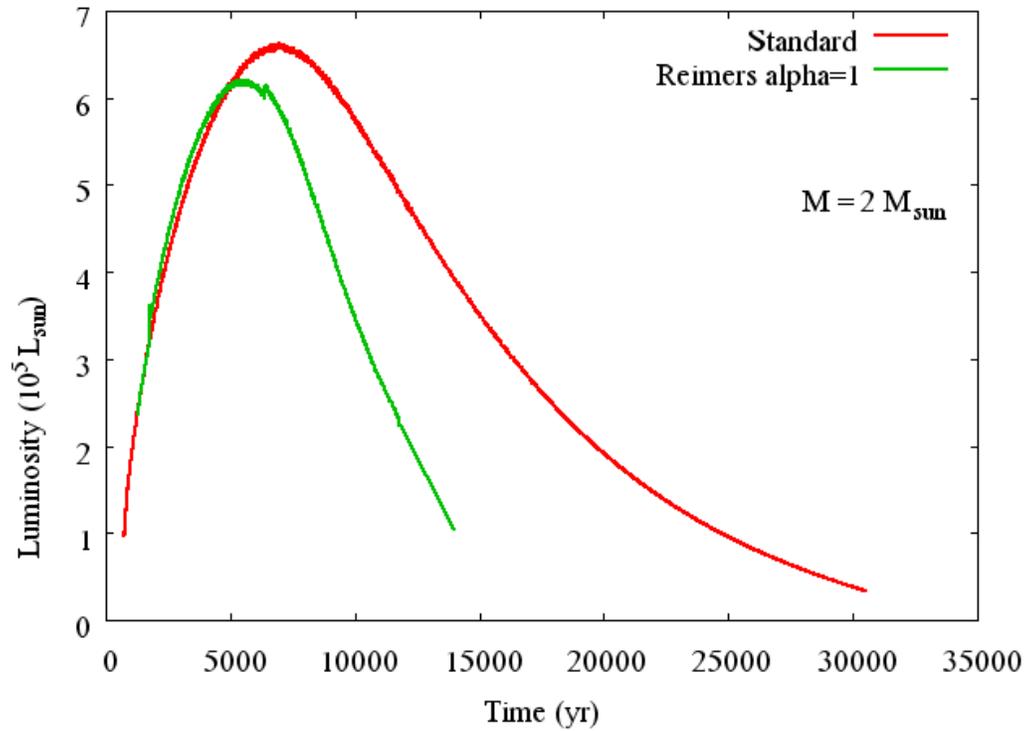

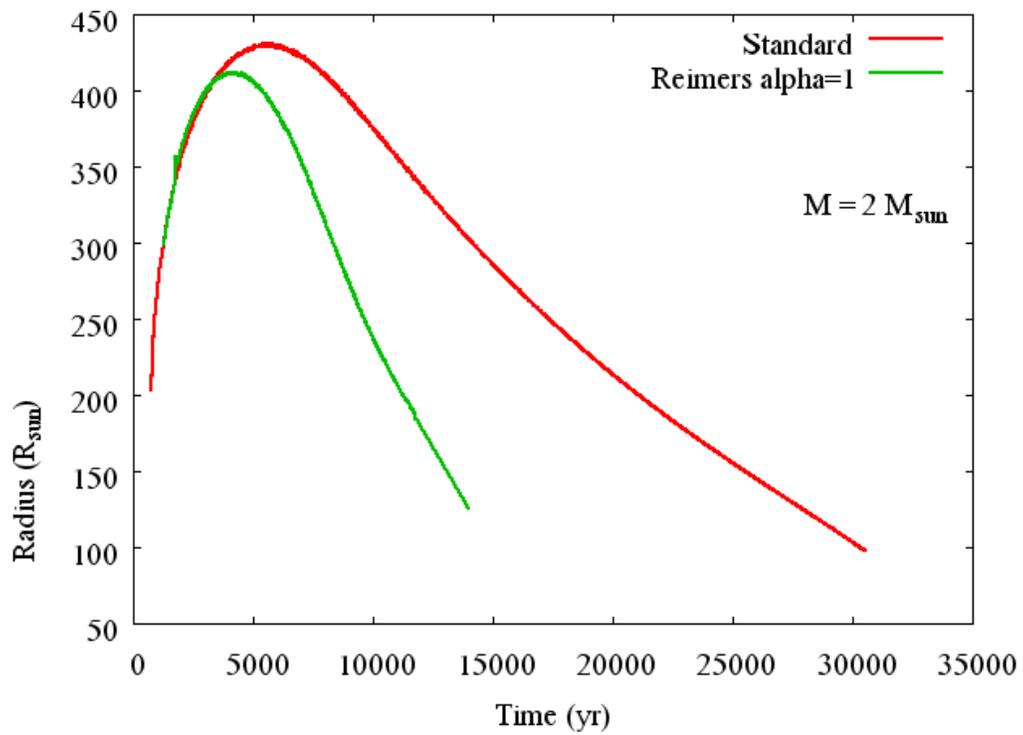

**Figure 3-40:** The time dependence of the luminosity and radius of a $2\,M_\odot$ helium star model during the hot bottom helium burning phase, with *Reimers'* mass loss formula.

Figure 3-41 shows the total mass as a function of time for α = *0.3, 1, 3,* as well as the edge of the helium exhausted core. For α = 1 the mass loss rate is initially high, but eventually it becomes comparable to the advancement of the helium burning shell, i.e. the rate of growth of the core. From there on it can be shown that additional mass loss is quite limited.

For α = *0.3* the amount of mass lost is rather small, while assuming α = *3* does reduce the mass below *Chandrasekhar's*. It is clear that within the range of uncertainty, one can not make a firm statement in this regard.

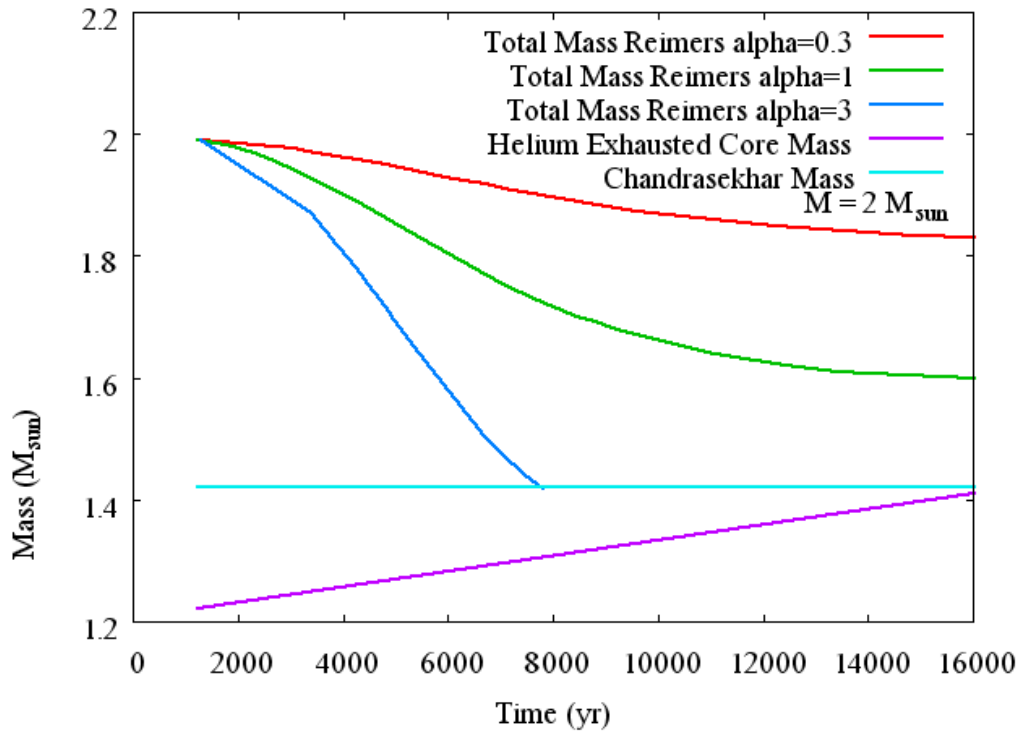

**Figure 3-41: The time dependence of the total mass and the mass of the helium exhausted core of a 2 M$_\odot$ helium star model during the hot bottom helium burning phase, with *Reimers'* mass loss formula using various values of the uncertainty factor α.**

We shall support our numerical result by the following analytical estimate. If we assume that the luminosity *(L)* of the star is approximately equal to the total nuclear energy generation rate *(Q)*, take *($Q_v$)* as the nuclear energy generated for unit mass of burned helium, $\dot{m}_c$ as the growth rate of the helium exhausted core, and *Y* as the mass fraction of helium, we can write:

*(3-7)* $$L(erg\sec^{-1}) \approx Q = Q_v \dot{m}_c Y$$

Converting into solar units and years, and substituting $Q_v \approx 6\ X\ 10^{17}\ erg\ gr^{-1}\ sec^{-1}$, and $Y \approx 1$, we get:

*(3-8)* $$\dot{m}_c(M_\Theta yr^{-1}) \approx \frac{L_\Theta{}^{yr}\!/_{\sec}}{Q_v M_\Theta}\left(\frac{L}{L_\Theta}\right) \approx 10^{-10}\left(\frac{L}{L_\Theta}\right)$$

Comparing the above estimate for the growth rate of the core to the mass loss rate (3-6), we get:

*(3-9)* $$\frac{\dot{m}_c}{\dot{M}} \approx \frac{250}{\alpha}\frac{M\!/_{M_\Theta}}{R\!/_{R_\Theta}}$$

For a total mass of *2 $M_\Theta$* and α = 1 we find that the growth rate of the core is higher than the mass loss rate as long as the radius of the star is less than *500* solar. As can be seen in Figure 3-40 the radius in our case is always below this, so the mass loss rate will always be below the core growth rate. Furthermore, the core has to grow by about *0.2 $M_\Theta$* to get to *Chandrasekhar's* mass, while as much as *0.6 $M_\Theta$* can be lost from the envelope before the mass is diminished below *Chandrasekhar's* mass. So indeed we confirm our numerical result, that mass loss should cause no problem using the *Reimers* formula with α = 1.

We also compared with other mass loss formulae from the literature, namely:

The formula given by *Vassiliadis & Wood 1993* for *Mira* variables relating mass loss to the pulsation period:

*(3-10)* $$\log \dot{M}(M_{\odot} yr^{-1}) = -11.4 + 0.0123 P(days)$$

Where $P$ is the period given by:

*(3-11)* $$\log P(days) = -2.07 + 1.94 \log R / R_{\odot} - 0.9 \log M / M_{\odot}$$

The observational fit given by *Lawlor and MacDonald 2006* for *Mira* variables relating mass loss to the pulsation period:

*(3-12)* $$\log \dot{M}(M_{\odot} yr^{-1}) = -22.8 + 6.32 \log P(days)$$

The fit to observational data for dust enshrouded *RSG*s and *AGB*s in the *LMC* by *van Loon et al. 2005*:

*(3-13)*

$$\log \dot{M}(M_{\odot} yr^{-1}) = -5.65(\pm 0.15) + 1.05(\pm 0.14) \log L / 10,000 L_{\odot} - 6.3(\pm 1.2) \log T_{eff} / 3500K$$

The fit to observational data for *WN* type *Wolf-Rayet* stars by *Nugis and Lamers 2000*:

*(3-14)* $$\log \dot{M}(M_{\odot} yr^{-1}) = -13.6 + 1.63(\pm 0.21) \log L / L_{\odot} + 2.22(\pm 0.63) \log Y$$

Table 3-1 below compares the *Reimers* mass loss rate to the above formulae along the evolutionary track of our *2 $M_{\odot}$* model from Figure 3-38. The points chosen are the beginning of the track and the maximum luminosity. We can see, that all the other formulae give smaller mass loss rates than the *Reimers* law, except the *Nugis &*

*Lamers* formula, which is slightly above the standard *Reimers* law near the luminosity peak, so we can state that our analysis using the *Reimers* law with the scaling factor α is a fairly conservative approach.

| Luminosity / Radius | Reimers | Vassiliadis & Wood | Lawlor & MacDonald | Van Loon et al. | Nugis & Lamers |
|---|---|---|---|---|---|
| $1\ 10^5$ / 200 | $8\ 10^{-6}$ | $1\ 10^{-10}$ | $4\ 10^{-10}$ | $3\ 10^{-6}$ | $4\ 10^{-6}$ |
| $6.5\ 10^5$ / 420 | $5.5\ 10^{-5}$ | $3.1\ 10^{-5}$ | $4\ 10^{-6}$ | $3\ 10^{-5}$ | $7.5\ 10^{-5}$ |

**Table 3-1: Comparison of mass loss rates given by various mass loss formulae for the case of the 2 M$_\odot$ helium star model.**

It is clear that should the mass loss take the model below Chandrasekhar's mass, runaway would be impossible. In order to get an adequate answer to this question, further analysis of the mass loss issue is needed.

### 3.5. Comparison to other programs

In order to validate our results, especially since, as mentioned in section 2.1.1, we treat convective zones as fully mixed and adiabatic, which might not be a satisfactory approximation in the envelope, we reproduced a helium star model using the *Tycho* evolution code (*Young & Arnett 2005*), which treats convection using the well known standard mixing length theory, as defined in *Kippenhahn and Weigert 1994*.

We will first underline the fact, that the mixing length model is merely an effective model for more complex physics, using an arbitrary mixing length parameter $\alpha_{MLT}$, defined as the ratio of the mixing length to the pressure scale height, which is not guaranteed to be constant over stellar mass or evolutionary state, and is usually calibrated to fit observational data or, more recently, multi-dimensional simulations of convection*.

In the formulation of the mixing length theory used by *Tycho*, the value fit to the Sun is $\alpha_{MLT} = 2.1$ (*Young & Arnett 2005*), while various works indicate values in the range $1.5 \leq \alpha_{MLT} \leq 2.7$ fit for red giant stars (*Young & Arnett 2005*, *Herwig 2005* and references therein). Accordingly we modified the value of $\alpha_{MLT}$ in this range.

We first calculated a *2.4 $M_\odot$†* helium star model with $\alpha_{MLT} = 1.4$ and *Reimers* mass loss formula (see section 3.4.2), from core helium burning, through off-center carbon burning, and up to the hot-bottom helium burning shell stage, when the model attained a carbon – oxygen core of about *1.35 $M_\odot$* with a luminosity over *100,000 $L_\odot$*.

The results of this simulation are clearly consistent with our reference *2 $M_\odot$* model calculated with the *ASTRA* code. This model also goes through off-center carbon burning, leaving behind a similar carbon abundance profile, which allows explosive ignition when *Chandrasekhar's* mass is reached. It also goes through a similar hot-bottom helium burning phase, however in this model the nuclear burning penetrates less into the convective envelope compared to our reference model, consequently the luminosity is lower for the same core mass.

---

\* Note also that different formulations of the mixing length model exist, resulting in different values of $\alpha_{MLT}$ that fit to the same evolutionary state.

† This value of the mass was chosen such that the off-center carbon ignition occurs at a similar *Lagrangian* mass as in our *2 $M_\odot$* model calculated with the *ASTRA* code.

Another comparison we made was to increase the mixing length parameter to $\alpha_{MLT} = 2.5$ at a certain stage of the abovementioned $\alpha_{MLT} = 1.4$ model. Comparing it to our reference model we can see that the structure of the model becomes very similar to our adiabatic convection reference model, i.e. the maximum of nuclear energy generation is located adjacent to the bottom of the convective region, and the luminosity consequently attains a similar higher value.

# 4. Discussion and Conclusions

In this work we set up to explore the hypothesis, that helium stars in a certain mass range can evolve to a carbon core explosion, similar to what is widely accepted as an explanation for the type I supernova phenomenon. This should happen, when their carbon – oxygen core grows thanks to the helium shell burning above the core. For this purpose, we undertook a thorough examination of helium stars in the mass range from Chandrasekhar's ($1.4\ M_\odot$) up to about $2.5\ M_\odot$ (above which the fate of the star is already different).

We found, that in the mass range of about $1.7 – 2.2\ M_\odot$, indeed this can happen. The exact mass range depends on several physical uncertainties, which we discussed above, such as the nuclear reaction rate and the resulting carbon mass fraction in the core, the treatment of convection, the *Urca* process and so on.

The main new insight we believe we gained is the crucial importance of an "early" off-center ignition of carbon. The problem this solves is, that since the temperature in the vicinity of the outward advancing helium burning shell increases to the point where the freshly produced carbon ignites, an aggressive carbon burning shell is created, which moves inward, and most probably exhausts all the carbon. Thanks to the previous episode of off-center carbon burning, however, a carbon depleted region stops this new shell from advancing towards the center. After the helium burning shell reaches *Chandrasekhar's* mass, the now *super-Chandrasekhar* mass carbon – oxygen core contracts, and the residual degenerate carbon at the center is ignited, resulting in a runaway similar to the classical *SN I* scenario.

Since the structure of the carbon – oxygen core of the helium stars of our interest is very similar to that of a carbon – oxygen star of the same mass, and their behavior is very similar to that of a mass accreting carbon – oxygen star, we also thoroughly examined the behavior of carbon – oxygen stars. We discovered that the models which ignite carbon off-center (in the mass range of about $1.05 – 1.18 \, M_{\odot}$, depending on the carbon mass fraction) present an interesting *SN I* progenitor scenario of their own, since whereas in the standard scenario runaway always takes place at the same density of about $2 \times 10^9 \, gr/cm^3$, in our case, due to the small amount of carbon ignited, we get a whole range of densities from $1 \times 10^9$ up to $6 \times 10^9 \, gr/cm^3$.

In comparison to the standard *SN I* scenario, our helium star scenario has the following distinctions:

1. The accretion rate is not a free parameter.

2. At the moment of explosion, if it will indeed occur, the star will possess an envelope of non-negligible mass, consisting of helium and carbon burning products, and possibly some oxygen burning products.

3. Similar to the case of carbon – oxygen stars above, they feature runaway at a variety of densities.

At this point we shall emphasize again, that we regard this scenario as a basis for an astrophysical model. If our results are indeed sound, their main contribution could be to help resolve the emerging recognition that at least some diversity among *SNe I* exists (see *Höflich et al 1998*), since runaway at various central densities is expected to yield various outcomes in terms of the velocities and composition of the ejecta, which should be modeled and compared to observations.

Several issues which were beyond the scope of this work call for further investigation:

1.  The question of mass loss from the envelope, which might decrease the mass of the star below *Chandrasekhar's* before helium burning ceases, and thus undermine our scenario, has been affirmatively answered using various available mass loss formulae. However, all the existing formulae have been observationally calibrated with stars of mass range and/or evolutionary stage different than those of our interest, and might not be suitable to our case. We also feel that the complicated issue of mass loss driven by pulsations should be more thoroughly investigated. For details see section 3.4.

2.  The hot bottom burning is a complex phenomenon, which has been extensively dealt with in the literature for hydrogen burning stars. The key question is if and how do the bottom of the convective envelope together with the burning shell move outward, and its answer lies in an reliable modeling of the convective mixing process, which is evidently a three dimensional phenomenon. A further complication involves the double (helium and carbon burning) shell, and the question of its stability. As is well known, in the case of hot bottom burning with a double hydrogen and helium burning shell, thermal pulsing arises. For details see section 3.3.

3.  For both the helium and carbon – oxygen case, a deeper treatment of the question whether thermal *Urca* can hinder the formation of a convective zone when electron capture on $Mg^{24}$ sets in is needed. For details see section 2.2.3.3.

4. Our work provides initial models, which should be used for detailed simulations of the explosive runaway. The real value of our results would be judged by fitting the results of the dynamical simulations to the observational data.

5. The mass distribution function of suitable candidates for our helium star scenario has yet to be established. As for our carbon – oxygen star scenario, according to *Liebert et al. 2005*, some 6% of the white dwarves have masses above *1 $M_\odot$* (and below *Chandrasekhar's* mass of *1.4 $M_\odot$*). Since the carbon – oxygen stars igniting carbon off-center lie between about *1.05 – 1.18 $M_\odot$*, i.e. span about ¼ of the above range, we can roughly estimate their incidence among the white dwarf population should be in the order of magnitude of 1%. Of course, a sounder estimate of the occurrence of this type of *SNe* needs a more thorough investigation, including modeling of binary evolution.

# Appendix A.    The Evolution Code

## A.1. *Structure of the code*

During each time step the code passes through the following stages:

1. The boundaries of the convective regions are determined. For each convective region the composition is mixed, and entropy is set to a common value, so that the total energy is conserved.

2. Since the entropy profile has been changed, the density profile is changed to fulfill the equation of hydrostatic equilibrium.

3. The size of the time step *dtime* is decided on, so that the thermodynamic parameters (density, temperature, entropy etc.) will not change by more than some preset fractions. This estimate is done based on the changes that had occurred during the previous step. Should the parameters change by more than some allowed fraction, the step is cancelled and repeated with reduced *dtime*.

4. For each zone the code calculates the energy production rate and the change of composition due to nuclear reactions and the energy loss rate due to neutrino production.

5. The entropy equation (i.e. energy conservation) is solved to get a new entropy profile, next the equation of state is used to calculate the pressure, temperature etc., from the density and entropy, and then the radiative energy flux is calculated. This is done iteratively until the obtained profile changes less than the required criterion. During this process the density is kept unchanged. Note that this means the reaction rates are calculated explicitly (i.e. set at the beginning of the time step), while the luminosity is calculated implicitly (i.e. solved for the values at the end of the time step).

6. After the entropy profile has been calculated, the density profile is changed to fulfill the equation of hydrostatic equilibrium.

## A.2. The equations of evolution

The equation of conservation of energy at a given point in the star can be written as:

(A-1)
$$dU + PdV = dQ$$

where $U$ is the internal energy per unit mass, $P$ is the pressure, $V$ the volume per unit mass, and $Q$ the heat absorbed by a unit mass, which can be written as:

(A-2)
$$dQ = \left( q - \frac{\partial L}{\partial m} \right) dt$$

Here $q$ is the rate of energy production by nuclear reactions minus the rate of energy losses by neutrinos per unit mass, and $L$ is the radiative energy flux.

Also, we can write the differential of the internal energy as:

(A-3)
$$dU + PdV = TdS + \sum_i \mu_i dX_i$$

Here $T$ is the temperature, $S$ the entropy per unit mass, the $X_i$ are the mass fractions in the composition, and the $\mu_i$ are the chemical potentials defined by: $\mu_i = \left( \frac{\partial U}{\partial X_i} \right)_{S,V}$

Eliminating equations (A-1) and (A-3) we get:

(A-4)
$$TdS = \left(q - \frac{\partial L}{\partial m}\right)dt + \sum_i \mu_i dX_i$$

This equation, which we will call the entropy equation, is then translated into a set of difference equations for each *Lagrangian* zone:

(A-5)
$$\frac{T\Delta m}{\Delta t}\Delta s + L_{out} - L_{in} - q\Delta m = 0$$

where $\Delta m$ is the mass of the zone, $T$ is the temperature in the zone, $\Delta t$ is the time step, $q$ is the rate of energy production in the zone, including the nuclear reactions, the neutrino losses and the term $\sum_i \mu_i dX_i$, $L_{in}$, $L_{out}$ the radiative energy flux entering and leaving the zone respectively, and $\Delta s$ the change in the entropy of the zone during the time step $\Delta t$. Given the entropy and density profile, we can calculate the temperatures $T$, the energy production rates $q$ and radiative fluxes $L$, and we seek the entropy changes $\Delta s$ that solve the equation. For reasons of numerical stability, we seek a solution of the entropy equation such that the energy production rates $q$ are calculated explicitly, i.e. at the beginning of the time step, while the temperatures $T$ and the radiative fluxes $L$ are calculated implicitly, i.e. at the end of the time step.

After we calculate the temperatures $T$, the energy production rates $q$ and radiative fluxes $L$ for the entropy and density profiles at the beginning of the time step, we use an iterative method to get the change in the entropy profile as follows. We solve the entropy equation by means of a *Newton – Raphson* type method, where we linearize the equation by writing for each zone $k$:

(A-6)
$$\frac{T\Delta m_k}{\Delta t}\Delta s_k + L_{out} - L_{in} - q\Delta m_k - \frac{\partial L_{in}}{\partial s_{k-1}}\delta s_{k-1} + \left(\frac{T\Delta m_k}{\Delta t} + \frac{\partial L_{out}}{\partial s_k} - \frac{\partial L_{in}}{\partial s_k}\right)\delta s_k + \frac{\partial L_{out}}{\partial s_{k+1}}\delta s_{k+1} = 0$$

We now solve for the $\delta s$ that stand for the corrections to the entropy changes $\Delta s$, and put $\Delta s = \Delta s + \delta s$. Then we update the entropy profile by putting $s = s + \Delta s$, and subsequently update the temperature profile $T$ and the radiative fluxes $L$, again solve equation (A-6), and so on until we satisfy the convergence criteria for the change in the entropy, temperature and radiative flux profiles between iterations.

As mentioned above, convective regions are treated as thoroughly mixed isentropic zones, so for each convective region the entropy equation can be written as:

$$(A\text{-}7) \qquad \left( \int_C T(m)dm \right) dS = \left( \left( \int_C q(m)dm \right) + L_1 - L_2 \right) dt + \sum_i \left( \int_C \mu_i(m)dm \right) dX_i$$

This treatment of convection thus eliminates the need to calculate the convective luminosity, which is usually a complex expression. Furthermore, in solving the entropy equation, we can treat each convective region as a single *Lagrangian* zone obeying a difference equation quite similar to equation (A-5):

$$(A\text{-}8) \qquad \frac{\sum_i T_i \Delta m_i}{\Delta t} \Delta s + L_{out} - L_{in} - \sum_i q_i \Delta m_i = 0$$

Here the summations go over all the *Lagrangian* zones included in the convective region in question.

It is to be noted that the density profile is kept constant during this iterative solution process. There is a possibility to also change the density to fulfill hydrostatic equilibrium after each iteration, but we prefer not to do this, since the solution of equation (A-6) uses derivatives of the radiative flux, which are calculated at constant density, and if the densities were to be changed during the iterative process, the latter equation would be a much poorer approximation to equation (A-5), and thus convergence would be much more difficult.

## A.3. The treatment of convection

As has been previously mentioned, we employ a somewhat simplified method for the treatment of convection in the code. This treatment is based on the aforementioned assumption, that the time scale of the convective mixing is much shorter than the time scale of the nuclear reactions and radiative transport that govern the entropy equation. Under the above assumption, we treat each convective region as a fully mixed isentropic zone, which is actually equivalent to using a very large mixing length coefficient in the widely popular mixing length theory treatment.

The actual algorithm consists of the following four stages:

1.  Determine the boundaries of the convective regions, and mix the entropy and composition of each convective region accordingly.

2.  Make a "*gedanken"* step. This is a regular time step going through the procedure of solving the entropy equation as described above, only that the entire star is assumed to be radiative. It is needed because otherwise a *Lagrangian* zone once declared convective could not turn to be radiative, since the criteria for convection are such, that a region of *Lagrangian* zones having the same entropy and composition would always be convective.

3.  Determine new boundaries of the convective regions, and mix the entropy and composition of each convective region accordingly. This is done exactly as in stage 1, with the exception that a *Lagrangian* zone declared as radiative in stage 1 is not allowed to be convective, since the gedanken step is meant to allow convective regions to shrink and not to grow.

4. Restore the initial entropy, density and composition profiles (from prior to stage 1), in order to cancel the changes made by the *gedanken* step, and mix again the entropy and composition of each convective region according to the new convective region boundaries determined by the *gedanken* step.

The algorithm for the determination of the convective region boundaries works as follows:

1. Starting from the innermost *Lagrangian* zone upwards, we check each zone for stability against convection with its outer neighbor, until we find the first unstable pair. A new convective region is created, and the pair is marked as the inner and outer boundary of the convective region.

2. We mix the first zone with its outer neighbor. Mixing means that the composition is averaged, and a common value of the entropy is found, such that without changing the density of the individual zones, the total internal energy is conserved.

3. We go on checking the current outermost *Lagrangian* zone of the convective region for stability against convection with its outer neighboring *Lagrangian* zone. If they are unstable, we add the outer neighbor to the convective region, and again mix all the *Lagrangian* zones included in the convective region. If they are stable we stop the process, and fix the outer boundary of the convective region.

4. Since during the process of finding the outer boundary of the convective region, the mixing process might have changed the composition and the entropy, it is possible that under the new conditions the innermost *Lagrangian* zone we started from is now unstable against convection with its inner neighbor. Therefore we now go through a process similar to that in the previous stage, only going inwards from the inner boundary of the convective region, until we can fix the inner boundary.

5.  After we fixed the inner and outer boundaries of the convective region, we repeat stages 1 through 4 starting from the *Lagrangian* zone immediately above the outer boundary of the previous convective region, until we have checked the entire star.

6.  At last, we again check for all the convective regions we found, that the *Lagrangian* zones at their boundaries are stable against convection with their neighbors outside the convective region. This should be done due to the mixing, which can change the thermodynamic conditions and thus the stability against convection. If we find any instability, all the convective regions are canceled (but the changes done by the mixing are preserved), and the process is restarted from stage 1.

The checking of stability against convection follows the *Schwarzschild* criterion, which can be written as follows:

*(A-9)*
$$\left( \frac{\partial T}{\partial P} \right)_{star} < \left( \frac{\partial T}{\partial P} \right)_{s,x}$$

For two adjacent *Lagrangian* zones we can write:

*(A-10)*
$$\frac{T_2 - T_1}{P_2 - P_1} < \frac{T(P_2, s_1, x_1) - T_1}{P_2 - P_1}$$

Or, if we assume that zone 2 is above zone 1, then $P_2 < P_1$, and the criterion we check for is:

*(A-11)* $$T_2 - T(P_2, s_1, x_1) > 0$$

## *A.4. The equations of hydrostatic equilibrium*

Under our quasi-static assumption, we can assume that any change of the entropy profile will cause an immediate adjustment of the density profile, such that the velocity of each *Lagrangian* zone vanishes and the equation of hydrostatic equilibrium is obeyed. Thus, in principle, we should solve the equation of hydrostatic equilibrium each time we alter the entropy profile – both during solution of the entropy equation and during the solution for the convective regions' boundaries. Since such a practice is both time-consuming and proves to have no effect on the results, we only solve the equation of hydrostatic equilibrium at the end of the time step (after solving the entropy equation), and after setting the boundaries of the convective regions (before rechecking the stability of the boundaries – step 6 in the description of the algorithm).

The equation of hydrostatic equilibrium can be written for *Lagrangian* coordinates as:

*(A-12)* $$4\pi r^2 \frac{\partial P}{\partial m} + \frac{Gm}{r^2} = 0$$

Here $P$ is the pressure, $m$ the *Lagrangian* mass, $r$ the radius, and $G$ the constant of gravity.

We replace the above differential equation by the appropriate difference equation:

$$(A\text{-}13) \qquad 4\pi r_i^2 \frac{P_{i+1} - P_i}{\frac{1}{2}(dm_{i+1} + dm_i)} + \frac{Gm_i}{r_i^2} = 0$$

where $P_i$, $P_{i+1}$ are the values of the pressure in two consecutive *Lagrangian* zones, $dm_i$, $dm_{i+1}$ are their respective masses, $r_i$ is the radius at the boundary between the two zones, and $m_i$ is the total mass from the center of the star to the boundary between the two zones, and $G$ the constant of gravity.

This equation we solve by means of a *Newton – Raphson* type method, where we linearize the equation by writing for each zone $i$:

*(A-14)*

$$\frac{8\pi r_i^4}{(dm_{i+1} + dm_i)}\left[ P_{i+1} - P_i + Gm_i - \frac{\partial P_i}{\partial r_{i-1}}\delta r_{i-1} + \left( \frac{4(P_{i+1} - P_i)}{r_i} + \left( \frac{\partial P_{i+1}}{\partial r_i} - \frac{\partial P_i}{\partial r_i} \right) \right)\delta r_i + \frac{\partial P_{i+1}}{\partial r_{i+1}}\delta r_{i+1} \right] = 0$$

We solve for the $\delta r_i$, correct the radii $r_i$, and repeat until we fulfill the criteria for convergence.

A certain complication arises regarding the treatment of the outer boundary, i.e. the last *Lagrangian* zone of the star, for which we cannot solve equation (A-14) as it is, because the values with index $i+1$ have to be defined. After experimenting with several methods, which give very similar evolutionary behaviors, but some of which are prone to numerical problems, we chose a simple formula, which works well in all cases we dealt with, so far as the outer zones are small enough.

For the last zone we take the external pressure as:

$$(A\text{-}15) \qquad P_{last+1} = \frac{Gdm_{last}M_{tot}}{8\pi r_{last}^4}$$

This is exactly the pressure we would get at the *Lagrangian* point $M_{tot}$ - $\frac{1}{2}dm_i$, if we solved the equation for hydrostatic equilibrium (A-12) for a very small $dm_i$, while assuming $P(M_{tot})=0$.

### A.5.  The outer boundary

The treatment of the outer stellar boundary follows a method described in *Kippenhahn et al. 1967*. We assume a small outer layer (equal to the outer half of the last zone treated by the evolution code) of the envelope to have constant luminosity, and solve for a hydrostatic profile with such a luminosity and effective temperature that would give a fit of the bottom of the profile to the conditions at the middle of the last zone treated by the evolution code.